\documentstyle[epsfig]{elsart}

\newcommand{\be}{\begin{equation}}
\newcommand{\ee}{\end{equation}}

\newcommand{\bbf}{\bf}
\newcommand{\ssl}{\sl}

\newcommand{\f}{w}
\newcommand{\T}{\mbox{\bf T}}

\newcommand{\bea}{\begin{eqnarray}}
\newcommand{\eea}{\end{eqnarray}}

\begin{document}
\bibliographystyle{plain}

\begin{frontmatter}
\begin{flushright}
ZU-TH 17/98\\
YITP-98-53  \\
DTP-MSU/31-98\\
hep-th/9810070
\end{flushright}

\vspace{3 cm}

\title{Gravitating Non-Abelian Solitons and
Black Holes with Yang-Mills Fields}

\author[MSV]{Mikhail S. Volkov} and
\author[DVG]{Dmitri V. Gal'tsov}
\address[MSV]{Institute  for  Theoretical    Physics,
University of Z\"urich--Irchel, \\
Winterthurerstrasse 190, CH--8057 Z\"urich,
Switzerland.\\ 
e--mail: volkov@physik.unizh.ch}
\address[DVG]{Department of Theoretical Physics, Moscow State University,
119899 Moscow, Russia, and Yukawa Institute for Theoretical Physics,
University of Kyoto, Kyoto 606, Japan.
e--mail: galtsov@grg.phys.msu.su}

\begin{abstract}
We present a review of gravitating
particle-like and black hole solutions with
non-Abelian gauge fields.
The emphasis is given to the description of the structure
of  the solutions and to the connection
with the results of flat space soliton physics.
We describe the Bartnik-McKinnon solitons
and the non-Abelian black holes arising in
the Einstein-Yang-Mills theory,
and consider their various generalizations.
These include axially symmetric and slowly rotating configurations,
solutions with higher gauge groups, $\Lambda$-term, dilaton,
and higher curvature corrections.
The stability issue is discussed as well.
We also describe the gravitating generalizations for
flat space  monopoles, sphalerons, and Skyrmions.
\end{abstract}

\vspace{5 mm}
%published in {\it Physics Reports}, {\bf 319}, Numbers 1--2, 
%1--83, October 1999
%\noindent
%PACS numbers: 04.70.Bw

\end{frontmatter}
published in {\it Physics Reports}, {\bf 319}, Numbers 1--2, 
1--83, October 1999 
\newpage
\tableofcontents
\newpage

\section{Introduction}

Non-Abelian solitons play an important role in gauge theories
of elementary particle physics \cite{Coleman75,Actor79,Rajaraman82}.
In most cases, the
effects of  gravity can safely be neglected, perhaps except for
the very heavy magnetic monopoles in some Grand Unified Theories.
Besides, it has recently become clear that solitons  are equally
important also in string theory, where  gravity is essential.
A large number of solutions for such gravitating lumps have been obtained
\cite{Duff95}, however, almost all of them  are  {\em Abelian}.
At the same time, the gauge group in string theory is quite complicated,
which suggests that non-Abelian solutions can be important as well.
From the point of view of General Relativity (GR), the theory
of gravitating non-Abelian gauge fields can be regarded as the most
natural generalization of the Einstein-Maxwell (EM) theory.
It is therefore reasonable to study gravitating gauge fields
and, in particular, to check whether
the standard electrovacuum results have natural generalizations.
In view of this, the
aim of the present review is to discuss
non-Abelian gravity-coupled solitons and black holes.
The concepts of solitons and lumps refer in this text
to any particle-like solutions of a non-linear field theory.
Such solutions are asymptotically flat, topologically
trivial and globally stationary, although not necessarily stable.

The first example of gravitating non-Abelian solitons was discovered by
Bartnik and McKinnon (BK)
in the four-dimensional Einstein-Yang-Mills (EYM) theory
for the gauge group SU(2) \cite{Bartnik88}.
This example can be regarded as canonical in the sense that solutions
in other models of gravitating
non-Abelian gauge fields studied so far inevitably
share a number of characteristic common features with the BK particles.
Soon after the BK discovery it was realized that, apart from solitons,
the EYM model contains also non-Abelian black holes
\cite{Volkov89,Kunzle90,Bizon90}. As these manifestly
violate the non-hair conjecture, they have attracted much
attention and stimulated a broad search for other black hole solutions
in models of four-dimensional Einstein gravity with non-linear
field sources. The results obtained have led to
certain revisions of some of the basic concepts of black hole physics
based on the uniqueness and no-hair theorems. Specifically,
it has been found that the violation of the no-hair conjecture
is typical for gravitating non-Abelian gauge theories,
especially for those models which admit solitons in the flat spacetime
limit.

Let us recall  (see \cite{Heusler96} for a more detailed account)
that the existing classification of black holes in GR
is based on Hawking's strong rigidity theorem, stating
that a stationary black hole is either static
(with non-rotating horizon) or 
axisymmetric \cite{Hawking73}.
This theorem uses only fairly general assumptions like the weak
energy condition (although some of them are being critically
revised \cite{Chrusciel94,Chruschiel96}).
For vacuum and electrovacuum black holes with non-vanishing surface
gravity at the horizon, Israel's theorems \cite{Israel67,Israel68}
ensure that staticity implies
spherical symmetry. It follows then that the static
Schwarzschild and Reissner-Nordstr\"om (RN) solutions,
respectively, are unique.
For zero surface gravity, static electrovacuum black
holes belong to the Majumdar--Papapetrou family \cite{HeuslerMJP}.
In the stationary case, a chain
of uniqueness theorems asserts that the regularity
of the event horizon and the asymptotic behaviour specified by mass,
angular momentum and Coulomb charges determine the solutions completely.
As a result, stationary electrovacuum black holes are necessarily
axisymmetric and should belong to
the Kerr--Newman family \cite{Carter73,Mazur82,Bunting83}.
Apart from the vacuum and electrovacuum cases, uniqueness
has been established  for
supergravity models containing in four-dimensions sets of
scalars parameterizing coset spaces
and the corresponding multiplets of the U(1) fields
\cite{Breitenlohner88,Breitenlohner98}.

Another well-known statement, the no--hair conjecture,
claims that the only allowed characteristics
of stationary black holes are those associated with the Gauss law,
such as mass, angular momentum and electric
(magnetic) charges \cite{Ruffini71,Bekenstein96}.
Thus, for example, it follows that
black holes cannot support external scalar fields,
since there is no Gauss law for scalars. On the other hand,
black holes with a spin 3/2-field, say, are  allowed, since they carry
a conserved fermion charge given by the Gauss flux integral
\cite{Aichelburg83}.
The conjecture has been proven for non-interacting
boson \cite{Bekenstein72,Bekenstein72a,Bekenstein72b}
and fermion fields \cite{Hartle71,Teitelboim72,Teitelboim72a}
of various spins, as well as for some special non-linear matter models
\cite{Adler78,Bekenstein95,Mayo96,HeuslerSIG,Heusler95,Sudarsky95}.

Although the uniqueness and no-hair  theorems had
actually been proven only for special types of matter,
the appealing  simplicity of these assertions
inspired a widespread
belief in their possible general validity. However,
after the discovery of the EYM black holes it became
clear that this is not the case.
First, the EYM black holes possess a short-ranged
external non-Abelian gauge field
and are not uniquely specified
by their mass, angular momentum and conserved charges.
The no-hair conjecture is therefore violated.
Second, static EYM black holes with non-degenerate horizon
turn out to be not
necessarily spherically symmetric
\cite{Kleihaus97b,Kleihaus97d}. This shows that
Israel's theorem does not generalize to the non-Abelian case.
A similar phenomenon has also been observed in
the EYM-Higgs model \cite{Ridgway95b}.
Next, the perturbative considerations suggest
that non-static EYM black holes with rotating horizon
do not necessarily have non-zero angular momentum
\cite{Volkov97,Brodbeck97a}, which shows that the Abelian staticity
conjecture \cite{Carter87} does not straightforwardly apply either
\cite{Heusler93a,Sudarsky93}.
In addition, the Frobenius integrability conditions for two commuting
Killing vectors are not automatically fulfilled for self-gravitating
Yang--Mills fields \cite{Heusler93a,HeuslerHPA}, and so
the Ricci circularity condition is not guaranteed.
The standard Lewis--Papapetrou parameterization of a stationary
and axisymmetric metric used for the uniqueness arguments
can therefore be too narrow. However,
explicit examples of the circularity violation are not known yet.
Finally, the three-dimensional
reduction of the Yang--Mills action in the presence of a Killing symmetry
does not lead to the standard sigma--model structure
essential for the uniqueness proof \cite{HeuslerLIVE,Galtsov98}.

All this shows that a number of very important features of
electrovacuum black hole physics cease to exists in the EYM theory,
and this seems to happen generically for models with
non-Abelian gauge fields.
At the same time, what might seem surprising from the
traditional point of view,
finds a natural explanation in the context of flat space
soliton physics \cite{Coleman75,Actor79,Rajaraman82}.
For example,
the existence of hairy black holes  can often be directly inferred
from the existence of solitons in flat spacetime.
For small values of Newton's constant the implicit function theorem
ensures that flat space solitons admit weakly gravitating, globally regular
generalizations -- weak gravity can be treated perturbatively.
It turns out  that for a large class of matter models the same argument
can be used to show the existence of solutions in a small neighbourhood
of the point $r_h=0$ in parameter space, where $r_h$ is the event
horizon radius \cite{Kastor92}. This implies that
a weakly gravitating lump can be further generalized to replace
the regular origin by a small black hole. The result is a hairy
black hole whose radius is considerably smaller
than the size of the lump
surrounding it. One can argue in the spirit of the no-hair
conjecture that a small black hole cannot swallow up a soliton
which is larger than
the black hole itself.  However, a larger black hole must be able to
do this, and indeed, the radius of a hairy black hole inside a soliton
usually cannot exceed some maximal value.

In view of the arguments above, the most surprising fact is
the existence of the BK particles, which have no flat space
counterparts, and of their black hole analogues, whose radius
can be arbitrary. The pure EYM theory is characterized by the
fact that all fields are massless, and the solutions of the theory
can therefore be thought of as bound states of two non-linear massless
fields. As a result, the EYM solitons and black holes exhibit
a number of special features which
distinguish them from solutions of other
gravitating non-Abelian gauge models.
For example, the BK particles provide the only known example
of solitons, whether gravitating or not,
which admit regular, slowly rotating generalizations \cite{Brodbeck97}.
Solitons of other models, such as the t'Hooft-Polyakov monopoles, say,
do not rotate perturbatively \cite{Brodbeck97a,Heusler98}, which means
that they either  cannot rotate at all or their angular
momentum assumes discrete values.
The EYM black holes are uniquely distinguished by their
peculiar oscillatory behaviour in the
interior region \cite{Donets97,Breitenlohner97}, which
is reminiscent of some cosmological models \cite{Belinskii70}.

On the other hand, some features of the EYM solutions
are shared by those of other non-linear models.
The BK solitons, for example, exhibit
a remarkable similarity \cite{Galtsov91a} with
the  well-known sphaleron solution of the Weinberg-Salam model
\cite{Klinkhamer84}, which
allows one to call the BK particles EYM sphalerons.
(Let us remind to the reader that the term ``sphaleron"
refers to the static saddle point solution in a gauge field theory
with vacuum periodicity \cite{Jackiw76}.
Sphalerons are characterized by half-integer values of the
Chern-Simons number of the gauge field and
can be reduced to any of the nearest topological vacua
via a sequence of smooth deformations preserving the boundary
conditions \cite{Manton83,Klinkhamer84}.
Such objects are likely to be responsible for the
fermion number non-conservation at high energies
and/or temperatures \cite{Kuzmin85}).
Other properties of EYM solitons and black holes,
such as their nodal structure and discrete mass spectrum,
are generic and shared by practically all known solutions
in models with gravitating non-Abelian gauge fields.
Such solutions can be thought of as eigenstates of non-linear
eigenvalue problems, which accounts for the discreteness of some
of their parameters.

The list of the non-linear models with gravitating non-Abelian gauge fields,
investigated during recent years, contains, apart form the pure EYM theory,
also various generalizations. These include the dilaton,
higher curvature corrections, Higgs fields, and
a cosmological constant.
The Einstein-Skyrme model has also been studied in detail.
All these theories admit gravitating solitons and hairy
black holes. The pure EYM theory has been studied most of all,
some of its features being typical for all other models.
For this reason, the central part of our review
will be devoted to a description of the basic EYM solutions as well
as their direct generalizations. However, other important solutions,
such as the gravitating monopoles and Skyrmions, will also be duly
described. We will concentrate  only on the most
important results, but our list of references is quite complete.
Several review articles on the related subjects are available
\cite{Gibbons91,Jetzer92,Moss94,%
Bizon94,Maison96,HeuslerHPA,Bekenstein96,HeuslerLIVE},
but overlaps with the present text are small.

The plan of the paper is as follows. In Sec.2 we give the basic definitions,
briefly discuss symmetries of the non-Abelian gauge fields,
and derive the reduced two-dimensional Lagrangians for the spherically
symmetric models. Sec.3 is devoted to the BK solutions, their
sphaleron interpretation and various generalizations.
Basic properties of the EYM black holes are discussed in Sec.4.
Sec.5 and Sec.6 contain a discussion of
stability of EYM solitons and black holes and
an analysis of their rotational excitations.
Gravitating solutions in models admitting solitons in flat space are
discussed in Sec.7. These include the t'Hooft-Polyakov monopoles,
the Yang-Mills-Higgs sphalerons, and the Skyrmions.
Some other important results not mentioned in the main text are
briefly discussed in Sec.8.
Most of the solutions described below are known only numerically,
although their existence has been established rigorously in some cases.
For this reason we present a number of figures and tables
describing the numerical results. All of them have been produced
with the  ``shooting to a fitting
point'' numerical procedure, described in \cite{Press92}.

\section{General formalism}

In this chapter the field-theoretical models, whose solutions will be
considered in the next chapters,  are introduced. These are the EYM theory
and its generalizations, including the dilaton, the Higgs fields,
as well as the Einstein-Skyrme model. We briefly discuss symmetry
conditions for gauge fields and derive the effective
two-dimensional Lagrangians for all models in the spherically
symmetric case.

\subsection{Einstein-Yang-Mills theory}
\setcounter{equation}{0}
The basic model which will be discussed below is the 
four-dimensional EYM  system for a compact, semi-simple
gauge group ${\cal G}$. The Lie algebra of ${\cal G}$  is characterized
by the commutation relations
$[\T_{a},\T_{b}]=i f_{abc} \T_{c}$
$(a,b,c=1,\ldots, {\rm dim}({\cal G}))$.
The basic elements of the model are $({\cal M},g_{\mu\nu},A)$, where
${\cal M}$ is the spacetime manifold with metric $g_{\mu\nu}$, and
the Lie-algebra-valued one-form is
$A\equiv A_{\mu}dx^{\mu}\equiv \T_{a} A^{a}_{\mu}dx^{\mu}$. 
We choose the standard action
\be                                              \label{1.1}
S_{\rm EYM}=
\int \left\{- \frac{1}{16\pi G}\,R -
\frac{1}{4{\rm K}{\sl g}^2}\,
 {\rm tr}\, F_{\mu\nu}F^{\mu\nu}\right\}
\sqrt{-g}\, d^4 x\, ,
\ee
where ${\sl g}$ is the gauge coupling constant,
$F_{\mu\nu}=\partial_{\mu}A_{\nu}- \partial_{\nu}A_{\mu}-
i[A_{\mu},A_{\nu}]\ \equiv\  \T_{a}F^{a}_{\mu\nu}\, $, and $K>0$
is a normalization factor: ${\rm tr}\,
(\T_{a}\T_{b})={\rm K}\,\delta_{ab}$.
For ${\cal G}$=SU(2) we choose
$\T_{a}=\frac{1}{2}\tau_{a},\; {\rm K}=1/2,\;
f_{abc}=\varepsilon_{abc}$
 with $\tau_{a}$ being the Pauli matrices.
We adopt the metric signature $(+---)$,
the spacetime covariant derivative will be denoted by $\nabla$, and
the Riemann and Ricci tensors are
$R^{\alpha}_{\ \beta\mu\nu}=
\partial_{\mu}\Gamma^{\alpha}_{\beta\nu}-\ldots\, $
and $R_{\mu\nu}=R^{\alpha}_{\ \mu\alpha\nu}$,
respectively.           
The gravitational constant
$G$ is the only dimensionful quantity in the action
(the units $\hbar=c=1$
are understood).

Apart from the general spacetime diffeomorphisms,
the EYM action is invariant
with respect to the gauge transformations of the gauge field  $A$:
\be                                              \label{1.5}
A_{\mu}\rightarrow {\rm U}(A_{\mu}+i\partial_{\mu}){\rm U}^{-1}\, ,
\ \ \ \ \
F_{\mu\nu}\rightarrow{\rm U}F_{\mu\nu}{\rm U}^{-1}\, ,\ \ \ \ \ \
g_{\mu\nu}\rightarrow g_{\mu\nu},
\ee
where ${\rm U}(x)\in \cal G$.
In addition, the Yang-Mills (YM) part of the action displays the conformal
symmetry 
\be                                              \label{1.6}
g_{\mu\nu}\rightarrow\Omega(x)\, g_{\mu\nu}\, , \quad
A_{\mu}\rightarrow A_{\mu}, \quad
F_{\mu\nu}\rightarrow F_{\mu\nu},
\ee
which is not, however, a symmetry of the Einstein-Hilbert action.
The variation of the action (\ref{1.1}) gives the Einstein equations
\be                                                 \label{1.7}
R_{\mu\nu}-\frac{1}{2}\,R\,g_{\mu\nu}=8\pi G\, T_{\mu\nu}
\ee
with the gauge-invariant YM  stress-energy tensor
\be                                               \label{1.8}
T_{\mu\nu}=\frac{1}{{\rm K}g^2}\, {\rm tr}\left(
-F_{\mu\sigma}F_{\nu}^{\ \sigma}+\frac{1}{4}\, g_{\mu\nu}\,
F_{\alpha\beta}F^{\alpha\beta}\right)\, ,
\ee
and the YM equations
\be                                              \label{1.9}
D_{\mu}F^{\mu\nu}=0\, ,
\ee
where the gauge covariant derivative is
$D_{\mu} \equiv \nabla_{\mu}-i[A_{\mu},\ \ ]$.
Owing to the conformal invariance the  stress tensor is traceless,
$T^\mu_\mu=0$.
 As a consequence of the gauge invariance, the dual field tensor
$\ast\!F_{\mu\nu}\equiv(1/2) \sqrt{-g}\,
\epsilon_{\mu\nu\rho\sigma}F^{\rho\sigma}\, \,
 (\epsilon^{0123}=1$) satisfies the Bianchi identities
\be                                              \label{1.12}
D_{\mu}\ast\!F^{\mu\nu}\equiv 0\, .
\ee

Asymptotically flat solutions in the theory have well-defined
ADM mass and angular momentum \cite{Arnowitt62,Abbott82}.
Similarly, one can define conserved
Lie-algebra-valued  electric and magnetic
charges \cite{Yang54,Abbott82a}. Note that the
straightforward definition \cite{Yang54}
\be                                             \label{1.12a}
Q=\frac{1}{4\pi}\oint_{S^{2}_{\infty}}\ast\!F\, ,\ \ \ \ \ \ \
P=\frac{1}{4\pi}\oint_{S^{2}_{\infty}} F\, ,
\ee
where the integration is over a two-sphere at
spatial infinity, is not gauge invariant in
the non-Abelian case \cite{Schlieder81,Chrusciel87},
and hence requires the gauge fixing.
This reminds of a similar situation
for the ADM mass, which was originally defined only for
distinguished coordinate systems \cite{Arnowitt62}.
On the other hand, in the presence
of a timelike Killing symmetry the ADM mass can be covariantly expressed
by the Komar formula \cite{Komar59}.  For the YM charges there exists
a similar gauge invariant construction \cite{Loos66} using
the Lie-algebra valued ``Killing scalar" $K$, where $D_\mu K=0$.
Now, it is important that these invariant definitions can
be generalized to the case where the configuration has no symmetries at all.
Indeed, the symmetries always exist in the asymptotic region,
and these can be used in order to put the surface integrals
into the covariant form.
As a result one can define the ADM mass \cite{Abbott82}
and conserved gauge charges \cite{Abbott82a} for an arbitrary isolated system
in the completely covariant and gauge-invariant fashion.
For example, for the gauge field given by
$A=\overline{A}+a$, where
$\overline{A}=i\overline{U}d\overline{U}^{-1}$,
and $a\to 0$ as $r\to\infty$, one can define
${\cal F}_{\mu\nu}=\overline{D}_{\mu}a_{\nu}-\overline{D}_{\nu}a_{\mu}$.
Here $\overline{D}$ is the covariant derivative with respect to
$\overline{A}$. The conserved and gauge
invariant (up to global gauge rotations)
charges $Q_a$ and $P_a$ are then given by (\ref{1.12a}) with $F$ replaced
by ${\rm tr}({\cal F}\,\overline{U}\T_a\overline{U}^{-1})$
\cite{Abbott82a}.

\subsection{Spacetime symmetries of gauge fields}

The issue of symmetry for non-Abelian gauge fields has been
extensively studied in
\cite{Romanov77,Bergmann78,Forgacs80,Harnad79,Harnad80,Harnad80a,%
Manton81,Gu81,Henneaux82,Coqueraux83,Basler85,Kubyshin89,Kubyshin89a,%
Bartnik90,Kunzle91,Brodbeck93,Brodbeck94,Bartnik97}.
The symmetry conditions for gauge fields
in the infinitesimal language were
first formulated in \cite{Bergmann78,Forgacs80}.
An EYM field configuration will respect spacetime isometries
generated by a set of Killing vectors $\xi_m$ if the metric
is invariant,
\be                                                    \label{2.1}
{\cal L}_{\xi_{m}}\, g_{\mu\nu}=0,
\ee
under the action of the isometry group,
while the corresponding change in the gauge field,
$A_{\mu}\rightarrow A_{\mu}-\epsilon^m\, {\cal L}_{\xi_{m}}\, A_{\mu}$,
can be compensated by a suitable gauge transformation,
$A_{\mu}\rightarrow A_{\mu}+\epsilon^m\, D_{\mu}W_m$, such that
\cite{Bergmann78,Forgacs80}
\be                                                   \label{2.4}
{\cal L}_{\xi_{m}}\, A_{\mu}=D_{\mu}W_{m}.
\ee
Here ${\cal L}$ stands for the Lie derivative.
In order to actually find $A_\mu$  for a given set of Killing vectors
$\xi_{m}$ the procedure is first to solve
the integrability conditions for (\ref{2.4}),
which gives the $W_{m}$'s, and then to solve (\ref{2.4}) for  $A_{\mu}$
\cite{Forgacs80,Basler85}.

Throughout this article we will assume the $2+2$ block-diagonal
form of the metric
\be \label{2+2}
ds^2=g_{\alpha \beta}dx^\alpha dx^\beta +h_{MN} dx^M dx^N,
\ee
where $x^\alpha\equiv\{x^0,x^1\}$, $x^M\equiv\{x^2,x^3\}$, and
$g_{\alpha \beta}$ depend only on $x^\alpha$.
In the case of spherical symmetry
the Killing vectors $\xi_m$ generate the so(3) algebra,
and the most general solution to (\ref{2.1})
can be parameterized as
\be \label{2.2}
ds^2=g_{\alpha\beta}dx^\alpha dx^\beta-
R^2\, (d\vartheta^2 + \sin^2\vartheta d\varphi^2),
\ee
where $R$ depends on
$x^\alpha\equiv \{t,r\}$.
The corresponding most general solution to Eq.(\ref{2.4})
for the gauge group SU(2)
is sometimes called Witten's ansatz
\cite{Dashen74,Witten77,Forgacs80,Bartnik90}:
\be                                                      \label{2.5}
A=a \, \T_r +i(1-{\rm Re}\;w)\,[\T_r, \,d\T_r]
+{\rm Im}\; w\, d\T_r\, .
\ee
Here the real one-form $a= a_\alpha dx^\alpha\equiv
a_0 dt +a_r dr$  and the complex
scalar $w$ depend only on $x^\alpha$.
The position-dependent gauge-group generators
\be                                                  \label{2.6}
\T_r=n^a\T_a,\ \
\T_{\vartheta}=\partial_{\vartheta}\T_r, \ \
\T_{\varphi}=
\frac{1}{\sin\vartheta}\, \partial_{\varphi}\T_r,
\ee
where
$n^a=(\sin\vartheta\cos\varphi,\sin\vartheta\sin\varphi,\cos\vartheta)$,
obey the standard commutation relations, such that
 $[\T_r,\T_{\vartheta}]=i\T_{\varphi}$.
 The ansatz (\ref{2.5})
is invariant under the U(1) gauge transformations
generated by U$=\exp(i\, \beta(t,r)\, \T_r)$, under which
\be                                                \label{2.7}
a_\alpha\rightarrow a_\alpha+\partial_\alpha {\beta},\quad
   w\rightarrow e^{ i\, \beta} w,
\ee
i.e., $a_\alpha$ and $w$ transform as a two-dimensional Abelian vector
and a complex scalar, respectively.
The four independent real amplitudes in (\ref{2.5}) contain
therefore one pure gauge degree of freedom.
One can choose $|w|$ and
$\Omega_\alpha=a_\alpha - \partial_\alpha(\arg \,w)$
as the three gauge-invariant combinations.
Under parity transformation,
$\vartheta\rightarrow\pi-\vartheta$,
$\varphi\rightarrow\pi+\varphi$, one has
$\T_r\rightarrow -\T_r$,
$\T_{\vartheta}\rightarrow \T_{\vartheta}$,
$\T_{\varphi}\rightarrow -\T_{\varphi}$.
As a result, the effect of parity on (\ref{2.5}) is
\be                                                   \label{2.8}
{\rm P}:\ \ \ a_\alpha\rightarrow\ -a_\alpha,\quad
           w\rightarrow\  w^* ,
\ee
and hence among the  four real amplitudes  one is
parity-even, and three are parity-odd.

After the  gauge transformation
with U=$\exp(i\, \T_2\, \vartheta)\exp(i\, \T_3\, \varphi) \,$
\cite{Arafune75},
an equivalent form of the ansatz (\ref{2.5}) in terms of the constant
group  generators $\T_a$ can be obtained:
\be                                                      \label{2.9}
A=a \, \T_3 +
{\rm Im}\; (w \T_+)\, d\vartheta-{\rm Re}\;
(w \T_+)\,\sin\vartheta\, d\varphi
+\T_3\cos\vartheta\, d\varphi,
\ee
where $\T_+=\T_1+i\T_2$.
The residual gauge freedom (\ref{2.7}) is now generated by
U$=\exp(i\, \beta (t,r)\,  \T_3)$.
The expression (\ref{2.9}) is sometimes
easier to use than  (\ref{2.5}), but
it contains the Dirac string singularity.
Another frequently used gauge is $w=w^*$, in which case
 the gauge field admits  the following useful representation:
\be                                                           \label{2.12}
A= a\, \T_r +
i\, \frac{1-\f}{2}\, {\rm  {U}}\, d\, {\rm  {U}}^{-1},
\ \ \ {\rm where}\ \ \
{\rm  {U}}=\exp(i\pi \T_r).
\ee
The corresponding field strength reads
\bea
F= &-&\T_r\, da\,-i\, dw \wedge [\T_r,d\T_r]    \nonumber \\
&-&w\,a\wedge d\T_r    
+ \T_r\,(w^2-1)\, d\vartheta\,\wedge \sin\vartheta\, d\varphi. \label{2.13}
\eea
Note that the components $F_{\alpha\varphi}$ and $F_{\alpha\theta }$
do not vanish.

 Unlike the situation in the Abelian case,
in order to achieve the temporal gauge
condition $a_0=0$ one has to make
a gauge transformation which renders the whole
configuration time-dependent. Therefore,
for static fields one cannot
set $a_0=0$, unless $\Omega_0=a_0 - (\arg \,w)^{\displaystyle .}=0$.
The second gauge-invariant combination,
 $\Omega_r=a_r - (\arg \,w)'$,
vanishes in the static case by virtue of the YM equations.
As a result, one can always reduce
$a_r$ to zero by a gauge transformation, and hence the
most general static,
spherically symmetric SU(2) YM fields can be parameterized by two real
functions: $a_0$ and $w$%                             
\footnote{In systems with a Higgs field the YM equations
do not always imply that $a_r$ vanishes in the static case.
The most general  spherically symmetric YM fields
include then $a_0$, $a_r$, and $\f=\f^\ast$.}.

In the static, purely magnetic case the SU(2) ansatz (\ref{2.9})
with $a={\rm Im}\,w=0$ can be generalized to the gauge
group SU(N) \cite{Wilkinson77,Kunzle91,Bartnik97}.
Such a generalization includes
instead of one function $\f$, $N-1$ independent real amplitudes
$\f_j(r)$, $j=1\ldots N-1$. The gauge field potential $A$ is
given by a Hermitean $N\times N$ matrix, whose non-vanishing
matrix element are
\bea                                                     
A_{j,j}=\frac12\,(N-2j+1)\cos\vartheta\, d\varphi,
\ \ \ \ j=1\ldots N,\ \ \                                  \nonumber \\
A_{j,j+1}=(A_{j+1,j})^\ast=\frac12\,\f_j\,\Theta,
\ \ \ \ j=1\ldots N-1.                                    \label{2:12a}
\eea
Here $\Theta=\sqrt{j(N-j)}\,(i\,d\vartheta+\sin\vartheta\, d\varphi)$
and the asterisk denotes complex conjugation.

In the static, axially symmetric case, the metric is given by%
\footnote{This metric fulfills the circularity condition and gives rise
to non-trivial EYM solutions. It is unclear, however, whether
other, non-circular solutions can exist as well.}
\be                                              \label{2:axial}
ds^2=e^\delta dt^2-e^\zeta d\varphi^2-e^\mu (d\rho^2+dz^2),
\ee
where $\delta$, $\zeta$, and $\mu$ depend on
$x^\alpha\equiv\{\rho,z\}$.
The purely magnetic SU(2) gauge field reads \cite{Rebbi80}
\be                                               \label{2:axial1}
A=a\,\T_{\varphi}+\{{\rm Re}\,w\,\T_{\rho}
+({\rm Im}\,w-\nu)\,\T_3\}d\varphi
\, ,
\ee
where the one-form $a=a_\rho d\rho+a_z dz$  and the complex scalar $w$
depend on  $\rho$ and $z$.
The group generators are given by
\be                                                      \label{2:axial2}
\T_\rho=
 \cos \nu\varphi\, \T_1+
 \sin \nu\varphi\, \T_2\, ,\quad
  \T_{\varphi}=\nu^{-1}
 \partial_{\varphi}\T_\rho\, ,
\ee
with integer $\nu$.
The field (\ref{2:axial1}) also has the residual U(1) invariance expressed
in the form (\ref{2.7}) and generated by
U$=\exp(i\, \beta(\rho,z)\, \T_\varphi)$.

Note that the ansatz in Eq.(\ref{2.9}) can be
generalized to describe spherically symmetric YM fields with
arbitrary winding numbers:
\be                                                     \label{3:8:1a}
A=a \, \T_3 +
{\rm Im}\; (w \T_+)\, d\vartheta-\nu\,{\rm Re}\;
(w \T_+)\,\sin\vartheta\, d\varphi
+\nu\,\T_3\cos\vartheta\, d\varphi.
\ee
Here an integer $\nu$ has the meaning of the Chern number of the U(1)
bundle \cite{Bartnik90}. However,
for $\nu\neq 1$ the YM equations impose the following condition
for non-vacuum fields: $\f=0$. This implies that all configurations
with $\nu\neq 1$ are {\em embedded Abelian}.
Since we are interested in non-Abelian solutions,
for which $\f$ does not vanish identically,
we shall always assume that  $\nu=1$ in (\ref{3:8:1a})%
\footnote{The change $\nu\to-\nu$ in (\ref{3:8:1a}) can be achieved
by a gauge transformation.}.

\subsection{Dimensional reduction and scalar fields}

We shall mainly be considering below spherically symmetric systems.
It is convenient to derive the field equations in the spherically
symmetric case by varying the effective two-dimensional action
(the Hamiltonian approach to the problem
was discussed in  \cite{Cordero76,Benguria77}).
Let us insert into the four-dimensional (4D) EYM action (\ref{1.1})
the ansatz (\ref{2.5}) for the gauge field
and the metric parameterized as
\be                                                     \label{2:32}
ds^2=
\sigma^2 N(dt+\alpha dr)^2-{1\over N}\, dr^2
-R^2\, (d\vartheta^2+\sin^2\vartheta\varphi^2) ,
\ee
with $\sigma$, $N$, $\alpha$, and $R$ depending only
on $x^\alpha\equiv\{ t,r\}$.
Integrating over the angles and dropping the total derivative,
the result is the two-dimensional (2D) effective action \cite{Brodbeck96c}
\be                                       \label{2:32a}
S=\int\left\{
m'-\alpha\dot{m}
+L_m \right\}\sigma\, d^2x,
\ee
where $m$ is defined by the relation
\be                                         \label{G}
(\partial R)^2\equiv\partial_\alpha R\,\partial^\alpha R
=\frac{2\kappa m}{R}-1.
\ee
From now on the coordinates $x^\alpha$ and all other 2D quantities
are assumed to be dimensionless, after the rescaling
$x^\alpha\rightarrow x^\alpha/{\rm L}$, where ${\rm L}$ is
a length scale. The dimensionless gravitational coupling constant
$\kappa$ in (\ref{G}) is proportional to $G/{\rm L}^2$.
For the scale-invariant YM theory ${\rm L}$ is
arbitrary. It is convenient to choose ${\rm L}=\sqrt{4\pi G}/{\sl g}$,
in which case $\kappa=1$.
The matter field Lagrangian $L_m$ in (\ref{2:32a}) then reads
\be                                       \label{2:22}
L_{\rm YM}=-\frac{R^2}{4}\, f_{\alpha\beta} f^{\alpha\beta}+
|Dw|^2
-\frac{1}{2R^2}\,(|w|^2-1)^2,
\ee
where $f_{\alpha\beta}=\partial_\alpha a_\beta -\partial_\beta a_\alpha$
and  $D_\alpha=\partial_\alpha-ia_\alpha$
are the 2D Abelian field strength and the gauge covariant derivative,
respectively. One can see that the residual
U(1) gauge symmetry (\ref{2.7}) arises naturally at the level
of the reduced action.

Note that, since the theory under consideration
is purely classical, the EYM length scale ${\rm L}$ and the corresponding
mass scale, M=$\kappa {\rm L}/G$, do not contain Planck's constant.
Restoring for a moment in the formulas the speed of light, $c$, one has
\be                                           \label{units}
{\rm L}=\sqrt{\frac{4\pi G}{c^4 {\sl g}^2}},\ \ \
{\rm M}=\sqrt{\frac{4\pi}{{\sl g}^2 G}}.
\ee
It is worth noting that,
up to the replacement $\sqrt{4\pi}/{\sl g}\to e$,
this corresponds to the classical system
of units based on $G$, $c$, and the charge $e$
that had been introduced in \cite{Johnson1881}
before Planck's units were discovered.
Dividing and multiplying by $\sqrt{\hbar c}$,
one obtains ${\rm L}=\sqrt{\alpha}\, {\rm L}_{\rm Pl}$ and
M=$\sqrt{\alpha}\, {\rm M}_{\rm Pl}$,
where ${\rm L}_{\rm Pl}$ and ${\rm M}_{\rm Pl}$ are Planck's length and
Planck's mass, respectively, and $\alpha=4\pi/\hbar c{\sl g}^2$.
Since one can expect the value of $\sqrt{\alpha}$ not to be
very far from unity, this shows that the
units (\ref{units}) are actually closely related to Planck's units.

{\bf a)}
Let us consider generalizations of the EYM theory with
additional scalar fields.
The first one is the EYM-dilaton (EYMD) model with the 4D matter action
\be                                              \label{2:23}
S_{\rm YMD}= \int \left\{\frac{1}{8\pi G}\,(\nabla\phi)^2
 -  \frac{1}{4{\rm K}{\sl g}^2}\,
{\rm e}^{2\gamma\phi}\,{\rm tr} \,F_{\mu\nu}F^{\mu\nu}\right\}
\sqrt{-g}\, d^4 x\, ,
\ee
where $\gamma$ is a parameter.  In the spherically symmetric case, with
$\phi=\phi(t,r)$, the 4D EYMD action reduces to the 2D form (\ref{2:32a})
with the matter Lagrangian $L_m$ given by
\be                                             \label{2:24}
L_{\rm YMD}=\frac{R^2}{2}\, (\partial\phi)^2+
{\rm e}^{2\gamma\phi} L_{\rm YM},
\ee
where $L_{\rm YM}$ is specified by (\ref{2:22}).
The length scale is now fixed, ${\rm L}=\sqrt{4\pi G}/{\sl g}$,
and $\kappa=1$.

{\bf b)}
Next, consider the EYM-Higgs (EYMH) model with the
Higgs field $\Phi\equiv\Phi^a\T_a$ in the adjoint representation of SU(2).
The 4D matter action is the sum of the YM term and
\be                                              \label{2:25}
S_{\rm H}=
\int \left(
\frac12\,{\rm tr}\,D_\mu\Phi\, D^\mu\Phi-\frac{\lambda}{4}\,
(\Phi^a\Phi^a-v^2)^2\right)
\sqrt{-g}\, d^4 x\, ,
\ee
where $D_\mu\Phi=\partial_\mu\Phi-i[A_\mu,\Phi]$.
The spherically symmetric Higgs field is
\be                                             \label{2:25a}
\Phi=v\phi\T_r,
\ee
$\phi=\phi(t,r)$ being a real scalar.
This gives the 2D matter Lagrangian                               
in (\ref{2:32a})                          
\be                                     \label{2:26}
L_{m}=L_{\rm YM}+\frac{R^2}{2}\,\,(\partial\phi)^2
-|w|^2\phi^2-\,\frac{\epsilon^2}{4}\,R^2\,\,(\phi^2-1)^2 .
\ee
Here the length scale is ${\rm L}=1/M_{\rm W}$,
$\epsilon=M_{\rm H}/\sqrt{2}M_{\rm W}$,
where $M_{\rm W}={\sl g}v$ and $M_{\rm H}=\sqrt{2\lambda}v$
are the vector boson mass and the Higgs boson mass, respectively.
The gravitational coupling constant is
$\kappa=4\pi G v^2$.

{\bf c)}
For the EYMH model with a complex
Higgs field $\Phi$ in the fundamental representation of SU(2)
the 4D Higgs field action is
\be                                              \label{2:27}
S_{\rm H}=
\int \left(
(D_\mu\Phi)^\dagger\, D^\mu\Phi-\frac{\lambda}{4}\,
(\Phi^\dagger\Phi-v^2)^2\right)
\sqrt{-g}\, d^4 x\, ,
\ee
where $D_\mu\Phi=\partial_\mu\Phi-iA_\mu \Phi$.
The spherically symmetric Higgs field is given by
\be                   \label{2:27a}
\Phi=v\,\phi\,\exp(i\xi\,\T_r)|a\rangle
\ee
  where
$\phi=\phi(t,r)$ and $\xi=\xi(t,r)$ are real scalars,
while $|a\rangle$ is a  constant unit vector of the 2-dimensional
complex vector space, $\langle a|a\rangle=1$.
The corresponding 2D matter field Lagrangian is
\be                                     \label{2:28}
L_m=L_{\rm YM}+R^2|Dh|^2
-\frac12\,\phi^2|w-{\rm e}^{i\xi}|^2
-\frac{\epsilon^2}{4}\,R^2\,\,(\phi^2-1)^2 ,
\ee
where $h=\phi\exp(i\xi/2)$ and
$D_{\alpha}=\partial_\alpha-i a_{\alpha}/2$,
the other parameters being
the same as in the triplet case.

{\bf d)}
Finally, we shall consider the SU(2) Skyrme model. The 4D
matter action is
\be                                            \label{2:29}
S_{\rm Sk}= \int {\rm tr}\,\left\{-\frac{f^2}{4}
A_\mu A^\mu+
\frac{1}{32e^2}
 \left( F_{\mu\nu}F^{\mu\nu}\right)\right\}
\sqrt{-g}\, d^4 x\, ,
\ee
where $A_\mu=U^\dagger\partial_\mu U$ and $F_{\mu\nu}=[A_\mu,A_\nu]$.
The spherically symmetric chiral field is specified by
\be                                     \label{2:30}
{\rm U}=\exp(2i\chi(t,r)\T_r),
\ee
which leads to the 2D Lagrangian
\be                                      \label{2:31}
L_{m}=\left(\frac{R^2}{2}\,+\sin^2\chi\right)(\partial\chi)^2
-\left(R^2\,+\frac12\sin^2\chi\right)\,\frac{\sin^2\chi}{R^2}.
\ee
The length scale is ${\rm L}=1/ef$, the gravitational coupling
constant being $\kappa=4\pi Gf^2$.

The Einstein-matter coupled field equations are obtained
by varying (\ref{2:32a}) with respect to $N$, $\sigma$, $\alpha$,
and the matter variables in $L_m$. After varying one can set
the shift function $\alpha$ to zero, but one cannot do this
before varying, since for time-dependent fields
one of the Einstein equations would be lost \cite{Brodbeck96c}.
In the static case, on the
other hand,  one can get rid of $\alpha$ from the very beginning.
If the Schwarzschild gauge condition $R=r$ can be imposed, which
is often the case, the
Einstein equations for static fields look especially simple.
Note that the 4D Einstein equations read in the dimensionless notation
\be
G_{\mu\nu}=2\kappa\, T_{\mu\nu}.
\ee
For all matter models described above the matter Lagrangian
in the static, purely magnetic case reduces to
\be                                                  \label{2:100}
L_m=-(NK+U),
\ee
where $K$ and $U$ depend only on the matter variables, and
$N=1-2\kappa m/r$. As a result, varying the reduced action (\ref{2:32a})
with respect to $m$ and $\sigma$ gives the
independent Einstein equations
\be                                            \label{2:101}
m'=NK+U\equiv r^2T^{0}_{0},
\quad\quad (\ln\sigma)'=2\kappa K\equiv\kappa\,\frac{r}{N}\,
(T^{0}_{0}-T^{r}_{r}),
\ee
from which the components of the
(dimensionless) stress tensor can be read off.
The matter field equation are obtained by varying $\sigma L_m$.
The dimensionless ADM mass is given by $m(\infty)$.
The dimensionful mass is $(\kappa {\rm L}/G)\,m(\infty)$.

We shall consider below time-dependent problems only for the EYM fields,
in which case  the full system of  equations
following from  (\ref{2:32a}) and (\ref{2:22}) reads
\be
\partial_{\alpha} (r^{2}\sigma f^{\alpha\beta})=
2\sigma\,{\rm  Im}\,(\f D^{\beta}\f)^{\ast},                \label{2:33}
\ee
\be
D_{\alpha}(\sigma D^{\alpha}\f)=
\frac{\sigma}{r^{2}}\,(|\f|^{2}-1)\f,                       \label{2:34}
\ee
\be
(\ln\sigma)' = \frac{2}{r} \left(\frac{1}{N^{2}\sigma^{2}}
|D_{0}\f|^{2}+|D_{r}\f|^{2}\right),                         \label{2:35}
\ee
\be
\dot{m}=2N\,{\rm Re}\,D_{0}\f\,(D_{r}\f)^{\ast},            \label{2:36}
\ee
\be
m'=-\frac{r^{2}}{4}f_{\alpha\beta}f^{\alpha\beta} +
\frac{1}{N\sigma^{2}}|D_{0}\f|^{2}+N|D_{r}\f|^{2}+
\frac{1}{2r^{2}}\,(|\f|^{2}-1)^{2}                          \label{2:37}
\ee
with the notation of (\ref{2:22}) and
the asterisk denoting complex conjugation.

\subsection{Static EYM fields. Embedded Abelian solutions}

For static EYM fields the results of
this chapter can be summarized as follows.
The most general spherically symmetric SU(2) gauge field
can be parameterized as
\be                                                      \label{2.40}
A=a_0\,\, \T_3\,dt +
\f\, (\T_{2}\, d\vartheta-\T_{1}\,
\sin\vartheta\, d\varphi)
+\T_3\cos\vartheta\, d\varphi,
\ee
and the spacetime metric is
\be                                                  \label{2.41}
ds^2=\sigma^2 N dt^2-\frac1N\,dr^2
-r^2\, (d\vartheta^2+\sin^2\vartheta\, d\varphi^2).
\ee
The amplitudes $a_0$, $\f=\f^\ast$, $m$, $N\equiv 1-2m/r$,
and $\sigma$ depend on $r$.
The EYM equations (\ref{2:33})--(\ref{2:37}) assume the form
\begin{eqnarray}
\left(\frac{r^2}{\sigma}\, a_0'\right)^{\prime}&=&
\frac{2}{\sigma N}\, \f^2\, a_0,                    \label{2.42}  \\
\left(\sigma N \f'\right)^{\prime}&=&\frac{\sigma}{r^2}\, \f(\f^2-1)-
\frac{a_0^2}{\sigma N}\, \f,                        \label{2.43}  \\
m'&=&\frac{r^2}{2\sigma^2}\, a_0^{\prime\, 2}+
\frac{w^2 a_0^2}{N\sigma^2}+
N\f'^2+\frac{1}{2r^2}(\f^2-1)^2\, ,                 \label{2.44}  \\
\frac{\sigma'}{\sigma}&=&  
\frac{ra_{0}^{\prime 2}}{N\sigma^2}+
\frac{2}{r}\, \f'^2 .       \label{2.45}
\end{eqnarray}
In the next two chapters we shall study solutions
to these equations. The simplest one is
\be                                                     \label{2.46}
a_0=0,\ \ \ \f=\pm 1,\ \  \ m=M,\ \ \ \sigma=1,
\ee
which describes the Schwarzschild metric and a
pure gauge YM field.
The simplest solution with a non-trivial gauge field
describes colored black holes
\cite{Bais75,Cho75,Wang75,Perry77,Kamata82}
\be                                                     \label{2.47}
a_0=a_0(\infty)+\frac{Q}{r},\ \ \f=0,\ \
N=1-\frac{2M}{r}+\frac{Q^2+1}{r^2},\ \ \sigma=1.
\ee
The metric is RN with the electric charge
$Q$ and unit magnetic charge.
This solution is {\em embedded Abelian} in the sense that the
SU(2) gauge field potential (\ref{2.40}) is a product of the potential
of the U(1) dyon, ${\cal A}=a_0\,dt+\cos\vartheta d\varphi$,
and a constant matrix.
The appearance of such a solution
is not surprising: since the SU(2) gauge group contains U(1)
as a subgroup, the set of solutions in the EYM theory
includes, in particular, the embeddings of all electrovacuum solutions.
These are often  found by integrating the EYM equations.
In this way, for example,
the colored black holes and their generalizations
with rotation and cosmological term
\cite{Aragone78,Kasuya82,Kasuya82a,Kasuya82b}, as well as
solutions with cylindrical and plane symmetries
\cite{Mondaini82,Mondaini83,Mondaini83a,Mondaini83b,Mondaini85}
have been obtained.
Note that the embedded Abelian
solutions are not completely equivalent to their U(1) counterparts.
For example,
the Dirac string for (\ref{2.40}),(\ref{2.47}) can be globally
removed by passing to the regular gauge
(\ref{2.5}) -- the SU(2) bundle is trivial.
In addition,  unlike their U(1) counterparts, the colored black holes
are unstable \cite{Lee92a}.

\section{Particle-like solutions}
\setcounter{equation}{0}

The rapid progress in the theory of gravitating solitons
and hairy black holes started after Bartnik and McKinnon had
discovered particle-like solutions for the EYM equations
(\ref{2.42})--(\ref{2.45}) \cite{Bartnik88}. This
came as a big surprise, since it is well known that when taken apart,
neither pure gravity nor YM theory admit particle-like solutions.
Indeed, the existence of stationary gravitational solitons
is ruled out by Lichnerowicz's theorem \cite{Lichnerowicz55}.
One can go farther than this and consider
 regular configurations with finite mass and arbitrary
time-dependence, as long as they do not radiate their energy to infinity.
In particular, system which are  exactly periodic in time
do not radiate, and such solutions are also ruled out \cite{Gibbons84}.
It is worth noting, however, that the vacuum theory of gravity admits,
``quzi-solitons", geons,
which exist for a long time as compared to the characteristic
period of the system before radiating away their energy \cite{Brill64}%
\footnote{Note though, that no rigorous existence proof for geons
has been given so far.}.

Another famous statement
``there are no classical glueballs"
asserts that the pure YM theory in flat space does not admit
finite energy non-singular solutions which do not radiate energy
out to infinity \cite{Coleman75,Deser76,Pagels77,Coleman77,Coleman77a}.
This can be readily seen in the static situation, since
the spatial components
of the stress-energy tensor $T_{\mu\nu}$
then satisfy  \cite{Deser76}
\be                                            \label{3.2}
\int_{\rm R^3}T_{ik}\, d^3 x=0   .
\ee
This is the consequence of the identity
$T_{ik}=\partial_j(T_{i}^{j}x_k)$ implied by the conservation
law $\partial_j T_{i}^{j}=0$. The physical meaning of (\ref{3.2})
is that the total stresses in an extended object must balance.
It follows then that the sum of the principal pressures,
$\Sigma_i p_i$, where $p_i=-T^{i}_{i}$ (no summation),
cannot have a fixed sign.  However, for the YM field
the scale invariance implies that
$\Sigma_i p_i=T_{00}>0$, which shows that the system is
purely  repulsive and the force balance is impossible.
The argument can be extended to exclude
solutions with arbitrary time-dependence, as long as they
do not radiate \cite{Pagels77,Coleman77}.
It is interesting that the
geon-type solutions in the pure YM theory are also excluded --
any field configuration initially confined in some
region falls apart in the time it takes light to cross the region
\cite{Coleman77a}. This can be viewed as another indication
of the purely repulsive nature of the gauge field.

Although for purely attractive or repulsive systems
the balance cannot exist, the situation changes
if the system includes interactions of
both types. The example is provided by
the YM-Higgs models (\ref{2:25}), (\ref{2:27}).
For the scalar Higgs field one has
$\Sigma_i p_i<0$, which corresponds to pure attraction.
However,  in a combined system with both gauge and Higgs fields
repulsion and attraction can compensate each other, which leads
to the existence of soliton solutions. These
will be considered in Sec.$7$ below. In a similar way, the
coupling of the repulsive YM field  to gravity can lead to a force balance,
as is illustrated by the BK example.

It should be stressed, however, that the presence
of both attractive and repulsive interactions is necessary but
not sufficient for the existence of equilibrium
configurations. This becomes especially clear when one takes into account
a number of the no-go results for the EYM system
\cite{Weder82,Deser84,Malec84,Malec87,Masood89,Ershov90}.
These, for example, rule out all non-trivial EYM solutions
in three spacetime dimensions \cite{Deser84},
and large classes of solutions in four dimensions.
In particular, all charged 4D EYM solitons are excluded \cite{Ershov90}.
In view of this one can wonder about the sufficient
condition which ensures the existence of the BK solutions.
Unfortunately, this condition is unknown.
Even though the existence of the BK solutions can be deduced
from a complicated analysis of the differential equations
\cite{Smoller91,Smoller93,Breitenlohner94}, it still
does not find a clear explanation in physical terms.
For example, the topological arguments, which proved to be very
useful in the YMH case, do not apply in the EYM theory.

It is interesting that,
even though the topology is different,
there exists a very useful analogy between the EYM theory and
the flat space doublet YMH model.
It has already been mentioned that
the BK particles are in some ways similar to the
sphaleron solutions of the Weinberg-Salam model \cite{Manton83}.
The resemblance
is two-fold. First, both sphalerons and BK solitons can be thought of as
equilibrium states of a pair of physical fields one of which is repulsive
(YM field) while another one is attractive (gravity or the Higgs field).
Secondly, both types of solutions relate to the top of the
potential barrier between distinct topological vacua of the gauge
field. In this sense one can call the BK particles EYM sphalerons
\cite{Galtsov91a}.

In view of their importance,  we shall describe
below in this chapter the BK  solutions and
the corresponding sphaleron construction in some detail.
We shall consider also the generalizations of the solutions due
to the dilaton, $\Lambda$-terms, as well as the solutions
for higher gauge groups and in the case of axial symmetry.
Although very interesting in their own right, these
exhibit essentially the same features as the BK solutions.

\subsection{Bartnik-McKinnon solutions}

These solutions are known numerically
\cite{Bartnik88,Kunzle90,Breitenlohner94},
their existence was established in
\cite{Smoller91,Smoller93,Breitenlohner94}.
 The gauge field potential and the spacetime metric
are chosen in the form (\ref{2.40}), (\ref{2.41}),
the field equations are given by (\ref{2.42})--(\ref{2.45}).
The electric field amplitude $a_0$
should be set to zero, since otherwise only the embedded Abelian
solutions are possible \cite{Galtsov89,Ershov90,Bizon92}.
As a result, the metric function $\sigma$ can be
eliminated from (\ref{2.43}) and (\ref{2.44}) \cite{Ray78}
and there remain only two independent equations
\bea                                                       \label{4.4}
&&N\f''+\left(\frac{2m}{r}-\frac{(\f^2-1)^2}{r^2}
\right)\frac{\f'}{r}=\frac{\f(\f^2-1)}{r^2},                \\
&&m'=N\f'^2+\frac{(\f^2-1)^2}{2r^2}.                          \label{4.5}
\eea
Given a solution to these,
$\sigma$ is obtained from
\be                                                         \label{4.5a}
\sigma=\exp\left(-2\int_r^{\infty}dr\,\frac{\f'^2}{r}\right).
\ee
Eqs.(\ref{4.4}), (\ref{4.5})  have singular points at the
origin, $r=0$,  at infinity, $r=\infty$, and at $r=r_h$, where
$N(r_h)=0$. Assuming that $N$ vanishes nowhere (no horizons),
consider the local power-series solutions to Eqs.(\ref{4.4}), (\ref{4.5})
in the vicinity of the singular points.
Requiring that curvature is bounded
one obtains at the origin
\bea
&&\f=1-br^2+
\left(\frac{3}{10}\,b^2-\frac{4}{5}\,b^3\right)r^4
+O(r^6),                                               \nonumber \\
&&N\equiv 1-\frac{2m}{r}=1-4b^2\, r^2+\frac{16}{5}\,b^3\,
r^4+O(r^6).                                             \label{4.6}
\eea
At infinity, using the results of \cite{Galtsov89,Ershov90,Bizon92},
the only nontrivial possibility  are configurations that
asymptotically approach the magnetically neutral solution (\ref{2.46}):
\bea
\f&=&\pm\left(1-\frac{a}{r}+\frac{3a^2-6aM}{4r^2}\right)
+O\left(r^{-3}\right),                                    \nonumber \\
m&=&M-\frac{a^2}{r^3}+O\left(r^{-4}\right).                  \label{4.7}
\eea

%%%%%%%%%%%%%%%%%%%%%%%%%%%%%%%%%%%%% 
\begin{figure} 
\hbox to\hsize{%\hss 
  \epsfig{file=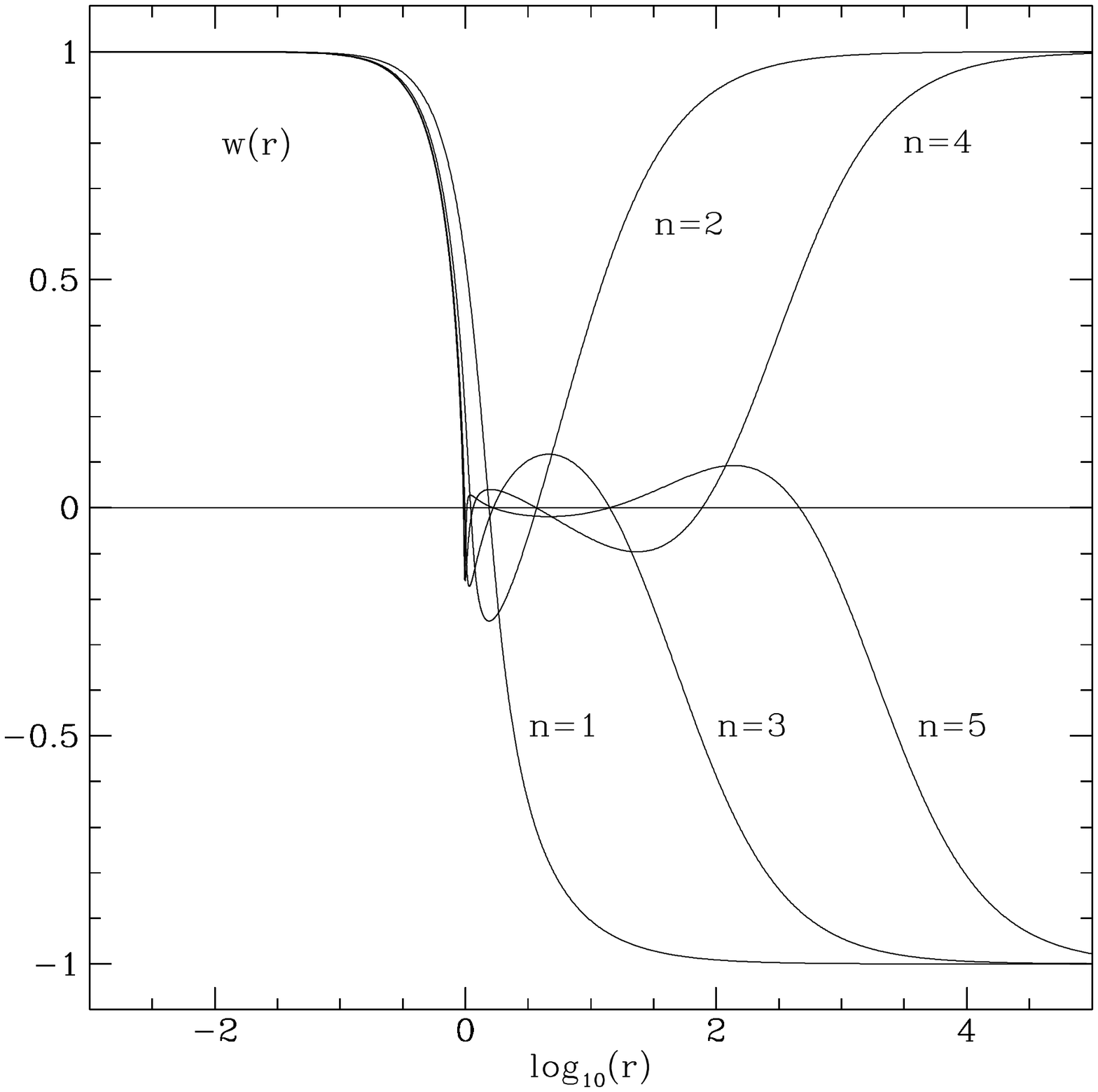,width=0.48\hsize,% 
      bbllx=1.8cm,bblly=5.5cm,bburx=20.0cm,bbury=20.0cm}\hss 
  \epsfig{file=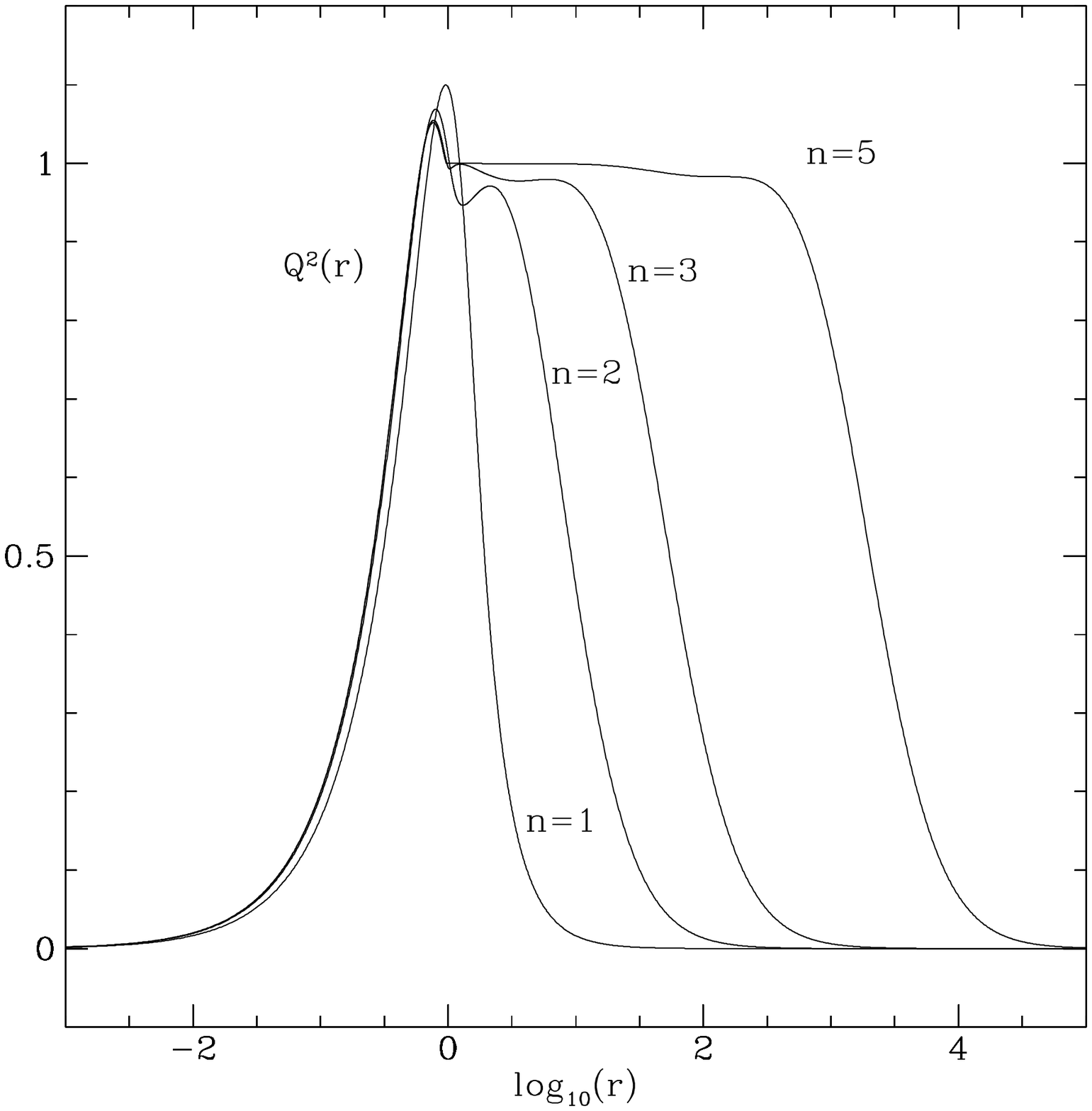,width=0.48\hsize,% 
     bbllx=1.8cm,bblly=5.5cm,bburx=20.0cm,bbury=20.0cm}%\hss 
  } 
\caption{$w(r)$ and the effective charge function $Q^2(r)$ 
for the lowest BK solutions. } 
\label{fig4:1} 
\vspace{3 mm}
\end{figure} 
%%%%%%%%%%%%%%%%%%%%%%%%%%%%%%%%%%%%% 

In these expressions $b$, $a$ and $M$ are free
parameters, $M$ being the ADM mass.
Applying the standard methods one can show
that the Taylor expansions in (\ref{4.6}) and (\ref{4.7})
have a nonzero convergence radius
\cite{Kunzle90,Breitenlohner94}.
Inserting these expressions into (\ref{4.5a}),
one obtains
\be
\sigma(r)=\sigma(0)+O(r^2),\ \ \ \ \ \ \ \
\sigma=1+O\left(r^{-4}\right)
\ee
at the origin and at infinity, respectively.

%%%%%%%%%%%%%%%%%%%%%%%%%%%%%%%%%%%%%
\begin{figure}
\hbox to\hsize{%\hss
  \epsfig{file=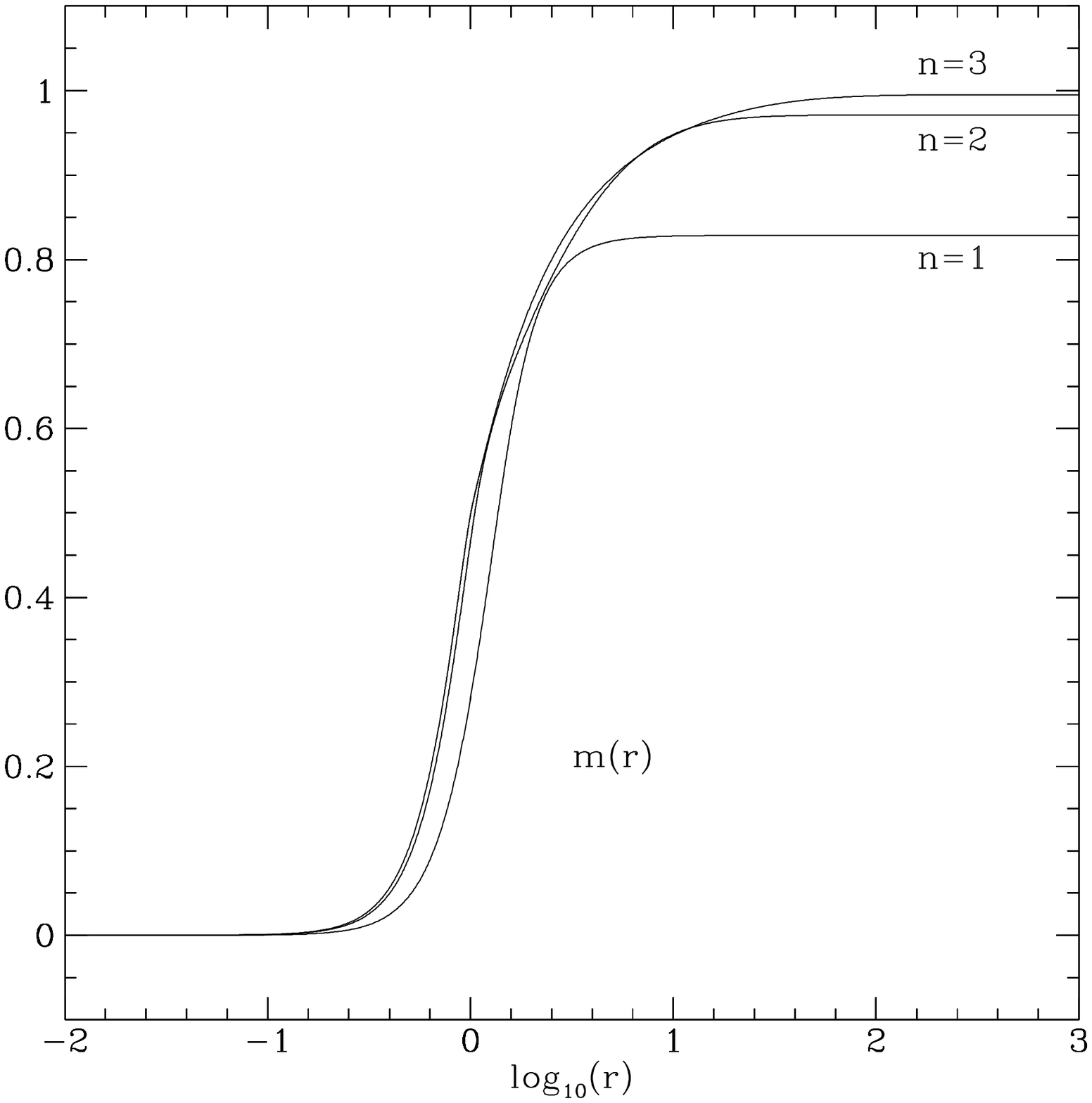,width=0.48\hsize,%
      bbllx=1.8cm,bblly=5.5cm,bburx=20.0cm,bbury=20.0cm}\hss
  \epsfig{file=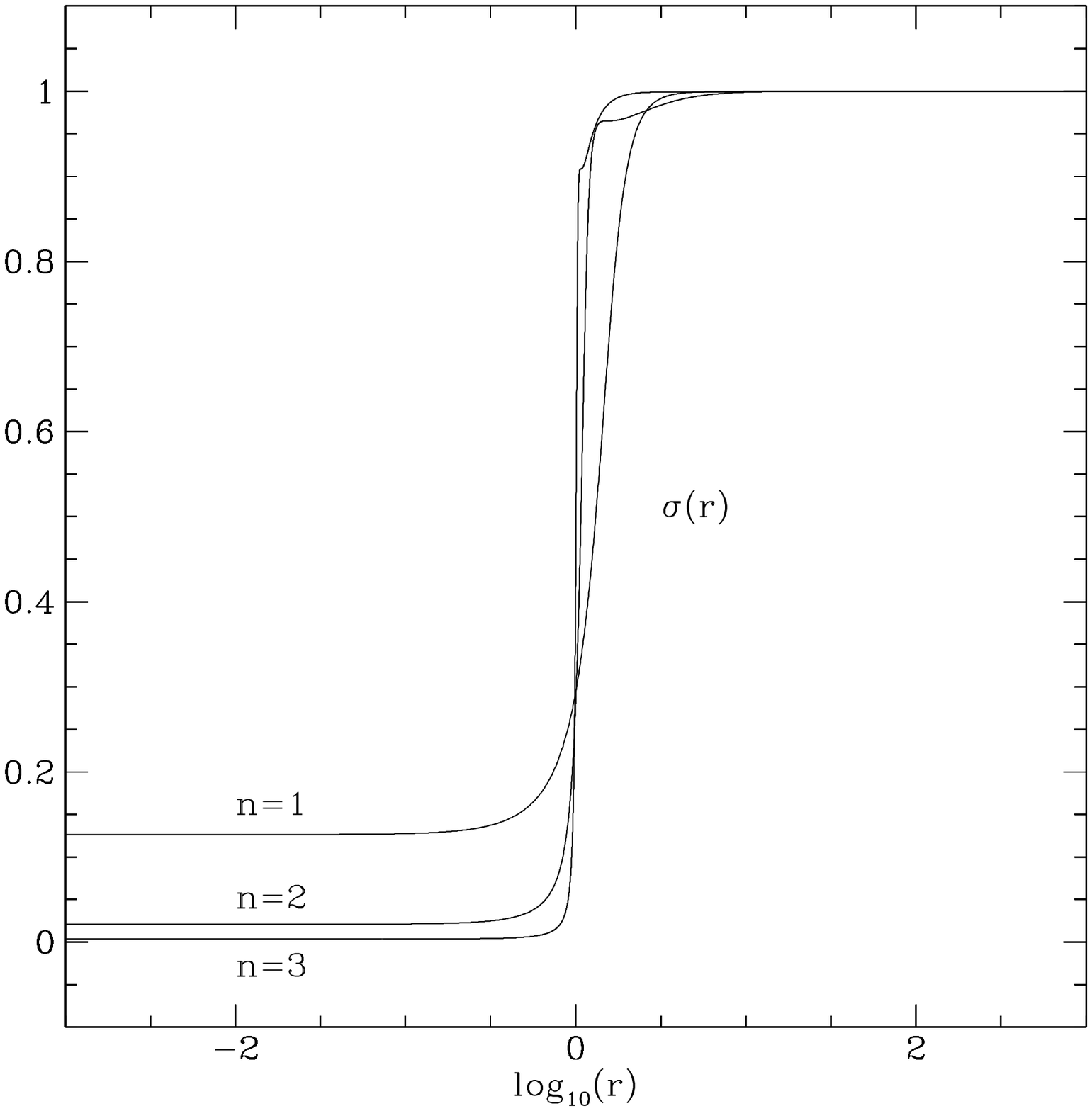,width=0.48\hsize,%
      bbllx=1.8cm,bblly=5.5cm,bburx=20.0cm,bbury=20.0cm}%\hss
  }
\caption{Metric  functions $m(r)$ and $\sigma(r)$
for the lowest BK solutions.
}
\label{fig4:2}
\end{figure}
%%%%%%%%%%%%%%%%%%%%%%%%%%%%%%%%%%%%%

The next step is to integrate
Eqs.(\ref{4.4}) and (\ref{4.5}) with the boundary conditions
(\ref{4.6}) and (\ref{4.7}).
The strategy
is to extend the asymptotics numerically
to the intermediate region, where the matching conditions
are imposed at some point $\bar{r}$ \cite{Press92}. 
There are three such conditions
for $\f$, $\f'$ and $m$. This agrees with the number of the
free parameters in (\ref{4.6}) and (\ref{4.7}).
It turns out that the matching
can be fulfilled, provided that the values of the
parameters $b$, $a$ and $M$
are restricted to a one-parameter family
$\{b_n, a_n, M_n\}$, where $n$ is a positive integer.
This gives an infinite family of globally regular solutions in the
interval $0\leq r < \infty$ labeled by $n=1,2,\ldots$.

The index $n$ has the meaning of the node number
for the amplitude $\f$: for the $n$-th solution $\f(r)$ has
$n$ nodes in the interval $r\in[0,\infty)$, such that
$\f(\infty)=(-1)^n$. For all solutions $\f$ is bounded
within the strip $|\f(r)|\leq 1$
(see Fig.\ref{fig4:1}).
This can be understood as follows:
Suppose that $\f$ leaves the strip, such that
$\f(r)>1$ for some $r>0$
(notice that the equations are invariant
under $\f\rightarrow -\f$).
Then, taking the boundary conditions into account,
$\f$ must develop a maximum outside  the strip:
$\f'(r_m)=0$, $\f(r_m)>1$, $\f''(r_m)<0$ for some $r_m>0$.
However, Eq.(\ref{4.4}) implies that in the region $\f>1$
an extremum can only be a minimum, which shows that the
condition $|\f|<1$ cannot be violated.

The behaviour of the metric functions $m$ and $\sigma$ is
similar for all $n$'s: they
increase monotonically with growing $r$ from
$m(0)=0$ and $\sigma(0)\equiv\sigma_n$ at the origin
to $m(\infty)=M_n$ and $\sigma(\infty)=1$ at infinity, respectively
(see Fig.\ref{fig4:2}). Here $M_n$ is the ADM mass of the solutions and
$\sigma_n$ is the time deceleration factor at the origin.

\begin{figure}
\hbox to\hsize{%\hss
  \epsfig{file=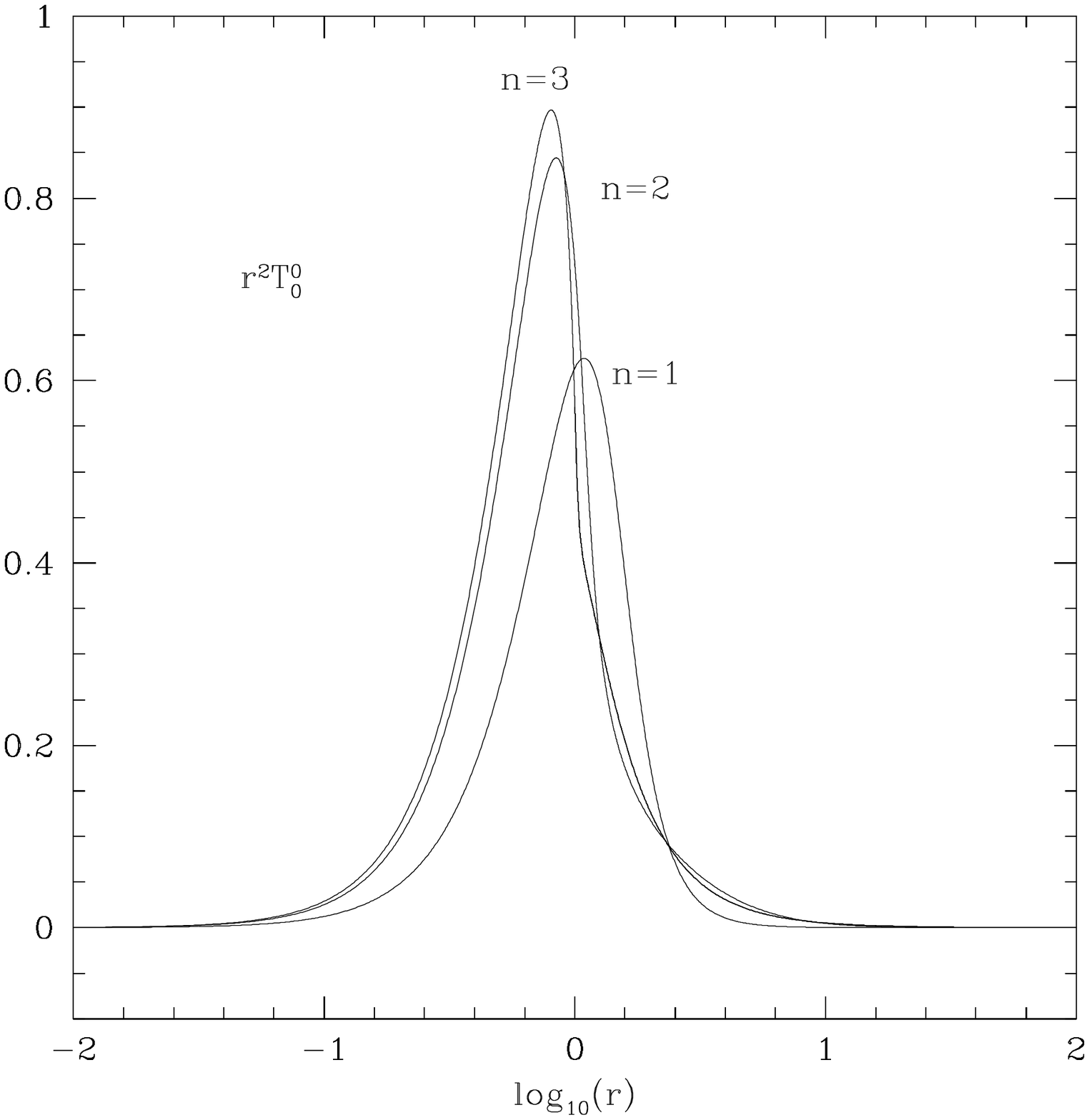,width=0.48\hsize,%
      bbllx=1.8cm,bblly=5.5cm,bburx=20.0cm,bbury=20.0cm}\hss
  \epsfig{file=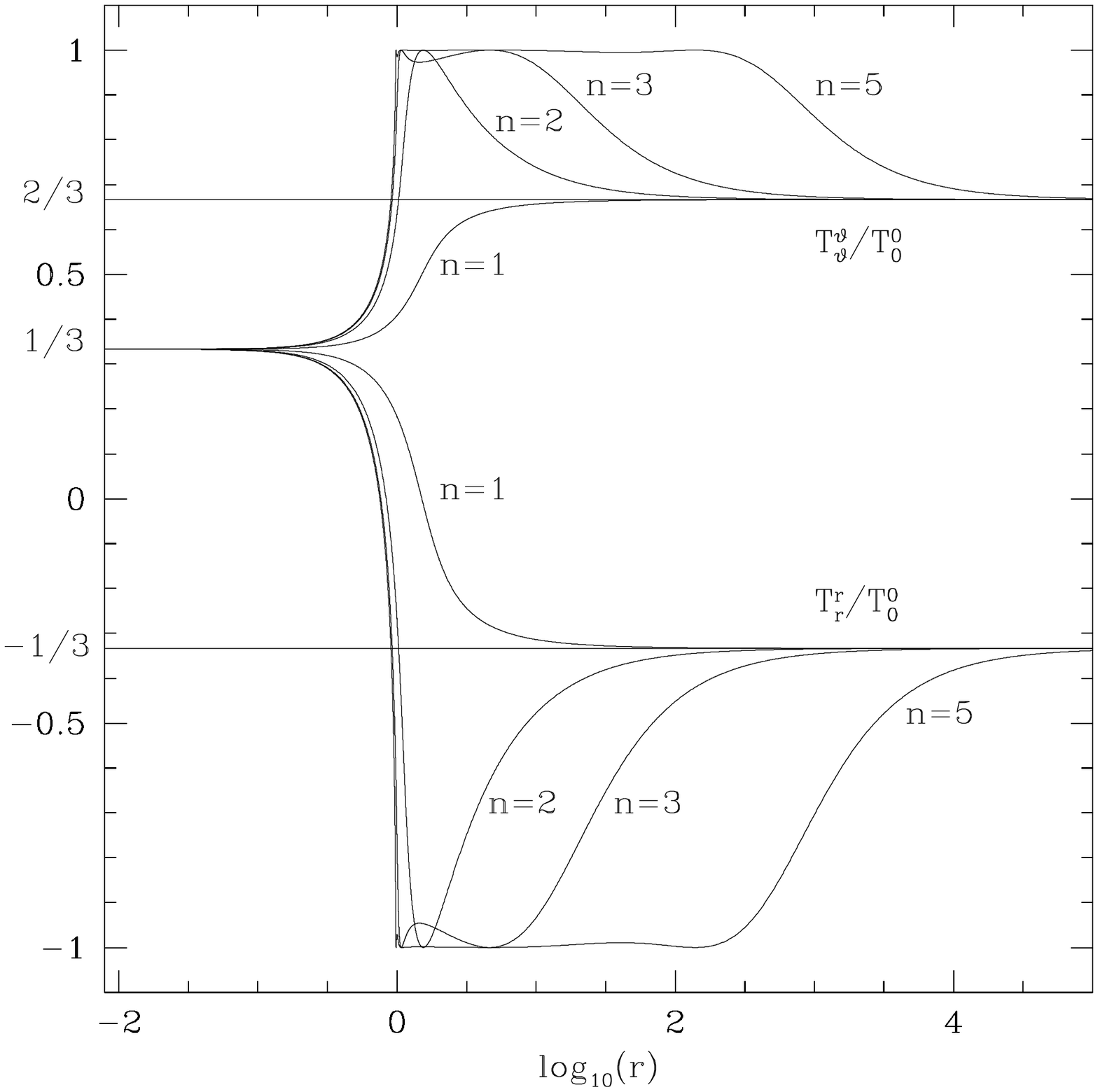,width=0.48\hsize,%
      bbllx=1.8cm,bblly=5.5cm,bburx=20.0cm,bbury=20.0cm}%\hss
  }
\caption{On the left: the  radial energy density
$r^2 T^0_0$ for the lowest BK solutions.
On the right: the equations of state
$T^r_r/T^{0}_{0}$ and $T^{\vartheta}_{\vartheta}/T^{0}_{0}$.
}
\label{fig4:3}
\vspace{3 mm}
\end{figure}

Qualitatively the solutions show three distinct
regions with the two transition zones; see Fig.\ref{fig4:1} --
Fig.\ref{fig4:3}.
This distinction is due to the characteristic behavior of the
effective charge function $Q(r)$ defined by the
relation  \cite{Bartnik88} (see Fig.\ref{fig4:1})
\be                                                     \label{4:charge}
N(r)=1-\frac{2M}{r}+\frac{Q^2(r)}{r^2}.
\ee

{\bf a)} In the interior region, $r\leq 1$, one has $Q(r)\approx 0$,
the functions $\f$, $m$ and $\sigma$
are approximately constant, and the metric becomes flat
as $r\rightarrow 0$. The stress tensor $T^{\mu}_{\nu}$
is isotropic
and corresponds to the equation of state $\rho =3p$
with $\rho\propto r^2$;
%\be                                              \label{4.8}
%T^{\mu}_{\nu}\propto r^2\times {\rm diag}(3,-1,-1,-1);
%\ee
note that this has the SO(4)-symmetry.
The transition zone  about $r= 1$ is marked by the high
energy density. The metric functions grow rapidly  in this zone
approaching their asymptotic values, while $Q(r)$ reaches
the unit value.

{\bf b)} In the near-field region, $1\leq r\leq R_0$,
one has $Q(r)\approx 1$,
the gauge field is well-approximated by that for the Dirac magnetic
monopole, $\f=0$, such that $\rho=-p_r=p_\vartheta$
with $\rho\propto r^{-4}$.
%\be                                                   \label{4.9}
%T^{\mu}_{\nu}\propto r^{-4}\times {\rm diag}(1,1,-1,-1),
%\ee
The metric is close to the extreme RN metric.
In the ``charge-shielding'' transition zone, $r\sim R_0$,
the effective charge decays to zero.

{\bf c)} Finally,  in
the far-field region, $r>R_0$, the solutions are
approximately Schwarzschild.
The gauge field strength
decreases as $1/r^3$, the components of the stress tensor are
$\rho=-p_r=3p_\vartheta/2$ with $\rho\propto r^{-6}$.
%\be                                                    \label{4.10}
%T^{\mu}_{\nu}\propto r^{-6}\times {\rm diag}(3,1,-2,-2)
%\ee
The polynomial decay of the functions reflects the massless nature
of the fields in the problem.

%%%%%%%%%%%%%%%%%%%%%%%%%%%%%%%%%%%%%
%%%%%%%%%%%%%%%%%%%%%%%%%%%%%%%%%%%%%

The above distinction becomes more and more pronounced
with growing $n$. The ADM mass  $M_n$ increases with $n$ from the
``ground state'' value $M_1=0.828$ to $M_{\infty}=1$,
rapidly converging to the upper limit (see Table 1).
The dimensionful mass is $\sqrt{4\pi/G{\sl g}^2}\,M_n$,
and the BK particles are therefore extremely heavy.
Notice that the discrete energy spectrum emerging in the problem
resembles that of some quantum system.
The analogy becomes even more striking if one thinks of
the node number $n$ as relating to some wavefunction.
The values of $\{b_n,a_n,M_n,\sigma_n\}$
for the lowest $n$ are given in Table 1, while those
for large $n$ can be
approximated by
\bea
&&b_n=0.706-2.186\,{\rm e}_n,\ \ \ \ \
M_n=1-1.081\,{\rm e}_n,\    \nonumber \\
&&a_n=0.259\,{\rm e}_n,\ \ \ \ \
\sigma_n=0.707\,{\rm e}_n            \label{4.11}
\eea
with e$_n=\exp(-\pi n /\sqrt{3})$ \cite{Breitenlohner94}.

\begin{table}
\caption{Parameters of the BK solutions.}
\vglue 0.4cm
\begin{tabular}{|c|c|c|c|c|c} \hline
$n$
& $b_n$
& $M_n$
&$a_n$
& $\sigma_n$   \\     \hline
1
&  $0.4537$
& $0.8286$
& $0.8933$
& $0.1264$ \\
2
&   $0.6517$
&   $0.9713$
&   $8.8638$
&   $0.0208$ \\
3
&   $0.6970$
&   $0.9953$
&  $58.929$
&  $0.0033$ \\
4
&   $0.7048$
&   $0.9992$
&  $366.20$
&  $0.0005$ \\
5
&   $0.7061$
&   $0.9998$
&  $2246.8$
&  $0.00009$ \\
\hline
\end{tabular}
\vspace{3 mm}
\end{table}

As $n\rightarrow\infty$,  more and more
nodes of $w$ accumulate in the first transition zone near $r=1$,
where $N$ develops a more and more deep minimum, closely
approaching the  zero value.
In the near-field zone at the same time, the amplitude and frequency
of the oscillations decrease with growing $n$, while
the size of the zone stretches exponentially, the whole configurations
approaching the extreme RN solution:
\be                                  \label{4.12}
\f=0,\ \ \ \sigma=1,\ \ \ N=\left(1-\frac1r\right)^2.
\ee
However, for finite $n$ the size of the near-field zone is finite,
and since $\f$ tends to $\pm 1$ in the far-field region, the
solutions are always neutral.

The limiting solution with $n=\infty$ can also be investigated
\cite{Breitenlohner94,Breitenlohner95,Smoller94,Smoller94a}.
This turns out to be non-asymptotically flat, which
can be qualitatively understood as follows.
The Schwarzschild coordinate system breaks down
in this limit, and one should use the general parameterization
(\ref{2:32}) with $R=R(r)$ (and $\alpha=0$).
The EYM field equations
(see Eqs.(\ref{3.5.1})--(\ref{3.5.3}) with $\Lambda=0$)
then can be reformulated as a dynamical
system.
This system admits a separatrix that starts
from the saddle point with $R=0$, $\f=N=1$, which
corresponds to the origin, and
after infinitely many revolutions ends up at
the focal point with $R=1$, $\f=N=0$, which corresponds to the
degenerate horizon.
The function $R$ thus changes only in the interval $[0,1]$.
In contrast, the phase space trajectories of
solutions with $n<\infty$ miss the focal point
and propagate further to the the region $R>1$, where they approach
(\ref{4.12}) for large $n$.
As a result, the configuration that the
BK solutions tend to for $n\rightarrow\infty$ is
the union of the oscillating solution
in the interval $R\in [0,1]$ and the extreme RN
solution (\ref{4.12}) for $R\in [1,\infty)$.
Such a configuration cannot be regarded as a single
solution globally defined for all $R\geq 0$.
For the oscillating solutions
the $t=$const. hypersurfaces are non-compact and in the limit
$R\rightarrow 1$ correspond to the cylinders R$^1\times S^2$.
The four-metric in this limit is the direct product of a unit
two-sphere and a unit pseudosphere \cite{Breitenlohner94}.

Other solutions to the EYM equations (\ref{4.4}),
(\ref{4.5}) with the regular boundary conditions (\ref{4.6})
at the origin are also non-asymptotically flat and belong to
the bag of gold type.
This case is in fact generic \cite{Breitenlohner94}.
The qualitative picture is as follows.
Using again the general parameterization (\ref{2:32})
with the gauge condition $\sigma=1$
the Einstein equation for $R(r)$ is
$R''=-2\f'^2/R$.
The regular boundary conditions at the origin imply that
$R(0)=0$, $R'(0)=1$. Since the second derivative
$R''$ is always negative, the first derivative $R'$ is always decreasing.
As a result, $R(r)$ tends to develop a maximum
at some finite $r$ where $R'$ vanishes. After this
the integral curve
for $R(r)$ bends down finally reaching zero 
at some finite $r$, where the curvature diverges.
This behaviour is generic. Only for the special
boundary conditions specified by $b=b_n$ in (\ref{4.6})
the integral curves of $R(r)$ bend down slowly enough to be able
to escape to infinity. For $b=b_\infty$, $R(r)$ develops an
extremum at $r=\infty$, $R(\infty)=1$. For all other values of
$b$ in (\ref{4.6}) $R(r)$ develops a maximum at a finite
value of $r$ leading to a bag of gold solution.

\subsection{Sphaleron interpretation}

The BK solutions admit an interesting
interpretation as EYM sphalerons
\cite{Galtsov91a,Sudarsky92,Gibbons93}.
This is based on some common features which they share
with the sphaleron solution
of the Weinberg-Salam model \cite{Manton83,Klinkhamer84}.
In fact, sphalerons exist also in other gauge
models \cite{Forgacs84,Bochkarev87,Mottola89}
with vacuum periodicity \cite{Jackiw76}.
In such theories the potential
energy is a periodic function of the winding number of the
gauge field. The minima of the energy,
which are called topological vacua,
are separated by a potential barrier of a finite height,
and ``sitting'' on the top of the barrier there is a classical
field configuration called sphaleron.
Since its energy determines the barrier height, the sphaleron
is likely to be important  for the barrier transition
processes when the system interpolates between distinct vacuum
sectors, at least in the electroweak theory
\cite{Kuzmin85,Rubakov96}.
The winding number of the gauge field changes during such
processes, which leads to the fermion number non-conservation
due to the axial anomaly.

In what follows we shall introduce the topological vacua
in the EYM theory, and show that the odd-$n$ BK field
configurations can be reduced to either of the neighbouring
vacua via a continuous sequence of static deformations
preserving the boundary conditions. 
The solutions therefore relate to the top of the potential barrier
between the vacua, which accounts for their sphaleron interpretation.   

\subsubsection{Topological vacua}
The sphaleron construction
for the BK solutions starts by defining the topological
vacua in the EYM theory as fields $\{g_{\mu\nu},A\}$
with zero ADM mass:
$g_{\mu\nu}=\eta_{\mu\nu}$,  $A=i{\rm U}d{\rm U}^{-1}$.
Here $\eta_{\mu\nu}$ is Minkowski metric on ${\rm R}^4$
\and U=U$(x^i)$.
Imposing the asymptotic condition at spatial infinity \cite{Jackiw76}
\be                                         \label{5.2}
\lim_{r\rightarrow\infty}{\rm U}(x^i)=1,
\ee
any U$(x^i)$ can be viewed as a mappings $S^3\rightarrow\ $SU(2),
and the set of all U's falls into a countable sequence
of disjoint homotopy classes characterized by an integer
winding number
\be                                            \label{5.3}
{\bf k}[{\rm U}]=\frac{1}{24\pi^2}\, {\rm tr}\int_{R^3}
{\rm U}d{\rm U}^{-1} \wedge {\rm U}d{\rm U}^{-1}
\wedge {\rm U}d{\rm U}^{-1}.
\ee
As a result, the vacuum fields split into equivalence classes
with respect to the winding number of the gauge field ${\bf k}[{\rm U}]$.
These are called topological vacua.
Note that the gravitational field is
assumed to be topologically trivial.

A representative of the $k$-th vacuum class,
$\{\eta_{\mu\nu},i{\rm U}_kd{\rm U}_k^{-1}\}$
with ${\bf k}[{\rm U}_k]=k$, can be parameterized by
\be                                    \label{5.4}
{\rm U}_{k}=\exp\{i\beta(r)\T_r\},\ \ \ {\rm where}\ \ \ \
\beta(0)=0,\ \ \beta(\infty)=-2\pi k.
\ee
The gauge field $i{\rm U}_{k}d{\rm U}_{k}^{-1}$
is given by Witten's ansatz
(\ref{2.5})  with $a=0$, $\f=\exp(i\beta(r))$.

Note that the asymptotic condition (\ref{5.2}) imposes
the following fall-off requirement
for the components $A_a$
of the vacuum gauge field
with respect to an orthonormal frame:
\be                                          \label{5.2a}
A_a=o(r^{-1})\ \ \ \ \ \ {\rm for}\ \ \ \ r\rightarrow\infty.
\ee
In what follows we shall demand that non-vacuum $A$'s
should obey the same fall-off condition.

The vacua with different $k$ cannot be continuously deformed into one
another within the class of vacuum fields subject to (\ref{5.2a}).
However, it is possible to join them
through a continuous sequence of
non-vacuum fields, $\{g_{\mu\nu}[\lambda], A[\lambda]\}$,
where the gauge field
$A[\lambda]$  obeys (\ref{5.2a})
for all values of the parameter $\lambda$, while
$g_{\mu\nu}[\lambda]$
is an asymptotically flat metric on ${\rm R}^4$.
The crucial point is that there are such interpolating sequences
that pass through the BK field configurations.

\subsubsection{The interpolating sequence}
Consider a family of static fields
\cite{Galtsov91a,Volkov94a}
\bea
ds^2&=&\sigma^2_\lambda(r) N_\lambda(r) dt^2
-\frac{1}{N_\lambda(r)}\, dr^2
-r^2\, (d\vartheta^2+\sin^2\vartheta\, d\varphi^2),     \label{5.5}  \\
A[\lambda]&=&i\frac{1-\f}{2}\,{\rm U}_{+}d{\rm U}_{+}^{-1}+
i\frac{1+\f}{2}\,{\rm U}_{-}d{\rm U}_{-}^{-1},            \label{5.6}
\eea
where $N_\lambda(r)=1-2\, m_\lambda(r)/r$, and
$\lambda\in [0,\pi]$.
The functions $m_{\lambda}(r)$, $\sigma_{\lambda}(r)$,
and ${\rm U}_{\pm}$ are given by
\bea
m_{\lambda}(r)&=&\frac{\sin^2\lambda}{\sigma_{\lambda}(r)}
\int_0^r\left(\f'^2+\sin^2\lambda\frac{(\f^2-1)^2}{2r^2}\right)
\sigma_{\lambda}(r)dr,                              \nonumber \\
\sigma_{\lambda}(r)&=&
\exp\left(-2\sin^2\lambda\int_r^{\infty} dr\, \f'^2/r\right), \nonumber \\
{\rm U}_{\pm}&=&\exp\{i\lambda (\f\pm 1)\T_r\},              \label{5.7}
\eea
where $\f$ coincides with the gauge amplitude for the $n$-th BK solution.
The boundary conditions for $\f$  ensure that, for any $n$,
\be                                                   \label{5.7a}
A_a[\lambda]=O(r)\ \ \ {\rm as}\ \ \ r\rightarrow 0,\ \ \ \ \ \
A_a[\lambda]=O(r^{-2})\ \ \ {\rm as}\ \ \ r\rightarrow\infty.
\ee
The metric (\ref{5.5}) becomes flat for $\lambda=0,\pi$.
The gauge field (\ref{5.6}) vanishes for $\lambda=0$,
whereas for $\lambda=\pi$ it can be represented as
\be                                                  \label{5.8}
A[\pi]=i{\rm U}d{\rm U}^{-1},\ \ \ {\rm with}\ \ \
{\rm U}=\exp\{i\pi(\f-1)\T_r\}.
\ee
Now, one has $\f(0)=1$, $\f(\infty)=(-1)^n$.
Comparing with (\ref{5.4}) one can see that for odd values of $n$
the pure gauge field (\ref{5.8}) has unit winding number,
whereas the winding number is zero if $n$ is even.
The field sequence (\ref{5.5}), (\ref{5.6}) therefore
interpolates between distinct topological vacua for $n$ odd,
and between different representatives inside the same vacuum sector
for $n$ even.

\subsubsection{The Chern-Simons number}
It is instructive to rederive the above result in a different way
\cite{Klinkhamer84,Galtsov91a,Moss92,Volkov94a}.
Consider an adiabatic time evolution along the family
(\ref{5.5}),(\ref{5.6}) by letting
the parameter $\lambda$ depend on time in such a way that
$\lambda(-\infty)=0$,  $\lambda(\infty)=\pi$.
Introduce the Chern-Simons current,
%\be                                        \label{5.9}
$$K^{\mu}=\frac{1}{8\pi^2}\,
\frac{\epsilon^{\mu\nu\alpha\beta}}{\sqrt{-g}}\,
{\rm tr}\,A_{\nu}\left(\nabla_{\alpha}A_{\beta}-
\frac{2i}{3}\,A_{\alpha}A_{\beta}\right),$$
whose divergence
$\nabla_{\sigma}K^{\sigma}=(1/16\pi^2)\,
{\rm tr}\ast\!F_{\mu\nu}F^{\mu\nu}$.
For any $\lambda(t)$ in (\ref{5.6}) the
temporal component of the gauge field is exactly zero
implying that all spatial components of $K^{\mu}$ vanish.
This gives
\be                                          \label{5.11}
\left. \int d^3
x\sqrt{-g}K^0\right|^t_{-\infty} =
\frac{1}{16\pi^2}\int_{-\infty}^t\int d^3 x \sqrt{-g} \, {\rm tr}
\ast\!F_{\mu\nu}F^{\mu\nu}.
\ee
The quantity on the left is $N_{\rm CS}[t]-N_{\rm CS}[-\infty]$,
where $N_{\rm CS}$ is the Chern-Simons number of the gauge field.
Since for a pure gauge field
$N_{\rm CS}$ coincides with the winding number,
one has $N_{\rm CS}[-\infty]=0$.
Substituting (\ref{5.6}) into (\ref{5.11})
and evaluating the integral on the right one finally arrives at
\bea
N_{\rm CS}[\lambda(t)]=
\frac{3}{2\pi}\int_{-\infty}^{t}dt\ \dot{\lambda}\sin^{2}\lambda
\int_{0}^{\infty}dr\ \f'(\f^{2}-1)                      \nonumber \\
=\frac{1-(-1)^n}{2\pi}\, (\lambda-\sin\lambda \cos\lambda). \label{5.12}
\eea
This shows that for odd values of $n$
the Chern-Simons number changes from zero to
one as $\lambda$ increases from 0 to $\pi$.
If $n$ is even then one has
$N_{\rm CS}=0$ for any $\lambda$, which corresponds
to a sequence of fields within one vacuum sector.

Now, for $\lambda=\pi/2$ the fields
(\ref{5.5})--(\ref{5.7})  exactly
correspond to those of the BK solutions.
The gauge field (\ref{5.5}) then
can be obtained from the BK field
via a gauge rotation with ${\rm U}=\exp\{i\pi(\f-1)\T_r/2\}$.
The metric function $m_{\lambda}(r)$ for $\lambda=\pi/2$
satisfies the Einstein equation (\ref{4.5}) with the
appropriate boundary condition and hence coincides with the BK
mass amplitude $m(r)$, while
$\sigma_{\lambda}(r)$
coincides with $\sigma$ in (\ref{4.5a}). As a result,
the configuration (\ref{5.5})--(\ref{5.7}) for $\lambda=\pi/2$
is the gauge rotation of the BK field.

Summarizing,  there are sequences of static fields
that interpolate between distinct vacuum sectors passing through
the BK configurations with odd $n$.
Using (\ref{5.12}), one can determine the
Chern-Simons number for these solutions: $N_{\rm CS}=1/2$.
This shows that the odd-$n$ BK solutions reside in between
the two vacua and can be reduced to each of them through a
continuous sequence of static deformations preserving the
boundary conditions. This accounts for their interpretation
as sphalerons.
Note also that these solutions admit a
zero energy fermion bound state  \cite{Gibbons93,Volkov94a,Brodbeck98}.
Moreover, the spectrum of the Dirac operator for fermions in the background
fields (\ref{5.5})--(\ref{5.7}) for odd values of $n$
exhibits the standard level crossing structure \cite{Volkov94a}.
As the parameter $\lambda$ varies from
zero to $\pi$, the lowest positive energy level crosses zero at
$\lambda=\pi/2$ and finally replaces the highest energy level of the
Dirac see. The fermion number therefore changes.
For even values of $n$ the BK solutions are topologically trivial,
since (\ref{5.12}) gives in this case $N_{\rm CS}=0$.

\subsubsection{The energy and the negative modes}
The above considerations show that the odd-$n$ BK solutions
relate to saddle points on the potential barrier
separating the neighbouring topological vacua.
For static {\it off-shell} fields in
GR the potential energy is the ADM mass --
provided that the configuration is asymptotically flat
and fulfills the initial value constraints.
Thus, for example, for a spherically symmetric,
globally regular static field configuration the
energy is
\be                                       \label{5.14}
M=\lim_{r\rightarrow\infty}m(r)=\int_0^\infty r^2 T^0_0 dr.
\ee
Here the second equality on the right is due to the initial
value constraint
$G_0^0=2T^0_0$ (in the dimensionless units chosen), since one has
$G^0_0=2m'/r^2$.
It is worth noting that for any value of $\lambda$
in (\ref{5.5})--(\ref{5.7}) the fields
do satisfy the initial value constraints,
from which the only non-trivial one is $G_0^0=2T^0_0$.
In addition, the $G_r^r=2T^r_r$ Einstein equation
is also fulfilled.  As a result, the energy  along the field sequence
(\ref{5.5})--(\ref{5.7}) is obtained by simply taking the limit $r\to\infty$
in $m_\lambda(r)$ in (\ref{5.7}).  The result
is given by  Eq.(\ref{5.18}) below (with $\beta=1$). The energy
vanishes for $\lambda=0,\pi$ and reaches
the maximum at the sphaleron position
$\lambda=\pi/2$ (see Fig.\ref{fig:ENERGYSURF}).

The last remark suggests that the BK solutions are unstable.
The corresponding negative mode is given by the derivative
of (\ref{5.5})--(\ref{5.7}) with respect to $\lambda$ at $\lambda=\pi/2$.
The two facts are essential: a) the energy
is maximal  at $\lambda=\pi/2$; b) the boundary conditions
at infinity (\ref{5.7a}) hold for any $\lambda$,
which ensures that the negative mode is normalizable.
The stability analysis presented in Sec.$5$ below gives
the following result:  the $n$-th BK solution
has $2n$ negative eigenmodes in the spherically symmetric
perturbation sector,
of which $n$ modes are parity-even and $n$ are parity-odd \cite{Volkov95a}.
The described above negative mode belongs to the odd-parity sector.

\subsubsection{The physical picture}
The fact that the BK solutions have more then one negative mode
reflects the important difference between the EYM
and the YMH sphalerons.
Sphalerons are usually associated with
non-contractible loops in the configuration space \cite{Manton83}.
Such loops can be obtained from the interpolating sequences
considered above by identifying the end points.
The maximal value of energy along a loop, $E_{\rm max}$, is minimized
over all loops to obtain $E_s\equiv\inf\{E_{\rm max}\}$.
If {\bf a)} $E_s$ exists and is positive; {\bf b)} there is a loop whose
$E_{\rm max}$ is equal to $E_s$, then there is a
saddle point solution called sphaleron \cite{Manton83}. By construction,
this has only one negative mode.
In reality, however, it is very difficult to show
that the conditions {\bf a)} and
{\bf b)} above hold, and this has never been done.
At the same time, these conditions are essential, since
the mere existence of non-contractible loops does not imply
the existence of any non-trivial solutions \cite{Manton83}.
For example, in flat spacetime pure YM theory there is
vacuum periodicity but there
are no static finite energy solutions at all.
For these reasons a somewhat weaker definition
is adopted in this text.  According to this, sphalerons
are only required to relate to the top of the potential
barrier between the vacua. As a result, they
are still topologically non-trivial, but can
have more that one negative mode, and their existence
is not guaranteed by the minimax argument.

%%%%%%%%%%%%%%%%%%%%%%%%%%%%%%%%%%%%%
\begin{figure}
\hbox to\hsize{%\hss
  \epsfig{file=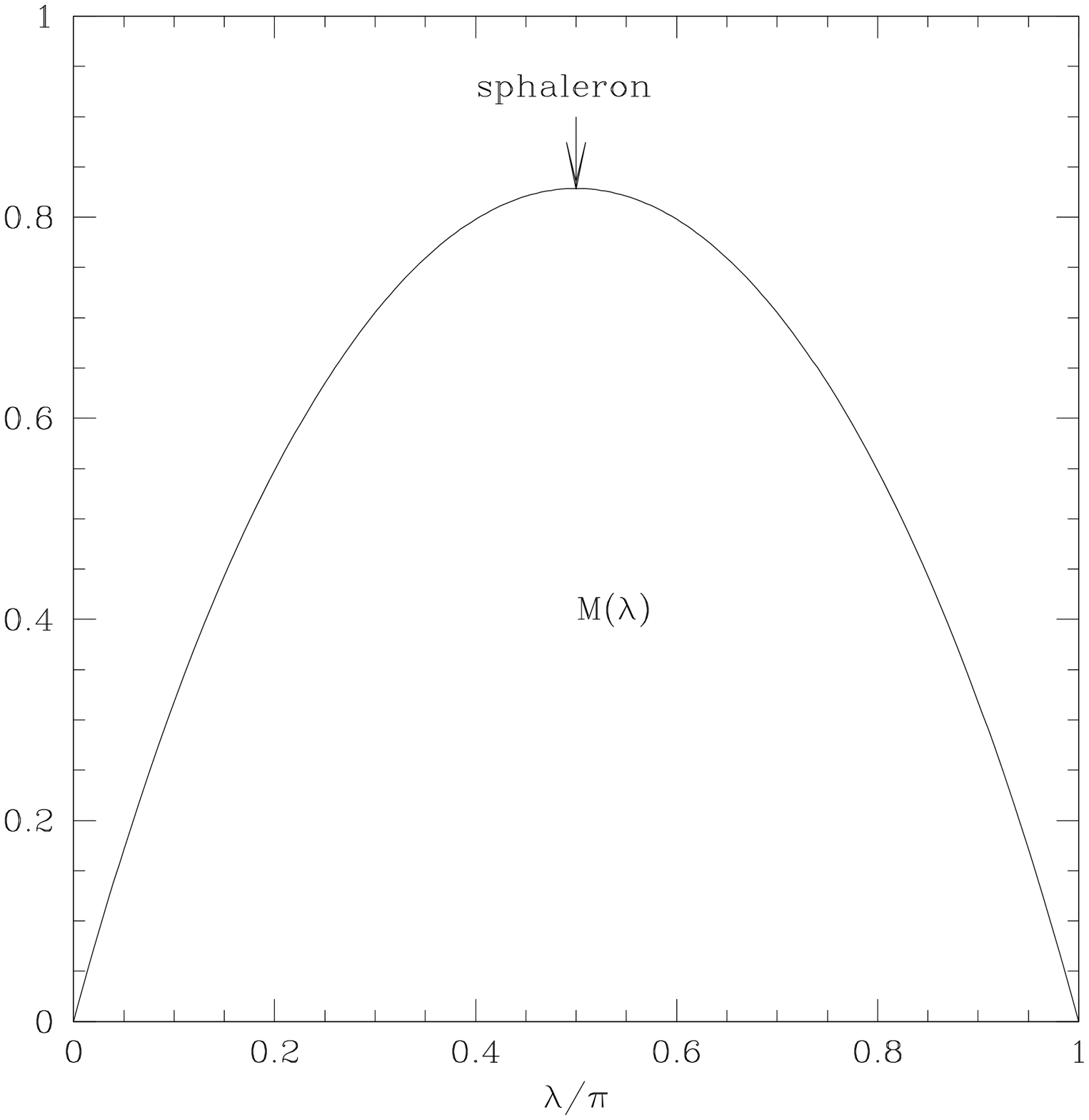,width=0.48\hsize,%
      bbllx=1.8cm,bblly=5.5cm,bburx=20.0cm,bbury=20.0cm}\hss
  \epsfig{file=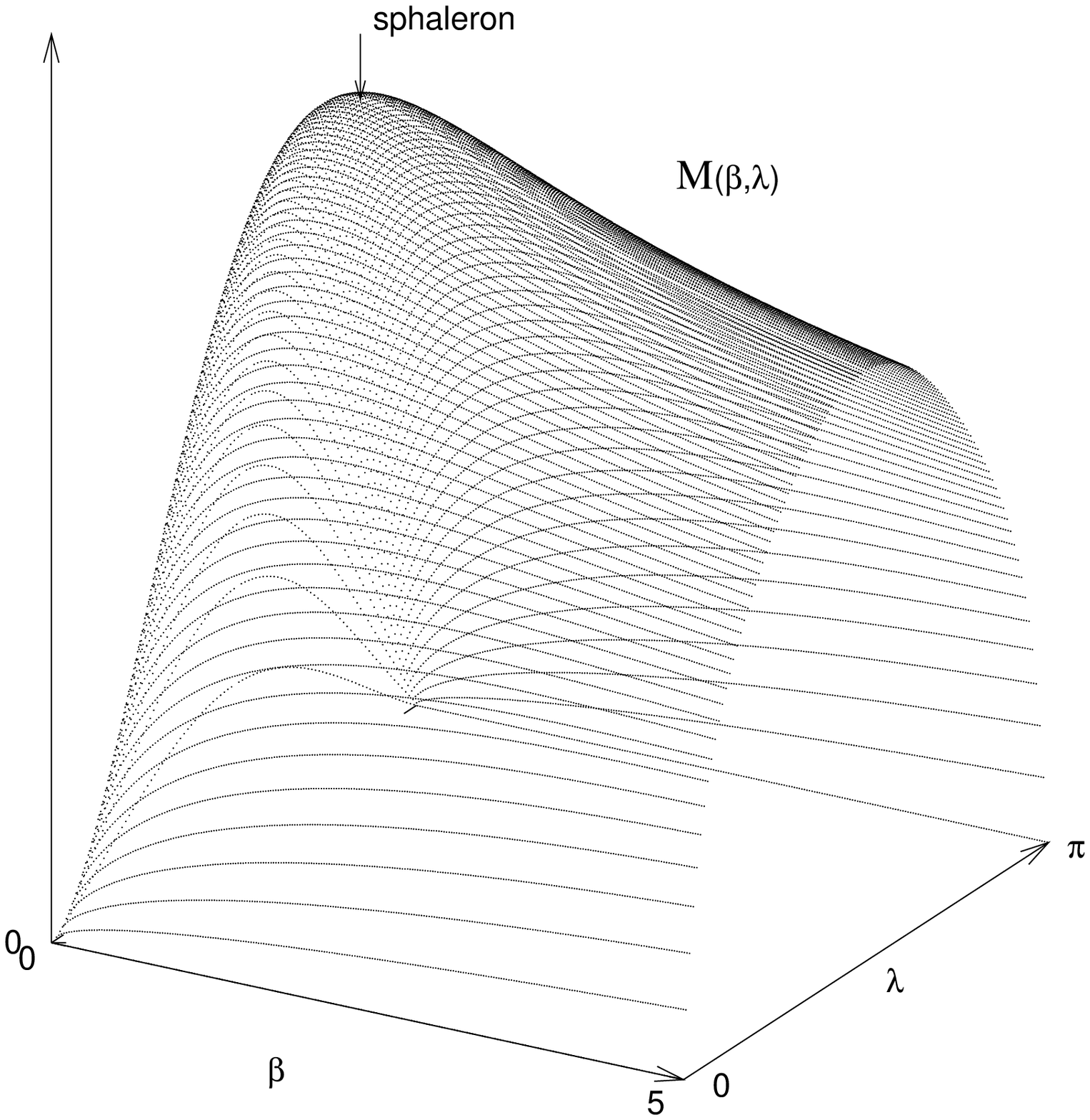,width=0.48\hsize,%
      bbllx=1.8cm,bblly=5.5cm,bburx=20.0cm,bbury=20.0cm}\hss
  }
\caption{On the left: The ADM mass along the field sequence
(\ref{5.5})--(\ref{5.7}) through the $n=1$ BK solution.
On the right: The potential energy (\ref{5.18}) in the vicinity
of this solution.
%The curve on the left
%corresponds to the section $\beta=1$.
}
\label{fig:ENERGYSURF}
\end{figure}
%%%%%%%%%%%%%%%%%%%%%%%%%%%%%%%%%%%%%

The 
following physical picture seems plausible:
The vacuum-to-vacuum transitions in the EYM theory
are suppressed at low energies, whereas the suppression is removed
at the energy of the order of mass of the $n=1$ BK solution.
In order to justify this picture, one should take into
account the fact that, since the $n=1$ BK solution has more than one
negative mode, its energy does not determine the minimal height
of the potential barrier between the vacua.
In fact, the latter is zero. In order to see this, it is
instructive to consider the interpolating sequence
(\ref{5.5})--(\ref{5.7}) through the $n=1$ BK solution,
and to generalize it by replacing
$\f(r)$ by $\f(\beta r)$,
where $\beta$ is a parameter. Since the boundary conditions
do not change, the new
sequence still interpolates between the distinct vacua,
but, unless $\beta=1$, no longer passes through the BK configuration.
For any $\beta$ the fields fulfill the $(00)$ and $(rr)$ Einstein
equations,
which allows one to define the energy as $m_\lambda(\infty)$, where
$m_\lambda(r)$ is given by (\ref{5.7}) with $\f(r)$ replaced
by $\f(\beta r)$. This gives
\bea
M(\beta,\lambda)=\beta\sin^2\lambda
\int_0^r\left(\f'^2+
\sin^2\lambda\frac{(\f^2-1)^2}{2r^2}\right)     \nonumber \\
\times\exp\left(-2\beta^2\sin^2\lambda
\int_r^{\infty}\f'^2\frac{dr}{r}\right)dr,           \label{5.18}
\eea
where $\f=\f(r)$ corresponds to the $n=1$ BK amplitude
(see Fig.\ref{fig:ENERGYSURF}).
For a fixed value of $\beta$ the energy is maximal at $\lambda=\pi/2$ and
vanishes as $\beta\rightarrow 0,\infty$,
so that the minimal barrier height is zero \cite{Volkov94,Volkov94b}.

In view of the last remark, one can wonder as to why the barrier
transitions in the EYM theory should be suppressed at low energies
\cite{Volkov94,Volkov94b}. Note, however, that in the low energy
limit the EYM theory reduces to pure YM theory, for which
transitions between distinct vacua are known to be strongly
suppressed, despite the fact that $\inf\{ {\rm E}_{\max}\}= 0$.%
\footnote{At finite temperature the total
transition amplitude in the YM theory can be large
due to the high number of thermal quanta participating in the
transitions, although for each individual quantum the amplitude is
small.}
On the other hand, the existence  of the energy scale
provided by the sphaleron mass suggests that the transitions
are unsuppressed above a certain energy threshold.
However, the evaluation of the path integral
is necessary in order to make any definite statements.

\subsection{Solutions with $\Lambda$-term}

A natural generalization of the BK solutions is provided
by including the $\Lambda$-term into the EYM equations
\cite{Volkov96a,Brodbeck96b,Torii95a}.
It is convenient from the very beginning to use the
general parameterization of the metric (\ref{2:32}) with
$\sigma=1$ and $\alpha=0$. The static EYM-$\Lambda$ equations then read
\bea      \label{3.5.1}
R''&=&-\frac{2\f'^2}{R} ,                           \\
(NR)'R'&=&2\left(N\f'^2-\frac{(\f^2-1)^2}{2R^2}\right) 
+1-\Lambda R^2 ,                                \label{3.5.2}   \\ 
(N\f')'&=&\frac{\f(\f^2-1)}{R^2}.               \label{3.5.3} 
\eea
There is also a first integral for these equations
\be                                             \label{3.5.3a}
(NR)'R'=2\left(N\f'^2-\frac{(\f^2-1)^2}{2R^2}\right)+1-\Lambda R^2.
\ee
The equations admit, for example, the de Sitter solution: $\f=\pm 1$, $R=r$,
$N=1-\Lambda r^2/3$. For $\Lambda=0$ the BK configurations fulfill
these equations.
The $n$-th BK soliton has the typical size $R_0(n)$,
for $R>R_0$ the metric being almost vacuum.
It follows that if the $\Lambda$-term is non-zero but very small,
$\Lambda\ll 1/R_{0}^2$, then the contribution $\Lambda R^2$ to the
energy density is negligible. For $R<R_0$ one therefore expects that
the solutions do not considerably deviate from the BK configurations.
In the region $R>R_0$, however, the effect of $\Lambda$ becomes
significant, which suggests that the metric approaches the de Sitter metric.
Hence -- for sufficiently small values of the cosmological constant --
the solutions are expected to resemble the regular BK solitons
surrounded by a cosmological horizon at $R\sim 1/\sqrt{\Lambda}$
and approach the de Sitter geometry in the asymptotic region.

%%%%%%%%%%%%%%%%%%%%%%%%%%%%%%%%%%%%% 
\begin{figure} 
\hbox to\hsize{%\hss
  \epsfig{file=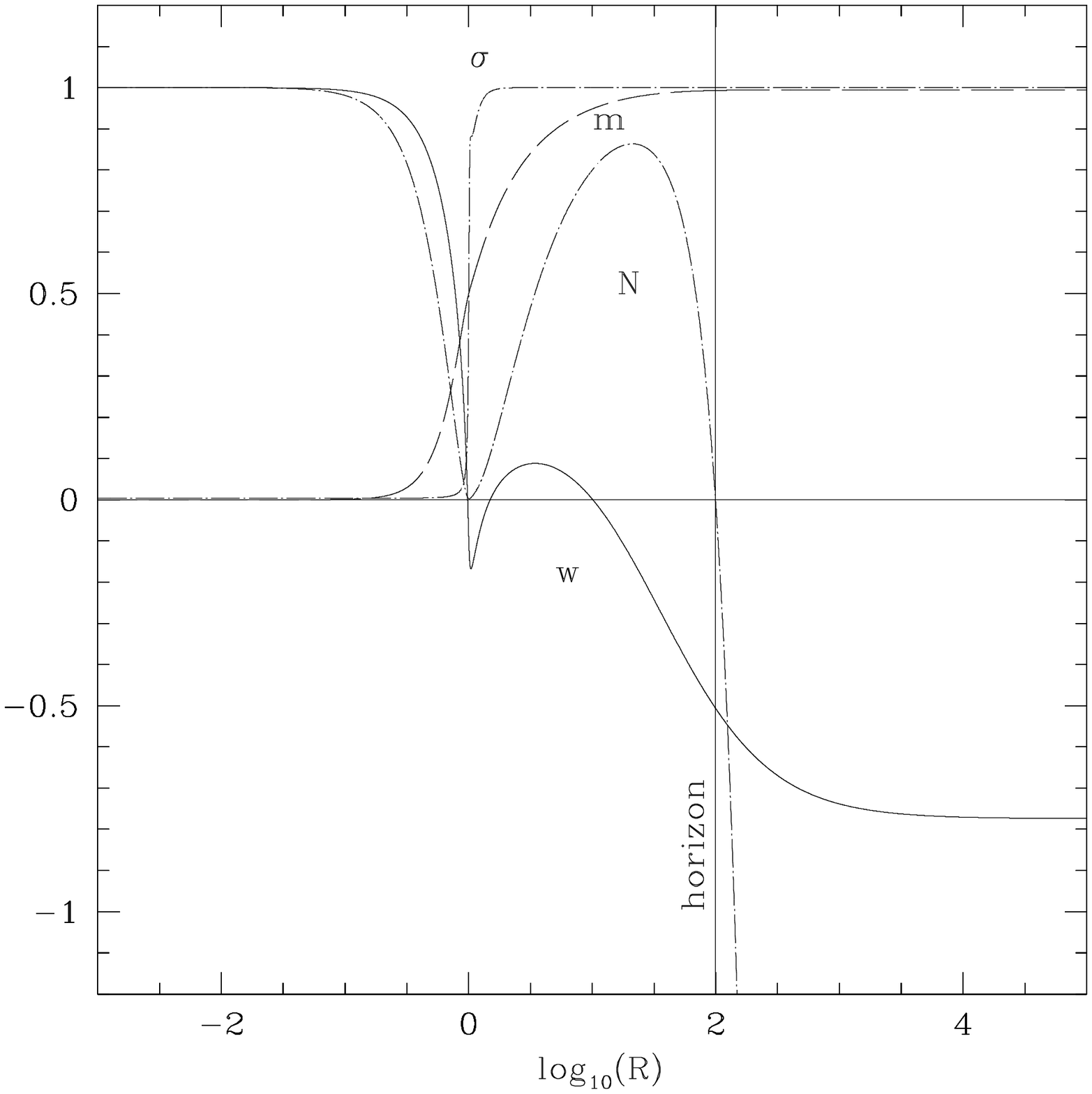,width=0.48\hsize,%
      bbllx=1.8cm,bblly=5.5cm,bburx=20.0cm,bbury=20.0cm}\hss
  \epsfig{file=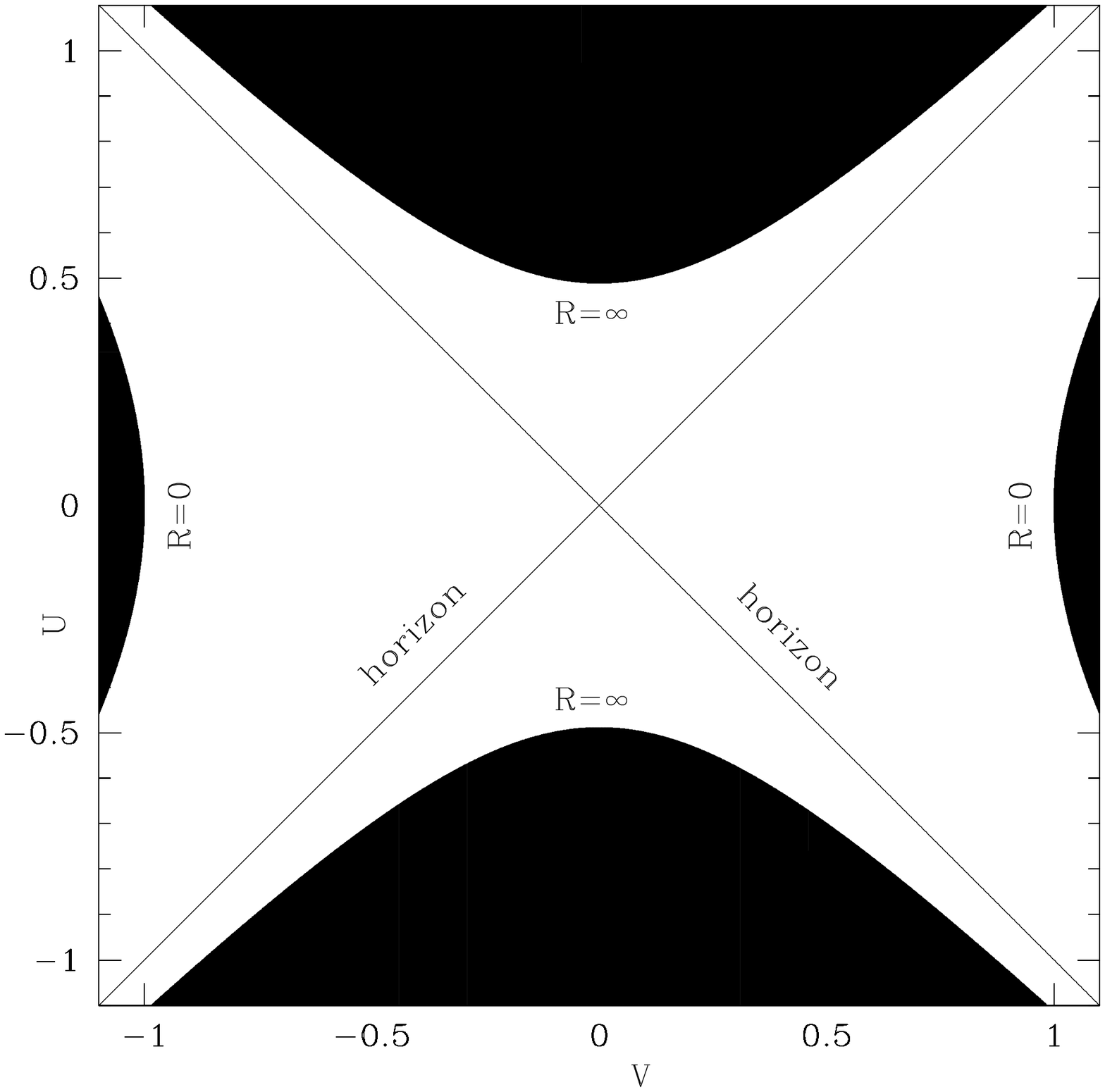,width=0.48\hsize,%
      bbllx=1.8cm,bblly=5.5cm,bburx=20.0cm,bbury=20.0cm}\hss
  }
\caption{On the left: amplitudes $w$, $N$, $m$, and $\sigma$
for the asymptotically de Sitter solution with $n=3$ and
$\Lambda=0.0003$. On the right, the conformal diagram for
this solution (the black regions should not be confused
with spacetime singularities). The definition
of the null coordinates $U$ and $V$ is given in \cite{Volkov96a}.
}
\label{fig:cosm}
\vspace{3 mm}
\end{figure}
%%%%%%%%%%%%%%%%%%%%%%%%%%%%%%%%%%%%%

The numerical analysis confirms these expectations \cite{Volkov96a}.
Starting from the regular boundary conditions at the origin
\bea                                          
&&\f=1-br^2+O(r^4),\ \ \ \ R=r+O(r^5),\ \          \nonumber \\
&&N=1+(4b^2-\Lambda/3)r^2+O(r^4),              \label{3.5.4}
\eea
the numerical procedure breaks down at some
$r_h>0$ where $N$ vanishes, which corresponds to the cosmological
horizon. The local power-series
solution at the horizon contains four
parameters: $r_h$, $R_h$, $R_h'$, and $w_h$:
\bea                                          
&&\f=w_h+w'_{h}x+O(x^2),\ \ \ \ R=R_h+R'_{h}x+O(x^2),\ \ \nonumber \\
&&N=N'_{h}x+O(x^2),                                \label{3.5.4a}
\eea
with $x=r-r_h$. Here $w_h$ and $N'_{h}$ are expressed in terms
of $R_h$, $R_h'$, and $w_h$ \cite{Volkov96a}.
The strategy is to extend numerically the local solutions
(\ref{3.5.4}) and (\ref{3.5.4a}) and impose the matching conditions
at some point $\bar{r}$,  where $0<\bar{r}<r_h$.
For any $\Lambda\ll 1$,
this leads to a family of solutions in the interval $0<r<r_h$,
which are parameterized by the node number $n=1,2,\ldots$.
For the $n$-th solution $\f$ oscillates $n$ times for $0<r<r_h$
(see Fig.\ref{fig:cosm}).

Since the boundary conditions at the cosmological horizon
are already fixed,
the next step is merely to integrate from the horizon outwards to extend
the solutions to the asymptotic region. For $r\to\infty$ one finds
\bea                                              \label{DS1}
&&\f=\f_\infty+\frac{a}{R}+O(R^{-2}),\ \ \ \ \
R=R'_{\infty}r+O(1),  \\
&& N=1-\frac{2M}{R}-\frac{\Lambda}{3}\,R^2+
\left((\f_{\infty}^2-1)^2
-\frac{2\Lambda}{3}\, a^2\right)\frac{1}{R^2}+O(R^{-3}),     \label{DS2}
\eea
where the parameters $w_\infty$, $a$, $R'_\infty$, and $M$ are
determined numerically. It turns out that $\f_\infty\neq\pm 1$,
which means that  the magnetic charge
does not vanish and in the string gauge (\ref{2.9}) is given by
\be                                           \label{3.5.5}
P=\frac{1}{4\pi}\oint_{S^2} F=(\f^{2}_{\infty}-1)\T_3.
\ee
The geometry  in the asymptotic region is RN-de Sitter.
It is interesting that the charge determined by the asymptotic
behaviour of $N$ does not coincide with the one in (\ref{3.5.5}).
Some numbers are: for the $n=1$
solution with $\Lambda=0.01$ one has $b=0.452$, $R_h=16.431$,
$\f_h=-0.944$, $\f_{\infty}=-1.001$, $M=0.821$ \cite{Volkov96a}.
 The spacetime is topologically non-trivial,
and the conformal diagram is qualitatively
identical to that for the de Sitter solution
(see Fig.\ref{fig:cosm}).
Similar to the de Sitter case, the diagram contains two
boundaries $R=0$, which now correspond to a pair of BK
solutions. One can think of the spacetime manifold as the
de Sitter hyperboloid slightly deformed by masses of  two
BK particles placed at the opposite sides of the spatial section.

%%%%%%%%%%%%%%%%%%%%%%%%%%%%%%%%%%%%%
\begin{figure}
\hbox to\hsize{%\hss
  \epsfig{file=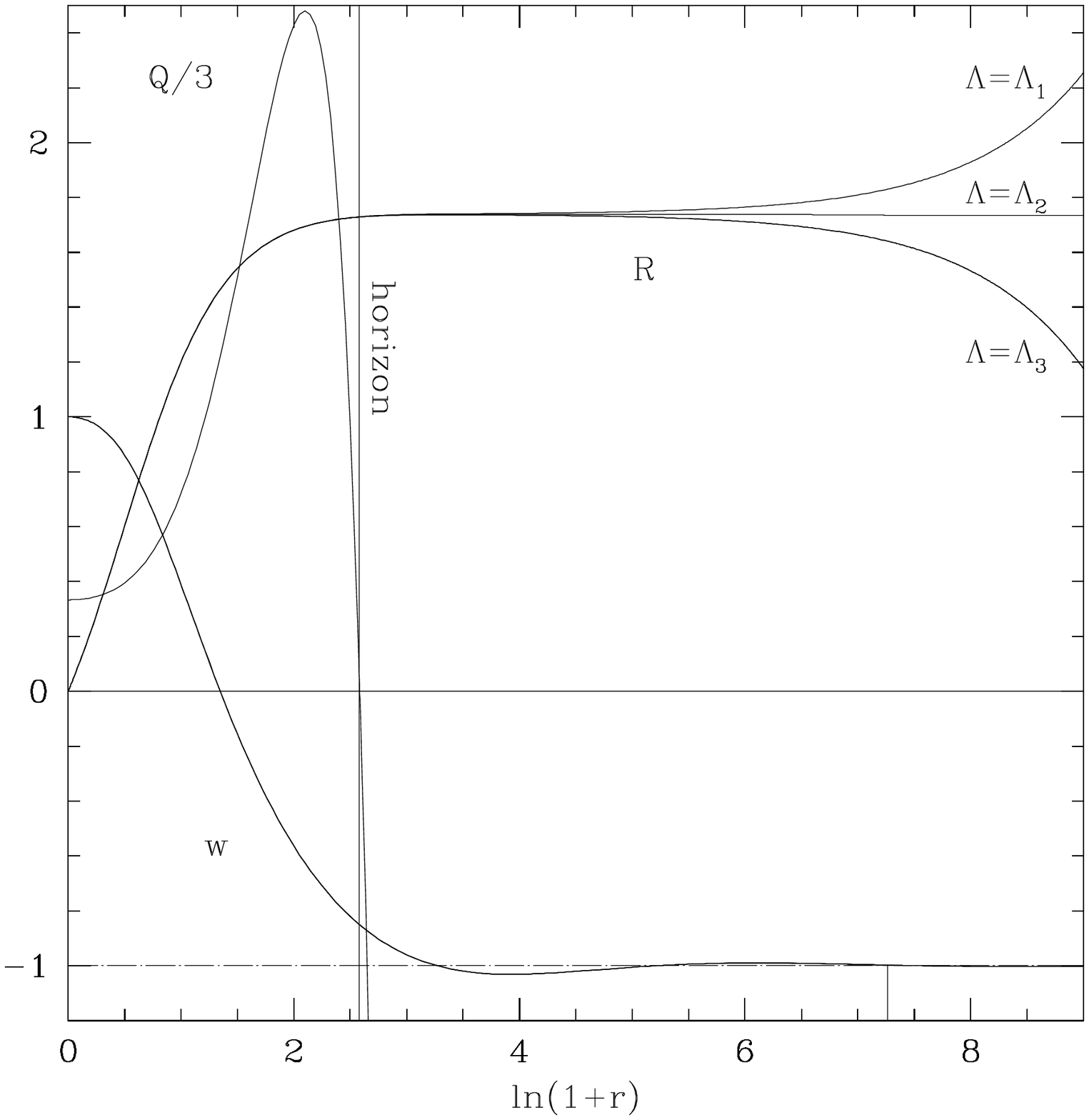,width=0.48\hsize,%
      bbllx=1.8cm,bblly=5.5cm,bburx=20.0cm,bbury=20.0cm}\hss
  \epsfig{file=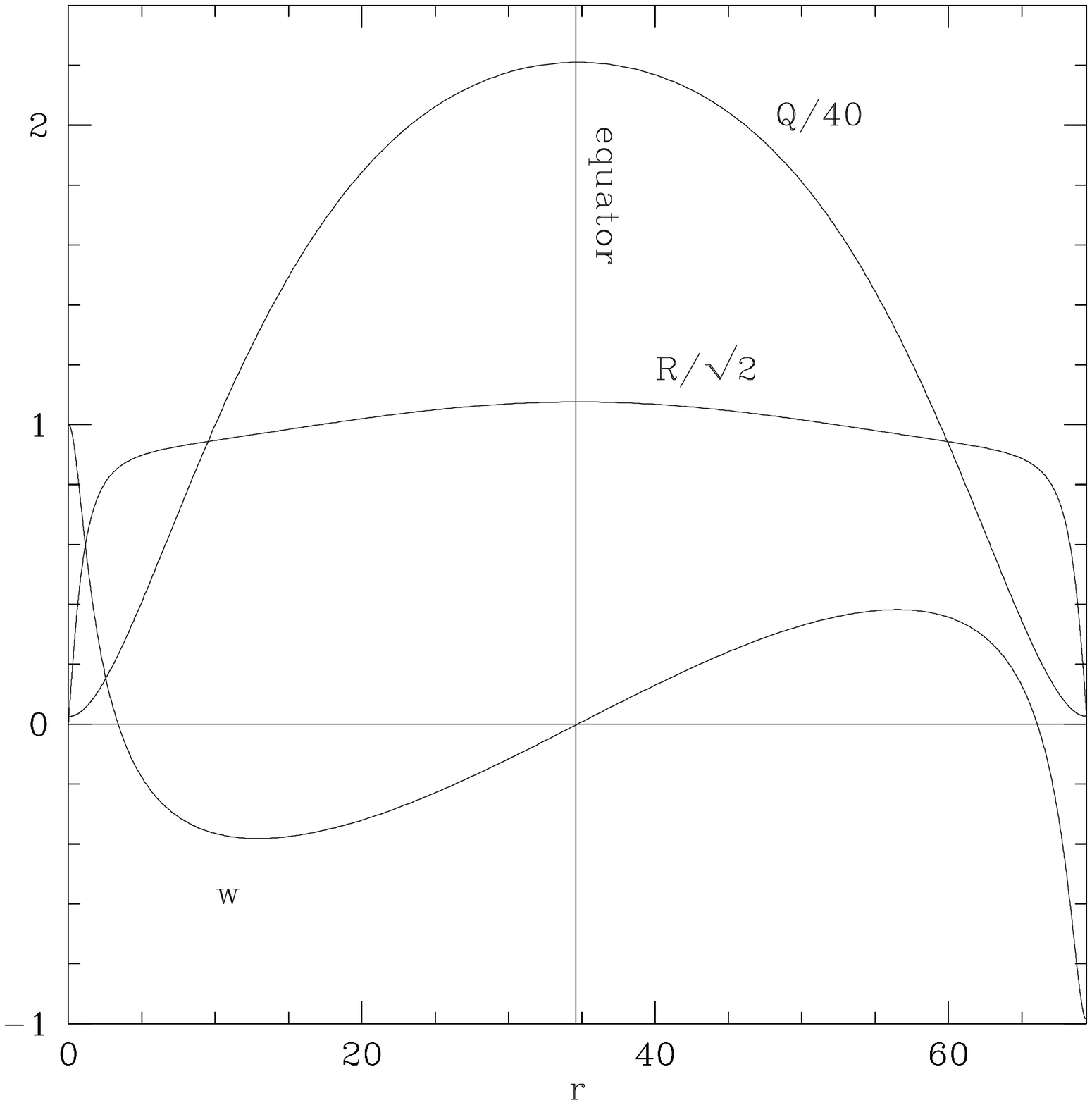,width=0.48\hsize,%
      bbllx=1.8cm,bblly=5.5cm,bburx=20.0cm,bbury=20.0cm}\hss
  }
\caption{
On the left: Change of the topology for the EYM-$\Lambda$
solutions. The $n=1$ solution for
$\Lambda=\Lambda_1=0.3304$ is asymptotically de Sitter.
The one for $\Lambda=\Lambda_2=0.3305$ is of the Nariai
type, while the one for $\Lambda=\Lambda_3=0.3306$
is of the bag of gold type. On the right: the regular
compact solution for $n=3$.
}
\label{fig:cosm1}
\vspace{3 mm}
\end{figure}
%%%%%%%%%%%%%%%%%%%%%%%%%%%%%%%%%%%%%

Let us consider now what happens for large values of $\Lambda$.
When $\Lambda$ grows,
the coefficient $R_{\infty}'$ in (\ref{DS1}) decreases until it vanishes for
some critical value $\Lambda_{\rm crit}(n)$. One has
$\Lambda_{\rm crit}(1)=0.3305$. For large $r$ the configurations then
approach the Nariai solution: $\f=\pm 1$, $R=1/\sqrt{\Lambda}$,
$N=-\Lambda r^2$.
For $\Lambda>\Lambda_{\rm crit}$,
$R(r)$ is no longer monotone: it reaches a maximum at some
finite $r_e$ and then starts decreasing until it vanishes
at some $r_{\rm sing}$, where $N$ diverges.
Such solutions belong to the bag of gold type, the `bag' containing
an event horizon and a spacetime singularity.
If $\Lambda$ continues to increase, the position of the maximum of
$R$ moves towards the origin, while the
horizon shifts towards the singularity. Finally, for
$\Lambda=\Lambda_{\rm reg}(n)$, the horizon merges with the singularity:
$r_h\rightarrow r_{\rm sing}$.
What remains is a completely regular manifold
whose spatial sections are topologically $S^3$ and have
the reflectional symmetry with respect to $r_e$.
One has $\Lambda_{\rm reg}(1)=3/4$, in which case the solution
is known analytically \cite{Ding93}
and describes the static Einstein universe
\be                                                  \label{3.5.6}
N=1,\ \ \ R=\sqrt{2}\sin(r/\sqrt{2}),\ \ \ \f=\cos(r/\sqrt{2}),
\ee
which corresponds to the standard metric on ${\rm R}\times S^3$.
For solutions with $n>1$ one has $3/4>\Lambda_{\rm reg}(n)>
\Lambda_{\rm reg}(\infty)=1/4$, and the spatial sections
are squashed three-spheres.

No regular solutions exist for $\Lambda>\Lambda_{\rm reg}(n)$.
In this case $N$ is everywhere positive and diverges at $r_{\rm sing}$
which is the position of the second zero of $R$. All EYM-$\Lambda$
solutions described above are unstable \cite{Brodbeck96b}.

\subsection{Stringy generalizations}

It is natural to wonder whether the BK solutions play any role in string
theory. The most popular stringy generalization of the EYM theory
is the EYM-dilaton theory (\ref{2:23}). One can also consider
a more general model with the (4D) Lagrangian
\be                                               \label{3:6:1}
{\cal L}=
- \frac{R}{4}
+\frac12(\nabla\phi)^2
- \frac14\, {\rm e}^{2\gamma\phi}\, (F_{a\mu\nu} F_a^{\mu\nu} -\beta
R_{GB})-U(\phi).
\ee
\noindent
Here
$R_{GB}=  R_{\mu\nu\lambda\tau}
R^{\mu\nu\lambda\tau} - 4 R_{\mu\nu} R^{\mu\nu} + R^2$
is the Gauss-Bonnet term, $\beta$
is a parameter, and $U(\phi)$ is the dilaton potential.
For $\gamma=1$, $U(\phi)=0$, $\beta=1$ this corresponds to the toroidal
compactification of the low-energy heterotic string effective action.
The axion can be set to zero provided that the gauge field is purely
magnetic. If $\gamma=1$, $\beta=0$, and $U(\phi)=(-1/8)\exp(-2\phi)$,
then the theory (\ref{3:6:1}) can be obtained via compactification
of ten-dimensional supergravity on the group manifold
\cite{Chamseddine98}.

In the static, spherically symmetric case the 2D matter field Lagrangian
following from (\ref{3:6:1}) is
\be                                        \label{3:6:2}
L_m=L_{\rm YMD}+2\gamma\beta\,{\rm e}^{2\gamma\phi}\,
\frac{\phi'}{\sigma^2}\,(N\sigma^2)'(1-NR'^2)-R^2U(\phi),
\ee
where $L_{\rm YMD}$ is given by Eq.(\ref{2:24}).
If $\beta=0$, the field equations in the gauge $R= r$  are:
\bea                                                 \label{3:6:3}
({\rm e}^{2\gamma\phi}\sigma N \f')'&= &  {\rm e}^{2\gamma\phi} \sigma\,
\frac{\f(\f^{2}-1)}{r^2}, \\
 (r^2\sigma N\phi')'&= &2\gamma \sigma
{\rm e}^{2\gamma\phi}\, \left( N\f'^2+\frac{1}{2r^2}(\f^2-1)^2 \right)
-r^2\sigma U'(\phi) ,                                \label{3:6:4}\\
m' &= & \frac{r^2}{2}\,N \phi'^2 +
{\rm e}^{2\gamma\phi}\, \left( N\f'^2+\frac{1}{2r^2}(\f^2-1)^2 \right)
+r^2 U(\phi) ,                                        \label{3:6:5}\\
(\ln \sigma)' &= & r\phi'^{2} + \frac{2}{r}\,
{\rm e}^{2\gamma\phi}\f'^{2}.                           \label{3:6:6}
\eea
We shall consider these equations for a number of special cases.

\subsubsection{EYMD theory}
Let us choose $U(\phi)=0$ and $\beta=0$.
Eqs.(\ref{3:6:3})--(\ref{3:6:6}) exhibit then the global symmetry
\be                                          \label{3:6:7}
\phi\rightarrow\phi+ \phi_0,\ \ \
r\rightarrow r\exp(\gamma\phi_0) ,
\ee
which allows one to set $\phi(0)= 0$.  The
local asymptotic solutions to (\ref{3:6:3})--(\ref{3:6:6})
are given by
\bea                                                \label{3:6:8}
&&\f=1-br^2+O(r^4),\ \  m=O(r^3),\ \ \phi= 2\gamma b^2\, r^2+O(r^4);\\
&&\f=\pm \left(1-\frac{a}{r}\right)+O(r^{-2}),\ \ m=M+O(r^{-1}), \nonumber \\
&&\phi=\phi_\infty+\frac{D}{r} +O(r^{-2}),\          \label{3:6:8a}
\eea
at the origin and at infinity, respectively.
Here $D$ is the dilaton charge. The symmetry (\ref{3:6:7}) implies
the existence of the conserved quantity
\be                                                \label{3:6:9}
N\sigma r^2\{\gamma\ln N\sigma^2-2\phi\}'
=C,
\ee
where $C$ is an integration constant.
In the regular case one has $C=0$, which leads to a first integral
of the field equations:
\be                                                \label{3:6:10}
g_{00}= \exp\left(\frac{2}{\gamma}\,(\phi-\phi_\infty)\right),
\ee
where $g_{00}\equiv N\sigma^2$.
This implies the relation between the dilaton charge and the ADM mass
\cite{Donets93b}:
\be                                                \label{3:6:11}
D= \gamma M.
\ee
It follows also that for the string theory value, $\gamma=1$,
the string metric  $ds_{s}^2= {\rm e}^{ 2\phi}ds^2$ is synchronous.

Numerical integration of Eqs.(\ref{3:6:3})--(\ref{3:6:6})
with the boundary conditions specified by
(\ref{3:6:8}), (\ref{3:6:8a}) 
shows that analogues of the BK solutions exist for any
value of $\gamma$ \cite{Donets93,Lavrelashvili93,Bizon93a}.
These EYMD solitons are also labeled by the node number $n$ of $\f$,
and the behaviour of the metric and the gauge field is qualitatively
the same as in the EYM case. The dilaton is a monotone
function.
In the limit where $\gamma\to 0$ the dilaton decouples
and one recovers the BK solutions. 
In the opposite limit, $\gamma\to\infty$, the gravitational
degrees of freedom decouple after the rescaling
\be                                               \label{3:6:11a}
\gamma\to\infty,\ \ \ \phi\to\phi/\gamma,\ \ \
r\to\gamma r.
\ee
Eqs.(\ref{3:6:3})--(\ref{3:6:6}) then reduce to those
of the flat spacetime YM-dilaton model:
\bea                                      \label{3:6:11b}
({\rm e}^{2\phi}\f')'&= &  {\rm e}^{2\phi}\,
\frac{\f(\f^{2}-1)}{r^2}, \\
(r^2\phi')'&= &2{\rm e}^{2\phi}\,
\left(\f'^2+\frac{(\f^2-1)^2}{2r^2}\right).
\label{3:6:11c}
\eea
Remarkably, these also admit regular solutions with the same
nodal structure as in the gravitating case
\cite{Lavrelashvili92,Bizon93}.
The dilaton can therefore play
a similar role like the Higgs field.
For all values of $\gamma$ the EYMD solutions have
$2n$ negative modes in the spherically symmetric perturbation sector.

Consider the modification
of the EYMD solutions due to the Gauss-Bonnet term \cite{Donets95}.
It turns out that when $\beta$ in (\ref{3:6:1}) is non-zero and
small, the qualitative structure of the solutions does not change.
However, if $\beta$ exceeds some critical value $\beta_{\rm cr}(n)$,
the solutions cease to exist. One has
$\beta_{\rm cr}(n)\leq \beta_{\rm cr}(1)=0.37$.
As a result, for the string theory value, $\beta=1$, there are no
regular particle-like solutions.
For $\beta\neq 0$ the scale symmetry (\ref{3:6:7})
still exists leading to the conserved quantity analogous
to that in (\ref{3:6:9}).
For regular solutions this implies the equality $M=\gamma D$,
but the relation (\ref{3:6:10}) no longer holds.

\subsubsection{Gauged supergravity}
For $\gamma=1$, $U(\phi)=-(1/8)\exp(-2\phi)$, and $\beta=0$
Eqs.(\ref{3:6:3})--(\ref{3:6:6}) are integrable
\cite{Chamseddine97,Chamseddine98}.
The action (\ref{3:6:1}) corresponds then to the truncation
of the $N=4$ SU(2)$\times$SU(2) gauged supergravity.
As a result, one can use the supersymmetry tools to derive
the following system of first order Bogomol'nyi equations:
\bea
N\sigma ^{2}&=&e^{2(\phi -\phi _{\infty})},  \label{3:6:13}   \\
N&=&\frac{1+w^{2}}{2}+e^{2\phi }\,\frac{(w^{2}-1)^{2}}{2r^{2}}
+\frac{r^{2}}{8} e^{-2\phi },  \label{3:6:14} \\
r\phi ^{\prime }&=&\frac{r^{2}}{8N}e^{-2\phi }\left( 1-4e^{4\phi }\,
\frac{(w^{2}-1)^{2}}{r^{4}}\right) ,  \label{3:6:15} \\
rw^{\prime }&=&-2w\frac{r^{2}}{8N}e^{-2\phi }\left( 1+2e^{2\phi }
\frac{w^{2}-1}{r^{2}}\right) ,  \label{3:6:16}
\eea
which are compatible with the field equations (\ref{3:6:3})--(\ref{3:6:6}).
The Bogomol'nyi equations are completely integrable and admit
a family of globally regular solutions
which can be represented in the form
\be                                        \label{3:6:17}
d{ s}^{2}=a^2\,
\frac{\sinh \rho}{R(\rho)}\left\{
dt^{2}-d\rho^{2}-R^{2}(\rho)(d\vartheta ^{2}+\sin ^{2}\vartheta d\varphi
^{2})\right\} ,
\ee
\be                                        \label{3:6:18}
R^{2}(\rho)=2\rho\coth \rho-\frac{\rho^{2}}{\sinh ^{2}\rho}-1,\ \
w=\pm \frac{\rho}{\sinh \rho},\ \
e^{2\phi }=a^2\, \frac{\sinh \rho}{2\,R(\rho)}.
\ee
Here $\rho\in [0,\infty)$, and the parameter $a$ reflects the
presence of the scale symmetry (\ref{3:6:7}).
These solutions are stable,
preserve 1/4 of the supersymmetries, and have unit magnetic charge.
When expressed in the Schwarzschild coordinates, the asymptotics
of the solutions are
\bea
&&N=1+\frac{r^2}{9a^2}+O(r^4),\ \
N\sigma^2=2e^{2\phi}=a^2+\frac{2r^2}{9}+O(r^4),\ \   \nonumber \\
&&w=1-\frac{r^2}{6a^2}+O(r^4);                     \label{3:6:19}\\    
&&N\propto\ln r,\ \ \ \ \
N\sigma^2=2e^{2\phi}\propto\frac{r^2}{4\ln r},\ \ \ \ \
w\propto\frac{4\ln r}{r^2};   
\label{3:6:20}                             
\eea
at the origin and at infinity, respectively.
The geometry is flat at the
origin, but it is not asymptotically flat.
The spacetime manifold is geodesically complete and
globally hyperbolic.

\subsection{Higher rank groups}

The simplest generalization of the SU(2) EYM(D) theory to higher
gauge groups is the SU(2)$\times$U(1) theory \cite{Galtsov92,Donets93a}.
However, unless the U(1) field vanishes, this does not admit
regular particle-like solutions, although black holes are possible.
The situation improves for larger groups. The SU(N) case
has been studied most of all. The existence of different embeddings
of SU(2) into SU(N) leads to several inequivalent expressions
for spherically symmetric gauge fields. Most of these
correspond to models with gauge groups being subgroups of SU(N),
and there is only one genuinely SU(N) ansatz
\cite{Wilkinson77,Kunzle91}.
In the static, purely magnetic case this is given by (\ref{2:12a}).
The EYM field equations read \cite{Kunzle94}
\bea                                               \label{3:7:1}
r^2Nw_{j}''+2(m-r{\cal P})w_j'&=&
{1\over 2}(q_{j}-q_{j+1})w_j, \\
m'&=&N{\cal G}+{\cal P},                           \label{3:7:2} \\
(\ln\sigma)'&=&2{\cal G}/r,                         \label{3:7:3}
\eea
where $j=1,\ldots ,N-1$ and $q_j= j(N-j)w_j^2-(j-1)(N-j+1)w_{j-1}^2+2j-N-1$
with $w_0= w_N= 0$. One has
\be
{\cal G}= \sum_{j= 1}^{N-1}j(N-j)\, w_j'^2,\quad
{\cal P}=
\frac{1}{4r^2}\sum_{j= 1}^{N}q_j^2.
\ee
Solutions to these equations for $N=3,4$ have been studied numerically
\cite{Kunzle94,Kleihaus95a,Kleihaus96,Kleihaus96a,Kleihaus97c,%
Kleihaus98a,Sood97}; see \cite{Mavromatos97} for the corresponding
existence proof. Since the results do not depend considerably on
whether the dilaton is included or not, we shall discuss
only the pure EYM case.
Consider the SU(3) theory, when there are two independent YM
amplitudes $w_1$ and $w_2$.
The local power-series solutions read
\bea
&&w_1=1-b_1r^2+b_2 r^3+O(r^4),\      \nonumber               \\
&&w_2=1-b_1r^2-b_2 r^3+O(r^4),\ \ \ m=O(r^3);      \label{3:7:4} \\
&&w_1=\pm \left(1-\frac{a_1}{r}\right)+O(r^{-2}),\ \nonumber \\
&&w_2=\pm \left(1-\frac{a_2}{r}\right)+O(r^{-2}),\ \ \
m=M+O(r^{-4})                                        \label{3:7:4a}                   
\eea
at the origin and at infinity, respectively. Notice the appearance
of the cubic terms in the expansions of the $w_j$'s at the origin.
The numerical matching with five shooting parameters, $b_j$, $a_j$ and $M$,
gives global solutions. For these the behaviour of $m$ and $\sigma$
is similar to that of the BK case, while the gauge field amplitudes
obey the condition $|w_j|\leq 1$ and are
characterized by two independent node numbers $n_1$ and $n_2$.
If $w_1=w_2\equiv w$, the equations
reduce to those of the SU(2) case -- up to the rescaling
$r\rightarrow 2r$, $m\rightarrow 2m$.
 As a result, there are
solutions with the node structure $(n,n)$, corresponding
to the embedded BK solutions.
There are, however, solutions with the same node structure
but with $w_1\neq w_2$, and they are slightly heavier.
Finally, there are solutions with arbitrary $(n_1,n_2)$, where
$n_1\neq n_2$. The novel feature is that one of the numbers can vanish,
while the corresponding amplitude is not constant.
The lowest value of the mass
$M(n_1,n_2)$ is $M(1,0)=2\times 0.653$, then comes
$M(2,0)=2\times 0.811$, followed by
the doubled mass of the embedded $n=1$ BK solution,
$M(1,1)=2\times 0.828$.
The next is the mass of the genuine SU(3) solution,
$M(1,1)=2\times 0.847$, and so on.
In the limit $n\to\infty$ the (0,n) and (n,n) sequences
display the oscillating behavior in the inner region, outside of which
the geometry tends to the extreme RN solution with magnetic charges
$P= \sqrt{3}$ and $P= 2$, respectively.
For comparison, in the SU(2) case the limiting solution is
characterized by $P=1$. If the dilaton is included,
then the limiting solution in the exterior region corresponds to the
extreme dilatonic black holes with magnetic charge $P$.

In the SU(4) case the nodal structure is determined by
$(n_1, n_2, n_3)$ \cite{Kleihaus98a}.
The diagonal sequence (n,n,n) contains three different types of solutions.
The first one, with $w_1= w_2= w_3$, corresponds to the $n$-th BK solution
rescaled by the factor $\sqrt{10}$.
For solutions of the second type two of the three $\f_j$'s are equal,
while for those of the third type all the $\f_j$'s are different.
For $n\to\infty$ these solutions approach in the exterior region
the extreme black hole configuration with the charge $P= \sqrt{10}$.
Other sequences tend in a similar way to the extreme solutions
with other charges, whose values
can be classified using the algebraic structure of SU(4).
A similar classification can also be carried out in the general SU(N)
case  \cite{Kleihaus97c}.
One should have in mind,
however, that for finite $n$ the regular solutions are always neutral.
For $n\to\infty$ they tend to a union of two solutions, one of which
is non-asymptotically flat and exists in the interval $0\leq r\leq P$,
while another one is the extreme RN solution in the region $r>P$.
For the regular SU(N) solution with dilaton
the equality $D= \gamma M$ still holds.

All known regular EYM solutions for higher gauge groups
are unstable \cite{Brodbeck94a,Brodbeck96a}.

\subsection{Axially symmetric solutions}

The EYM(D) field equations in the axially symmetric case
are obtained by using the ansatz ({\ref{2:axial1})
for the gauge field and that in ({\ref{2:axial}) for the metric.
These are rather complicated partial differential equations
for the seven real amplitudes in ({\ref{2:axial}), ({\ref{2:axial1}).
Due to the residual gauge invariance of ({\ref{2:axial1}),
the number of independent amplitudes is six.
The equations admit interesting solutions generalizing
the BK particles to higher winding numbers
\cite{Kleihaus97a,Kleihaus97e,Kleihaus98}.
Similar generalizations exist also
in flat space for monopoles \cite{Rebbi80} and sphalerons
\cite{Kleihaus94,Kleihaus94a}. The basic idea can be illustrated
as follows. The axial ansatz ({\ref{2:axial1}) for the gauge field
covers, in particular, the spherically symmetric case. Specifically,
choosing in ({\ref{2:axial1}) $\nu=1$ and
\bea
&&w(\rho,z)=\frac{i}{2}
\left\{1+\tilde{\f}(r)
+{\rm e}^{-2i\vartheta}(1-\tilde{\f}(r))\right\},   \nonumber  \\
&&a_\rho(\rho,z)+ia_z(\rho,z)
=\frac{\tilde{\f}(r)-1}{r}{\rm e}^{-i\vartheta}      \label{3:8:1}
\eea
with $z+i\rho=r$e$^{i\vartheta}$ and $\tilde{\f}(r)=\tilde{\f}^\ast(r)$
and omitting the tilde sign, the expression reduces to the
spherically symmetric ansatz (\ref{2.5}) with $a_0=a_r=0$.
In a similar way, the axially symmetric line element ({\ref{2:axial}) can
be used in the spherically symmetric case. As a result, given a static,
spherically symmetric solution one obtains a solution for
the axial EYM equations with $\nu=1$.
The next step is to use this solution
as the starting point in the numerical iteration scheme with
$\nu=1+\delta\nu$. The iterative field configurations
must satisfy certain regularity conditions at the symmetry axis
and at infinity.  The iterations converge,
and repeating the procedure
one obtains in this way solutions for arbitrary
$\nu$. The physical values of $\nu$ are integer. This finally gives the
generalized BK solutions characterized by a pair of integers $(n,\nu)$,
where $n$ is the node number of the amplitude $\f$ in ({\ref{2:axial1}).

For $\nu\neq 1$ the solutions are not spherically symmetric.
The contours of equal energy density $T^{0}_{0}$ are 2-torii
and squashed 2-spheres. The mass of the solutions, $M(n,\nu)$,
increases with $\nu$. For example, for $n=1$
one has $M(1,1)=0.828$ (the BK solution), $M(1,2)=1.385$,
$M(1,3)=1.870$; while for $n=2$ one finds $M(2,1)=0.971$, $M(2,2)=1.796$,
$M(2,3)=2.527$ and so on \cite{Kleihaus98}. If the winding number $\nu$
is fixed while the node number $n$ tends to infinity, then the mass
approaches the value $M(\infty,\nu)=\nu$. In this limit the solutions
exhibit a complicated oscillating behaviour in the interior region,
outside of which region the configurations approach the extreme RN
solution with the magnetic charge $\nu$.

One can think of the solutions with higher winding numbers as
describing nonlinear
superpositions of $\nu$ BK particles aligned along the symmetry axis.
This is supported by the fact that their Chern-Simons number,
which is computed in the same way as for the BK
solitons, is equal to $\nu/2$.
The solutions can be generalized
by including a dilaton field for an arbitrary value
of the coupling constant $\gamma$. It turns out that
the metric-dilaton relation (\ref{3:6:10})  holds
in the axially symmetric case too.
This implies the relation (\ref{3:6:11}) between mass and the
dilaton charge.
For $\gamma\rightarrow\infty$ gravity switches off,
but the solutions survive in this limit  \cite{Kleihaus97}.
Although the corresponding stability analysis has not been
carried out yet,
it is very likely that all these EYMD solitons are unstable.

It remains unclear whether the procedure described above gives all solutions
in the static, purely magnetic, axially symmetric case.
It is possible that configurations violating the
circularity condition can exist as well.
In particular, it is unclear whether the BK solutions exhaust all
possibilities for $\nu=1$.
%Note that the inverse
%is, strictly speaking, not true, since
%the spherically symmetric gauge field specified by Eq.(3:8:1a})
%can have arbitrary winding number.
%However, the solutions with $\nu>1$ are
%necessarily embedded Abelian and hence cannot be regular.

\section{Non-Abelian black holes}
\setcounter{equation}{0}

The intuitive physical idea behind the original no-hair conjecture
\cite{Ruffini71} apparently was that only exact physical
symmetries like gauge symmetries can survive
in a catastrophic event like a gravitational collapse.
Associated with gauge fields there are conserved charges, which can
be measured using the Gauss flux theorem. On the other hand,
all physical quantities which  are not coupled to the corresponding
gauge fields, like baryon number, are not strictly conserved.
As a result, they  either  disappear during the collapse or become
unmeasurable. The proof of this conjecture was given for a number
of special cases. In such a proof it is not the process of collapse
that is usually considered but the result of it --
a stationary black hole spacetime with certain matter fields.
It turns out that black holes cannot support
linear hair for scalar
\cite{Chase70,Bekenstein72,Teitelboim72},
spinor \cite{Hartle71,Teitelboim72a} and
massive vector \cite{Bekenstein72a,Bekenstein72b} fields;
see \cite{Greene93,Bekenstein95,Bekenstein96,Mayo96,Heusler96,Mavromatos96a}
for a more recent discussion.
A similar proof was given also for some non-linear matter models
\cite{Adler78,Bekenstein95,Mayo96,HeuslerSIG,Heusler95,Sudarsky95}.

An interesting confirmation of the no-hair conjecture was found
in the fermionic sector of the N=2 supergravity,
where there are black holes with
an external spin-3/2 gravitino field \cite{Aichelburg83}.
The latter, being subject to the Gauss law, gives rise to the
fermionic supercharge of the black hole, and this does not contradict
the conditions of the conjecture.
Note that in the bosonic sector supergravity black holes
generically support external scalar fields
associated with dilaton and
moduli fields \cite{Gibbons82}, and these are not
subject to a Gauss law. A similar example
is provided by black holes with conformal scalar field
\cite{Bocharova70,Bekenstein74}.
However, in all these cases the scalars
are completely parameterized by black hole mass and gauge
charges  and hence cannot be considered as independent --
sometimes they are called secondary hair.
In addition, the Abelian supergravity
black holes are subject to the uniqueness theorems
\cite{Breitenlohner88,Breitenlohner98}.
All this solidified the belief in the
universal validity of the no-hair conjecture.

The first example of the manifest violation of the conjecture
was found in the EYM theory%
\footnote{Hairy black holes in the Einstein-Skyrme model
had been described in \cite{Luckock87} before the EYM solutions were found,
which paper, however, has remained almost unknown.}.
In the early study of the EYM system it was observed that
for any solution $(g_{\mu\nu},{\cal A_\mu})$
of the Einstein-Maxwell theory there is a solution
$(g_{\mu\nu},A_{\mu})$ of the EYM theory with the same metric and
$ A_\mu=\T {\cal A}_\mu (x)$,
where $\T$ belongs to the Lie algebra of the gauge group \cite{Yasskin75}.
Therefore, either directly solving the EYM equations or
starting from the Kerr-Newman configurations, a number
of embedded U(1) black hole solutions for the EYM field equations
had been found \cite{Bais75,Cho75,Wang75,Perry77,Kamata82,Kasuya82,%
Kasuya82a,%
Kasuya82b}.
These were called {\em colored} black holes \cite{Perry77}%
\footnote{Unfortunately, despite the fact that they are neutral,
the same name stuck later to the non-Abelian EYM black holes.}.
In the spirit of the no-hair and uniqueness conjectures,
it was assumed for some time that these solutions exhaust all
stationary EYM black holes. A partial confirmation
of this assumption was given
by the ``non-Abelian baldness theorem"
proven for the SU(2) gauge group \cite{Galtsov89,Ershov90,Bizon92}:
all static EYM black hole
solutions with finite colour charges are the embedded Abelian ones.
This assertion, however, left open a possibility to have essentially
non-Abelian EYM black holes in the {\em neutral sector}, and
such solutions  were found in \cite{Volkov89,Kunzle90,Bizon90}.
For these black holes the only parameter subject to the
Gauss law is their mass,
and for a given value of the mass there can be several different solutions.
The no-hair conjecture is therefore violated.

The existence of one example triggered a broad search for other  hairy
black holes in various non-Abelian models. This has
led to their discovery  inside magnetic monopoles
\cite{Ortiz92,Lee92,Breitenlohner92,Breitenlohner95a},
Skyrmions \cite{Droz91,BizonSkyrm}, as well as in a number of other systems.
For the reader's convenience we list here the families
of hairy black holes known up to date:
%%%%%%%%%%%%%%%%%%%%%%%%%%%%%%%%%%%%%%%%%%%%%%%%%%%%%%%%
\begin{enumerate}
\item Black holes with pure Yang-Mills fields.
Apart from
static and spherically symmetric SU(2) EYM black holes
\cite{Volkov89,Volkov90,Kunzle90,Bizon90},
this family includes also static and
axially-symmetric solutions \cite{Kleihaus97b,Kleihaus97d},
stationary and non-static configurations \cite{Volkov97,Brodbeck97},
solutions for higher gauge groups
\cite{Galtsov92,Kleihaus95a,Kleihaus97c},
and the generalizations including a dilaton
and the higher curvature terms
\cite{Donets93,Lavrelashvili93,Torii93,Donets93a,ONeill94,Torii97,Tamaki97,%
Kleihaus96,Kleihaus96a,Kleihaus98a,Sood97,%
Kanti96,Kanti97a,Alexeyev97}.

\item Black holes with Yang-Mills-Higgs fields. These are
magnetically charged solutions with the triplet Higgs field
\cite{Ortiz92,Lee92,Breitenlohner92,Breitenlohner95a,Aichelburg93},
their analogues for a more general matter model
as well as the generalizations to
the non-spherically symmetric case
\cite{Ridgway95b,Ridgway95,Weinberg95},
and the neutral solutions with the doublet Higgs field
\cite{Greene93,Maeda94,Torii95,Tachizawa95}.

\item Skyrme black holes
\cite{Luckock87,Droz91,HeuslerHPA93,BizonSkyrm,Kleihaus95}.

\end{enumerate}
In view of this variety of hairy black holes
one can wonder as to how to reconcile their existence
with the original arguments behind the no-hair conjecture.
One possibility, which can probably be applied for unstable
solutions, is to argue that hairy black holes cannot appear via
gravitational collapse.
Another option, which is suggested by the examples
of stable black holes inside solitons, is to
assume that the topological charge
cannot disappear during the collapse. This is supported
by the fact that in all known examples the size of black holes
inside solitons is bounded from above, otherwise no hairy
solutions exist.  It seems then that a black hole cannot
swallow up a topological object with the typical size
exceeding that of the black hole.

\subsection{EYM black holes. The exterior region}

The EYM black hole solutions satisfy the same equations
as the regular BK solutions
\cite{Volkov89,Volkov90,Volkov90a,Kunzle90,Bizon90}:
\bea                                                       \label{6.1}
&&N\f''+\left(\frac{2m}{r}-\frac{(\f^2-1)^2}{r^2}
\right)\frac{\f'}{r}=\frac{\f(\f^2-1)}{r^2},                \\
&&m'=N\f'^2+\frac{(\f^2-1)^2}{2r^2}.                          \label{6.2}
\eea
The boundary conditions at infinity are still given by (\ref{4.7})
\be                                                       \label{6.3}
\f=\pm \left(1-\frac{a}{r}\right)+O\left(r^{-2}\right),\ \ \
m=M+O\left(r^{-3}\right),
\ee
which are now supplemented with the requirement that there is
a regular event horizon  at $r=r_h>0$:
\be                                                        \label{6.4}
N(r_h)=0, \ \  \ N(r)>0\ \ \ {\rm for} \ \ r>r_h.
\ee
The regularity assumption implies that all curvature invariants
at $r=r_h$ are finite. The local power-series solution to
(\ref{6.1}), (\ref{6.2})
in the vicinity of the horizon reads
\bea                                                       \label{6.5}
\f&=&\f_h+\frac{\f_h(\f_h^2-1)}{N_h'}\,
(r-r_h)+O((r-r_h)^2),          \\
N&=&\frac{1}{r_h}\left(1-\frac{(\f_h^2-1)^2}{r_h^2}\right)(r-r_h)
+O((r-r_h)^2),                                             \label{6.5.1}
\eea
It follows from
the second condition in (\ref{6.4}) that  $r_h^2>(\f_h^2-1)^2$.

For a given $r_h>0$  Eqs.(\ref{6.1})--(\ref{6.5.1}) define a non-linear
boundary value problem in the interval $r_h\leq r<\infty$.
The following simple argument
is in favour of the existence of a non-trivial
solution at least for $r_h\gg 1$. Passing everywhere to the radial
coordinate $x\equiv (r-r_h)/r_h$, some terms in the equations
acquire the factor $1/r_h^2$ and can be omitted for large
$r_h$, if only they are
bounded for $x\geq0$.
As a result, the Einstein equation (\ref{6.2}) decouples
and admits the solution $m=r_h/2$ corresponding to the
Schwarzschild metric.
The remaining YM equation (\ref{6.1}) reads
\be                                                      \label{6.6}
\left(\frac{x\f'}{x+1}\right)'=\frac{\f(\f^2-1)}{(x+1)^2} ,
\ee
where the differentiation is with respect to $x$.
The following solution to this equation is known \cite{Boutaleb79}:
\be                                                    \label{6.7}
\f=\pm\, \frac{2x-1-\sqrt{3}}{2x+5+3\sqrt{3}}\, .
\ee
This describes
{\it a regular YM hair\ } on the Schwarzschild background.
This shows that the non-trivial
solutions to the full system of equations are likely to exist,
at least for  $r_h \gg 1$.

%%%%%%%%%%%%%%%%%%%%%%%%%%%%%%%%%%%%%
\begin{figure}
\hbox to\hsize{%\hss
  \epsfig{file=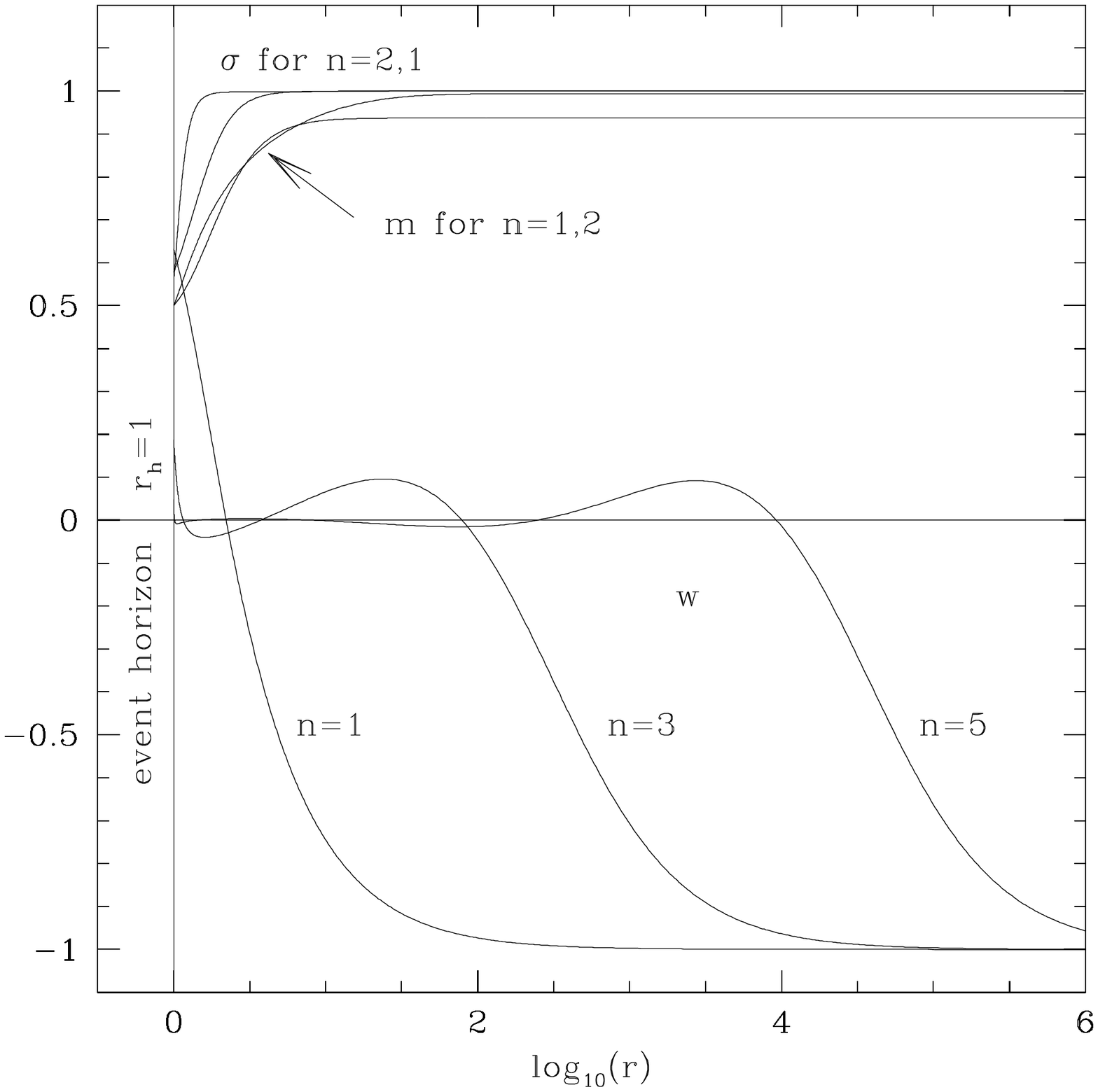,width=0.48\hsize,%
      bbllx=1.8cm,bblly=5.5cm,bburx=20.0cm,bbury=20.0cm}\hss
  \epsfig{file=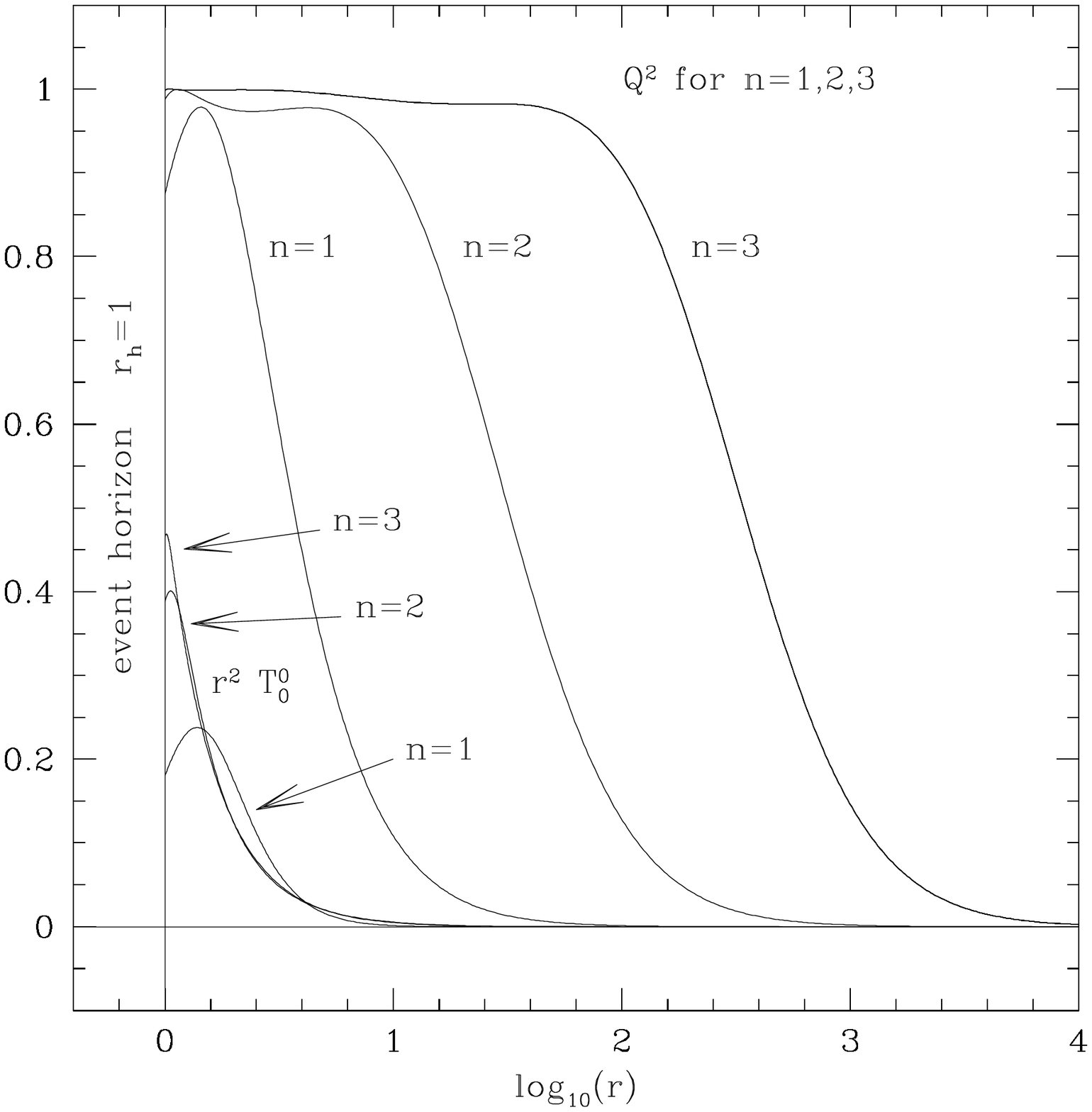,width=0.48\hsize,%
      bbllx=1.8cm,bblly=5.5cm,bburx=20.0cm,bbury=20.0cm}\hss
  }
\caption{On the left: the amplitudes $w$, $m$ and $\sigma$
for the non-Abelian black hole solutions with $r_h=1$.
On the right: the effective charge $Q^2$ and the radial
energy density $r^2T^0_0$.
}
\label{fig6:w}
\vspace{3 mm}
\end{figure}
%%%%%%%%%%%%%%%%%%%%%%%%%%%%%%%%%%%%%

In order to obtain the solutions to the full problem
(\ref{6.1}), (\ref{6.2}) the procedure is to
numerically extend the asymptotics (\ref{6.3}), (\ref{6.5})
to the intermediate region. The matching conditions for
$\f$, $\f'$ and $m$ determine then the three parameters
$M$, $a$ and $\f_h$ in (\ref{6.3}), (\ref{6.5}).
As a result, for any given $r_h>0$, one finds a
sequence of the global solutions in the interval
$r_h\leq r<\infty$. These are
parameterized by the node number $n$ of $\f$.
The non-Abelian black holes can therefore
be labeled by a pair $(r_h,n)$, where $r_h> 0$, $n=1,2,\ldots\, $.
The existence of these solutions
has been established in \cite{Smoller93a,Breitenlohner94}.

For any $(r_h,n)$
the behaviour of the amplitudes $\f$, $m$ and $\sigma$
in the exterior region
is qualitatively similar to that for the regular BK solutions
(see Fig.\ref{fig6:w}). The amplitude $\f$ starts from some value
$0<\f_n(r_h)<1$ at the horizon and after $n$ oscillations around zero
tends asymptotically to $(-1)^n$. Considerations similar to
those used in the regular case show that one has
$|\f(r)|\leq 1$
everywhere outside the horizon.
The metric functions $m$ and $\sigma$
increase monotonically with growing $r$ from
$m(r_h)=r_h/2$ and $\sigma(r_h)=\sigma_n(r_h)$
to $m(\infty)=M_n(r_h)$ and $\sigma(\infty)=1$, respectively.
In the asymptotic region, $r\rightarrow\infty$, the geometry
is Schwarzschild with the mass  $M\equiv M_n(r_h)$
depending on $r_h$ and $n$.
Since the YM field strength decays asymptotically as $1/r^3$,
the Gauss flux integral vanishes and
the node parameter $n$
cannot be associated with any kind
of YM charge of the black hole.

One can qualitatively distinguish between two regions
in the parameter space of the solutions:
{\bf a)} $r_h>1$, and {\bf b)}  $0<r_h\leq 1$.
This is due to the comparison
of the ``bare'' mass of the black hole
and its total ADM mass. The bare mass is $m(r_h)=r_h/2$.
Since $m(R)$ can be thought of as the total energy,
including the gravitational binding energy, confined in the region
$r<R$, the bare mass is the total energy trapped inside
the event horizon. Using (\ref{6.2}), the ADM mass $M\equiv m(\infty)$
can be represented
as the sum of the bare mass and the ``dressing'' mass
\be                                                    \label{6:mass}
M=\frac{r_h}{2}+\int_{r_h}^{\infty}r^2\, T^0_0\, dr,
\ee
the second term on the right being the contribution of the
matter distributed outside the horizon.

{\bf a)} $r_h>1$. In this case
the matter term in Eq.(\ref{6:mass})
is small, $M_n(r_h)\approx r_h/2$,  and the YM field
almost does not influence
the geometry outside the horizon.
The YM field energy increases
slightly with growing $n$,  the ADM mass being
(see Fig.\ref{fig6:mass})
\be                                                 \label{6:mass1}
M_1(r_h)\leq M_n(r_h)<
M_{\infty}=\frac{r_h}{2}+\frac{1}{2r_h}.
\ee
Here $M_{\infty}$ coincides with the mass of the RN
black hole with unit charge. The following approximation
is very good for $n\geq 2$ and $r\geq 1$:
$M_n(r_h)\approx M_{\infty}$ (see Fig.\ref{fig6:mass}).
This is due to the fact that for large values of $n$
the amplitude $\f$ oscillates in a small vicinity of zero.
The charge function $Q$ defined by Eq.(\ref{4:charge}) is then close to
unity, while metric is approximately RN.
For $n\rightarrow\infty$ the metric converges pointwise to the RN metric.

For $r_h\gg 1$ the matter term in Eq.(\ref{6:mass})
is of the order of $1/r_h$ and the
back reaction of the YM field on the spacetime
geometry becomes negligible. The solutions in this limit reduce
to the bound states of the YM field on the fixed Schwarzschild geometry.
These are described by Eq.(\ref{6.6}), whose solution for $n=1$
is given by Eq.(\ref{6.7}), while those for $n>1$ are known
only numerically.

{\bf b)} $0<r_h\leq 1$. As $r_h$ decreases, the effect of the YM field
on the geometry becomes more and more pronounced.
For $r_h\sim 1$ the ``bare'' and ``dressing'' masses
are of the same order of magnitude, such that
$M_n(r_h)\sim r_h$. For solutions with $r_h<1$
the role of the YM field is  dominant. The solutions then
look like small black holes dressed
in the YM ``coat''. As $r_h$ approaches zero
the external field configuration becomes more and more close to
that for the regular BK soliton with the same value of $n$.
The ADM mass for
$r_h<1$ varies within the following range:
\be                                                 \label{6:mass3}
M_1=0.828\leq M_n(r_h)<M_{\infty}(r_h)=1.
\ee
Here $M_1$ is the mass of the ground state BK solution
(see Fig.\ref{fig6:mass}). For any $r_h<1$, $M_n(r_h)$
converges to the unit value for $n\rightarrow\infty$.
In this limit an infinite number of nodes of $\f$ accumulates
near $r=1$. As a result, similar to the situation in the BK case,
the solutions
tend for $n\rightarrow\infty$ to a union of an oscillating solution
in the interval
$r_h<r<1$ and the extreme RN solution for $r>1$.

In the limit
$r_h\rightarrow 0$ the event horizon shrinks to zero and the
black hole solutions converge pointwise for $r>0$ to the regular
BK configurations.

%%%%%%%%%%%%%%%%%%%%%%%%%%%%%%%%%%%%%
\begin{figure}
\hbox to\hsize{%\hss
  \epsfig{file=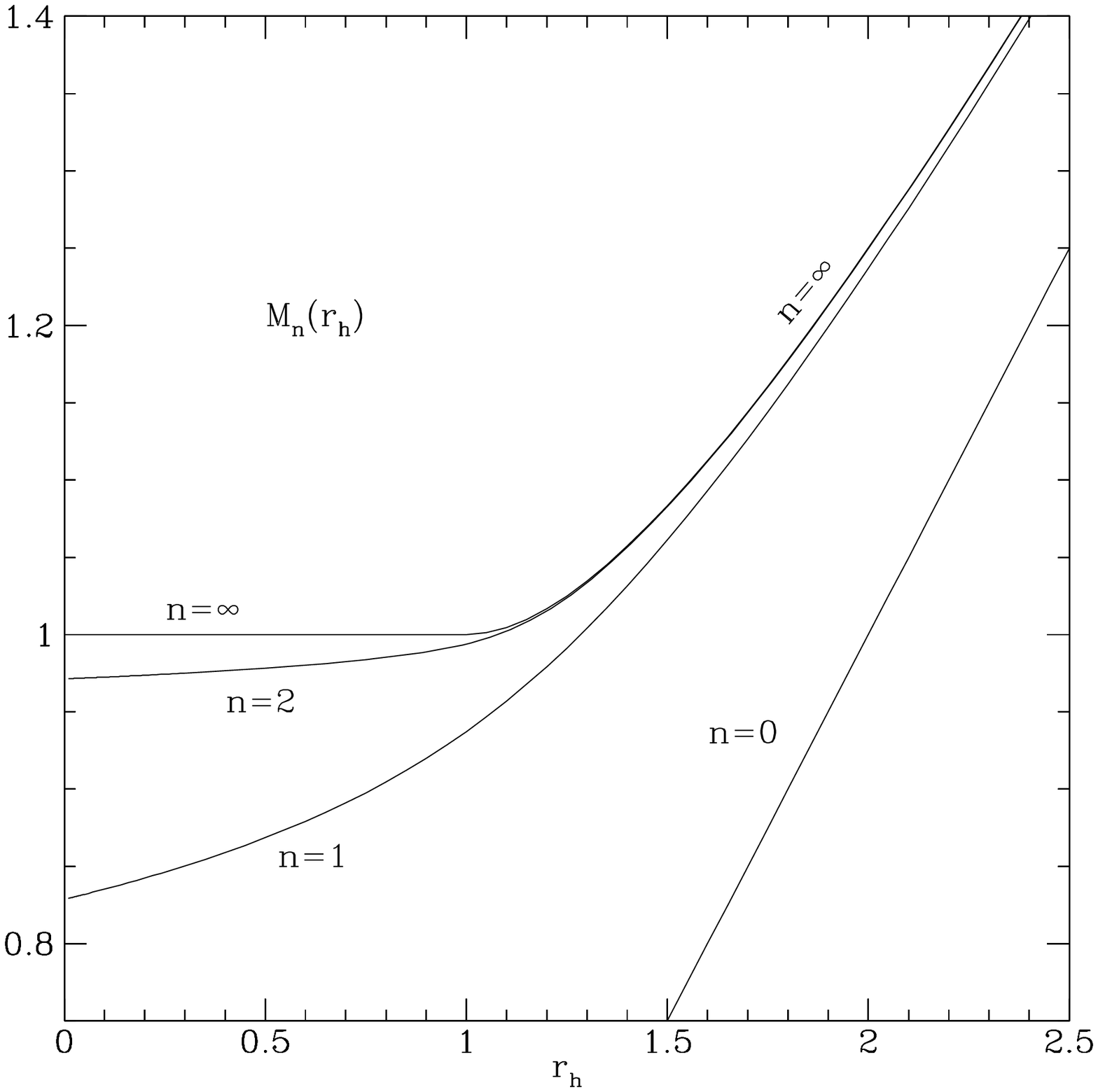,width=0.48\hsize,%
      bbllx=1.8cm,bblly=5.5cm,bburx=20.0cm,bbury=20.0cm}\hss
  \epsfig{file=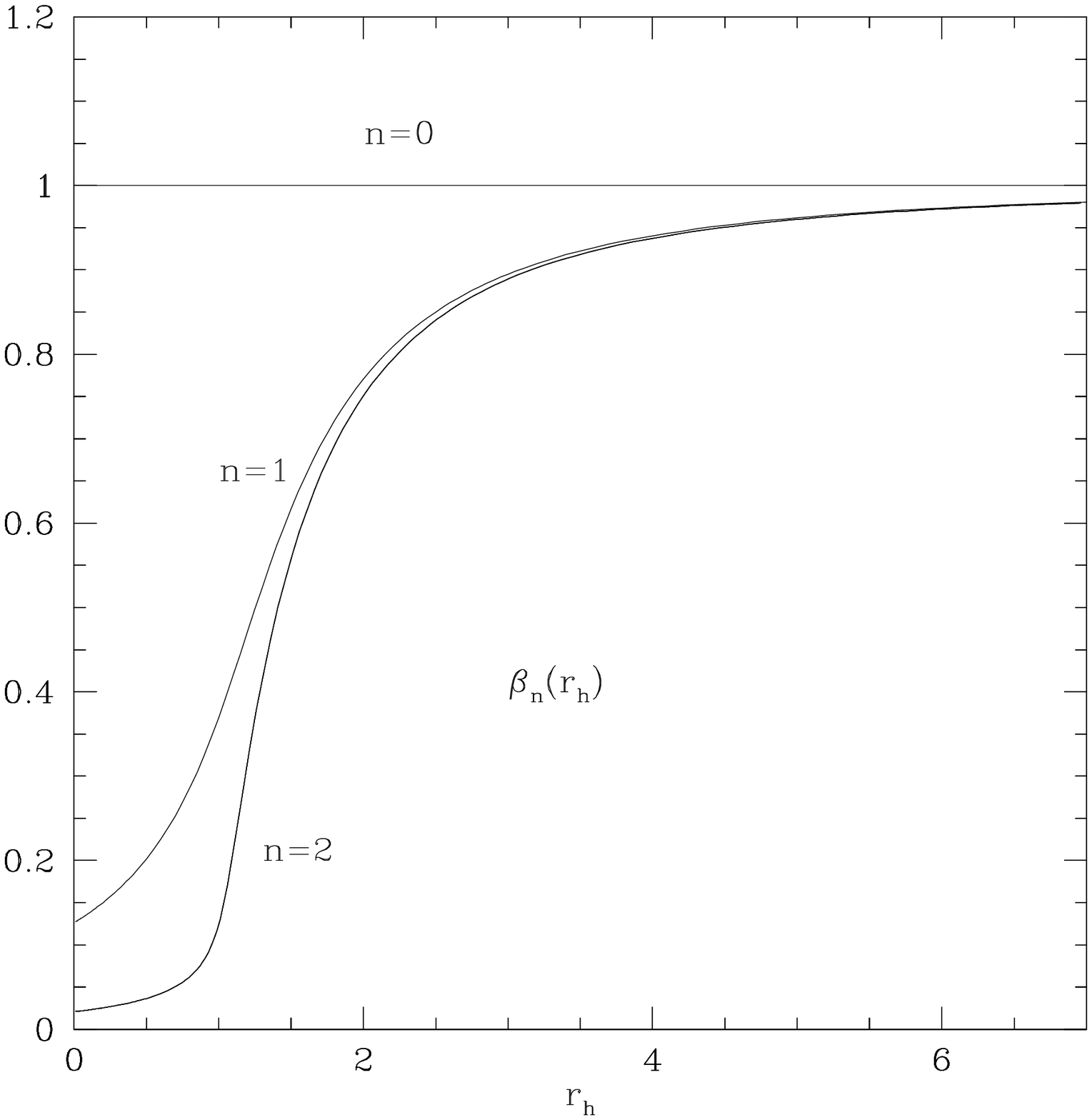,width=0.48\hsize,%
    bbllx=1.8cm,bblly=5.5cm,bburx=20.0cm,bbury=20.0cm}%\hss
  }
\caption{
The ADM mass $M_{n}(r_h)$ and the
temperature screening coefficient $\beta_{n}(r_h)$
versus the event horizon radius. For $n=0$ solutions
are Schwarzschild.
}
\label{fig6:mass}
\vspace{3 mm}
\end{figure}
%%%%%%%%%%%%%%%%%%%%%%%%%%%%%%%%%%%%%

The first law of black hole physics for the EYM black holes
was considered in \cite{Heusler93a,Heusler93}, the thermodynamics
was discussed in \cite{Moss92a,Torii93}.
The Hawking temperature of the EYM black holes can be
computed by analytically continuing the metric to the
imaginary time and requiring the absence of the conical
singularity. The result is $T=\kappa_{sg}/2\pi$, where the surface gravity
$\kappa_{sg}=\sigma_h N_h'/2$ can vanish only for the extreme RN
solution \cite{Smoller95a,Smoller96zero}. 
Using (\ref{6.5.1}) one obtains
\be
T=\beta_n(r_h)\,\frac{1}{4\pi r_h},
\ee
where the second factor on the right is the temperature
of the Schwarzschild black hole with radius $r_h$, while
$\beta_n(r_h)=\sigma(r_h)(1-(\f_{h}^2-1)^2/r_{h}^2)$
(see Fig.\ref{fig6:mass}). The presence of $\beta_n(r_h)$
in this formula leads to the existence of a (narrow) region in the
parameter space, $r_1(n)<r_h<r_2(n)$,
for which the temperature increases with growing $M$
and hence the specific heat is positive \cite{Torii93}.

All EYM black holes are unstable
and have $2n$ negative modes in the spherically symmetric
perturbation sector (see Sec.5 below). 
Note that for black holes the analogy
with sphalerons does not exist. The reason is very deep --
black holes are associated with thermal states and not with
excitations over pure vacuum states. For black holes one cannot
define classical YM vacua even formally, since
the topology is different and pure gauge fields are no longer
characterized by integer winding numbers.

\subsection{EYM black holes. The interior structure}

From the conceptual point of view the problem  of finding the solutions
in the interior region $r<r_h$ is simple. One should merely
integrate the field equations (\ref{6.1}),
(\ref{6.2}) inwards starting from the event horizon, since the
boundary conditions (\ref{6.5}) at $r_h$ are already specified
by the exterior solutions. The result, however,
turns out to be quite bizarre \cite{Donets97,Breitenlohner97}.
Therefore, before passing to the numerical analysis,
some preliminary analytical considerations can be useful.

\subsubsection{Special solutions}
It turns out that
the interior solutions generically do not
have an inner horizon. Indeed, supposing that such a horizon
exists at some $r_{-} <r_h $,
the solution in the vicinity of $r_{-}$
is given by (\ref{6.5}) with $r_h$ and $w_h$ replaced by
$r_{-}$ and some $w_{-}$, respectively. One can extend this solution
outwards to match the
one that propagates inwards from $r_h$. Now, in order
to fulfill the three matching conditions for $\f$, $\f'$ and $m$
one needs at least
three free parameters. At the same time, only $r_{-}$ and $w_{-}$ can
be used for this, since the solution coming from $r_h$
is completely specified by the boundary conditions at $r_h$.
One can use $r_h$  as the third matching parameter.
As a result, it might happen that for some special values of
the event horizon  an inner horizon exists too. However, for
an arbitrary $r_h$ the matching is impossible.
This shows that the non-Abeian black holes
generically do not possess inner horizons
\cite{Galtsov97a,Galtsov97b,Breitenlohner97a,Breitenlohner97b}
and therefore have a spacelike singularity, as
suggested by the strong cosmic censorship hypothesis.

Next, one can study the power series solutions to
(\ref{6.1}) and (\ref{6.2}) in the vicinity of the singularity
$r=0$. One finds three distinct types
of local solutions, but none of them are generic.
First, there are local solutions with a
Schwarzschild type singularity:
\bea                                                       \label{7.1}
\f&=&\pm\left(1-b r^2+b^2\,\frac{8b-3}{30m_0}\, r^5\right)+O(r^6), \\
m&=&m_0\,(1-4b^2r^2+8b^4r^4)+2b^2r^3+O(r^6),              \label{7.1a}
\eea
with arbitrary $m_0$ and $b$.
Since the inner horizon is generically absent, a solution
of this type
should match the one propagating inwards from $r_h$.
However, the number of free parameters is not enough for this.

Next, there are solutions with
a RN-type singularity:
\bea                                        \label{7.2}
\f&=&\f_0+\frac{\f_0}{2(1-\f_{0}^2)}\,r^2
+c\,r^3\, +O(r^4),\ \  \\
m&=&-\frac{(\f_{0}^2-1)^2}{2r}+
m_0 +O(r^2).   \label{7.2a}
\eea
Here the number of free parameters is three:
$m_0$, $c$, and
$\f_0\neq \pm 1$,
 which agrees with the number of
the matching condition. However, one has
$N\rightarrow +\infty$
for $r\rightarrow0$,
such that $N$ is positive near the singularity.
This requires the presence of an inner horizon,
which generically does not exist, hence this type of
singularity is also generically impossible.

Finally, there is a branch of local solutions
containing only one free parameter, $\f_0$ \cite{Donets97,Breitenlohner97}:
\bea                                           \label{IM}
 w&=&w_0 \pm r - \frac{w_0}{2(w_0^2-1)}\,r^2 +O(r^3),  \\
 m&=&\frac{(\f_{0}^2-1)^2}{2r} \pm 2w_0(w_0^2-1)+O(r). \label{IM1}
\eea
For these the singularity is spacelike, and
$N\propto -(\f_{0}^2-1)^2/r^2$ as $r\to 0$, which formally corresponds
to a RN metric with ``imaginary charge''.
The  geometry is  conformal to ${\rm R}^2\times S^2$.
It is worth noting that solutions of this type
were found  \cite{Page91} in the study of
the  phenomenon of mass inflation \cite{Poisson90}.

Summarizing, none of local power-series
solutions contain enough free parameters
to fulfill the matching conditions.
Note also that the possibility
of having a singularity at a finite value of $r$ can be ruled out
\cite{Smoller97}.

To recapitulate, solutions in the interior region do not generically
possess inner horizons and do not exhibit a power law behaviour
in the vicinity of the singularity. One can wonder then 
how these solutions look like. The answer is provided by
a numerical  analysis.  First  of  all,  it  turns  out  that  the
special solutions described above can indeed be obtained by fine tuning 
of the parameters -- apart from those in (\ref{IM}), (\ref{IM1}).
For $n=1$ and $r_h(1)=0.6138$ the singularity is
Schwarzschild-type, such that there is no inner horizon.
For all $n>2$ there are special solutions which have RN-type singularity
with an inner horizon at some $r_{-}(n)$.
For example for $n=2$ one has $r_h(2)=1.2737$ and $r_{-}(2)=0.0217$;
while for $n=3$ one finds $r_h(3)=1.0318$ and $r_{-}(3)=0.0894$.
It seems that there are also other special solutions
with the Schwarzschild-type singularity,
but these are very difficult to obtain numerically
\cite{Breitenlohner97}.

\subsubsection{Generic case}
Coming to the
{\it generic} behaviour of the solutions in the interior region,
this
turns out to be quite unusual. Specifically, choosing an arbitrary
external solution, with $n=r_h=1$, say, and integrating
inwards from the event horizon,
the metric functions $m$
and $\sigma$ and the derivative $\f'$ exhibit in the interior region
violent oscillations, whose amplitude and frequency grow
without bounds as the system approaches the singularity.
At the same time, since everything happens
at the very short scale, the amplitude $\f$ is almost
constant in the internal region.

During a typical oscillation cycle the qualitative
behaviour of the solution is as follows (see Fig.\ref{fig7:mass}).
The $k$-th cycle ($k=1,2\ldots)\, $
starts at the $k$-th local minimum of the mass function $m$
at some $r_k$ where $m(r_k)$ is very close to zero. Then, as
$r$ {\it decreases} from $r_k$ to some $R_k<r_k$, $m$ grows
exponentially fast until it reaches a very large value at $R_k$.
At the same time $\sigma$ decreases
by many orders of magnitude. In the region $r<R_k$
the mass function reaches a horizontal plateau where it
is almost constant, $m\approx M_k\equiv m(R_k)$;
similarly for $\sigma$: $\sigma\approx \sigma_k\equiv \sigma(R_k)$.
The plateau stretches as far as many orders of magnitude of $r$
towards the singularity, during which period
the geometry is approximately
Schwarzschild with a very large mass $M_k$. Then, at
the end of the plateau, $m$ catastrophically falls down reaching a very
deep minimum at some $r_{k+1}$, after which the next oscillation
cycle starts.

%%%%%%%%%%%%%%%%%%%%%%%%%%%%%%%%%%%%%
\begin{figure}
\hbox to\hsize{%\hss
  \epsfig{file=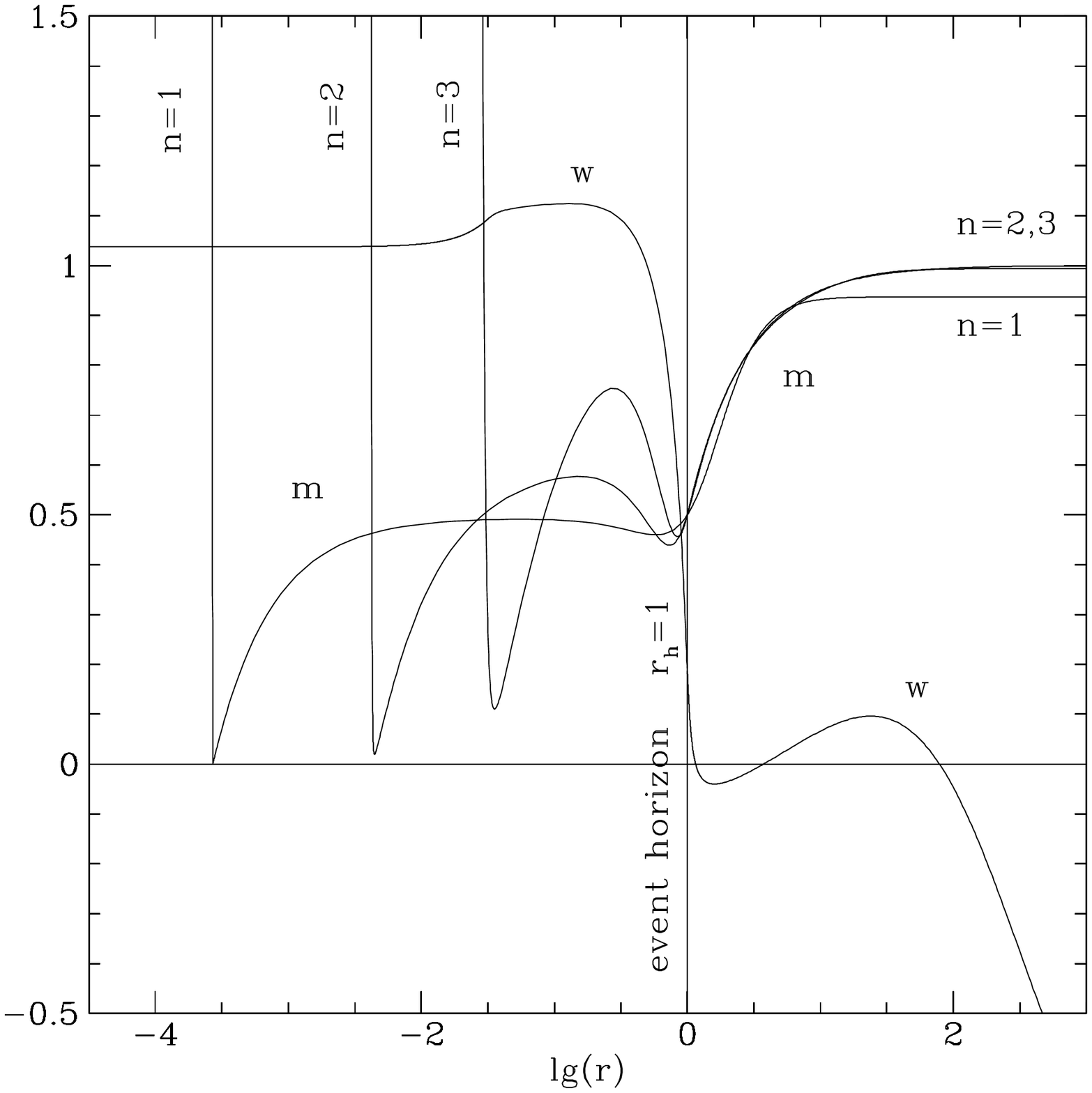,width=0.48\hsize,%
      bbllx=1.8cm,bblly=5.5cm,bburx=20.0cm,bbury=20.0cm}\hss
  \epsfig{file=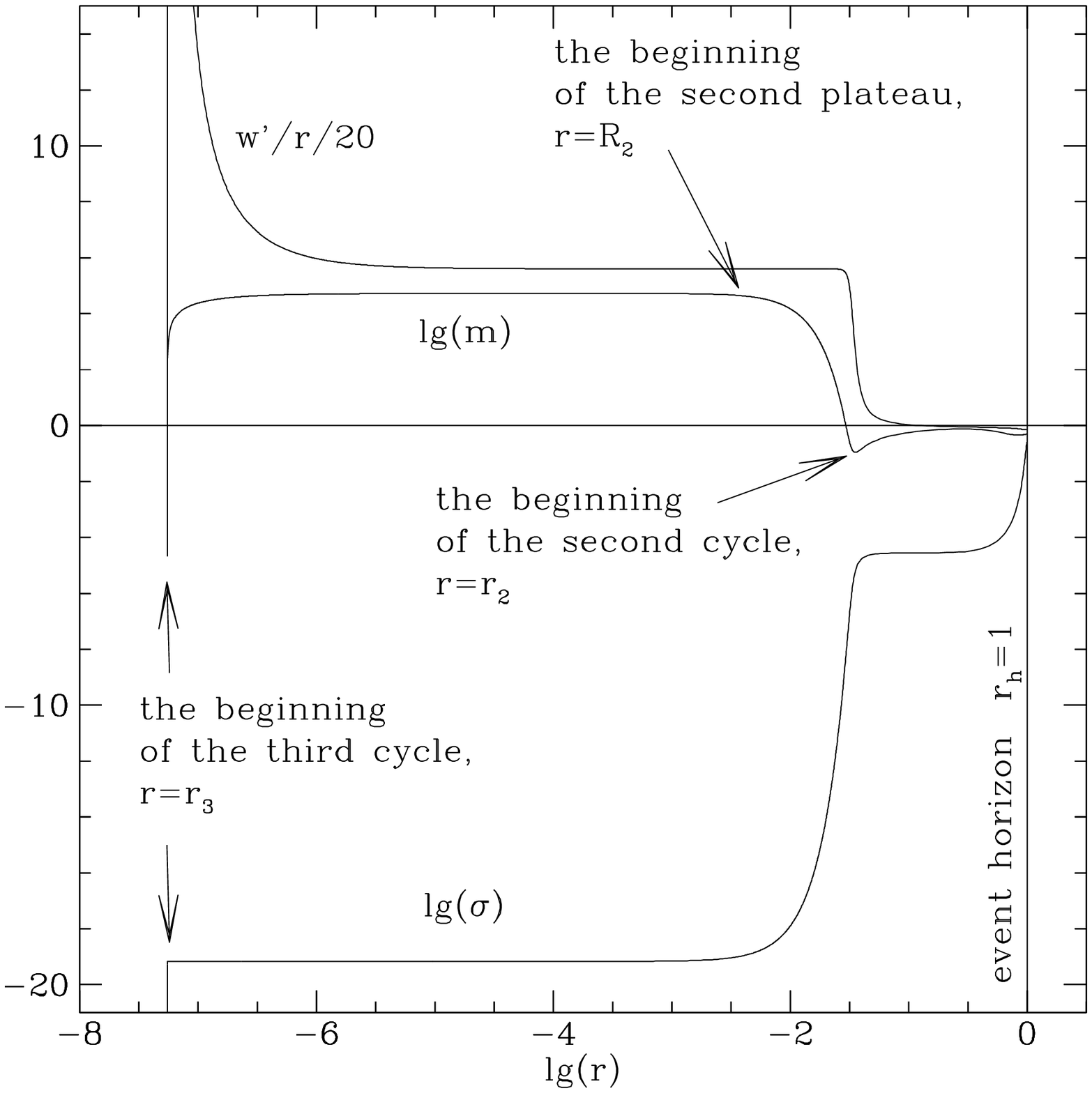,width=0.48\hsize,%
    bbllx=1.8cm,bblly=5.5cm,bburx=20.0cm,bbury=20.0cm}%\hss
  }
\caption{On the left: the mass function $m$ for the interior solutions
with $r_h=1$ and $n=1,2,3$. The amplitude $w$ is shown for the solution
with $n=3$.
On the right: the second oscillation cycle for the $n=3$ solution.
}
\label{fig7:mass}
\vspace{3 mm} 
\end{figure}
%%%%%%%%%%%%%%%%%%%%%%%%%%%%%%%%%%%%%

The oscillation amplitude for each subsequent cycle
is exponentially large compared to that for the preceding one.
As a result, it is extremely difficult to follow numerically more
than the first two or three cycles.
It turns out, however, that a simple analytic approximation
allows one to qualitatively understand the generic behavior of the system
for $r\rightarrow 0$
\cite{Donets97,Breitenlohner97}. Such an approximation is based on the
truncation of the field equations (\ref{6.1}) and (\ref{6.2})
due to the numerical observations that for $r\ll r_h$
one has 
\bea
&&w\approx {\rm const.}\neq \pm 1,            \nonumber \\
&&\frac{(w^{2}-1)^2}{r^2}\gg 1,               \nonumber \\
&&\frac{(w^2-1)^2\f'}{r}\gg w(w^2-1).         \label{7.3a}
\eea
Neglecting in Eqs.(\ref{6.1}) and (\ref{6.2})
the small terms compared to the large ones the equations
reduce to the following dynamical system:
\bea
&&\dot{x}=e^y-1,                      \nonumber \\
&&\dot{y}=1+e^y-2e^{2x},               \label{7.3}
\eea
where%
\footnote{The sign of $\f'$ is definite for $r\ll r_h$, and so one can
choose  $\f'>0$ utilizing the symmetry $\f\rightarrow -\f$.}
\be                                                 \label{7.4}
e^x=\f',\ \ \ \ \ e^y=-\frac{(\f^2-1)^2}{Nr^2}
\ee
and the differentiation is with respect to $\tau=-\ln(r/r_0)$,
with  $r_0$ being a constant.
We are interested in the behaviour of the solutions
for $\tau\rightarrow\infty$. First of all, let us notice that
there is only one critical point of the system,
$(x,y)=(0,0)$, which is an attractive center for $\tau\rightarrow-\infty$.
As a result, the behaviour of the solutions for
large and negative values of $\tau$ is known:
the trajectories starting close to the critical point
spiral outwards when $\tau$
increases (see Fig.\ref{fig7:phase}).
Now, the crucial fact is that
this spiraling motion
can never stop. One can show that, starting from
any point on the plain, the trajectory cannot escape to infinity
for finite $\tau$ and performs the full revolution
around the center within finite $\tau$. The spiraling behaviour
is therefore  generic and persists forever.
Note that the critical point around which the spiraling
occurs corresponds to $\f=\f_0$, $N=-(w_{0}^{2}-1)^2/r^2$.
For small $r$ 
this agrees with the local solution
(\ref{IM}), (\ref{IM1}).
Dynamically, however, this solution
cannot be reached. Indeed, the trajectories approach the critical point
only for $\tau=-\ln(r/r_0)\rightarrow -\infty$, such that
$r\rightarrow +\infty$, contradicting the assumption
that $r$ is small and that $r<r_h$.

%%%%%%%%%%%%%%%%%%%%%%%%%%%%%%%%%%%%%
\begin{figure}
\hbox to\hsize{%\hss
  \epsfig{file=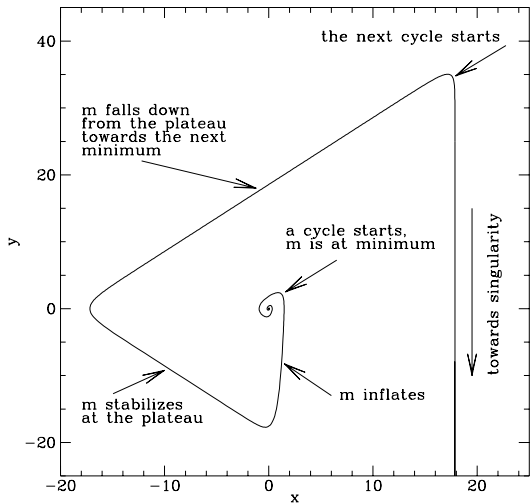,width=0.48\hsize,%
      bbllx=1.8cm,bblly=5.5cm,bburx=20.0cm,bbury=20.0cm}\hss
  }
\caption{The solution to the dynamical system (\ref{7.3}).}
\label{fig7:phase}
\end{figure}
%%%%%%%%%%%%%%%%%%%%%%%%%%%%%%%%%%%%%

The solution to the dynamical system (\ref{7.3}) shown in
Fig.\ref{fig7:phase}
illustrates nicely the various phases of a typical oscillation cycle
described above. After each revolution, the picture in Fig.\ref{fig7:phase}
repeats itself at an exponentially blown up scale \cite{Donets97}%
\footnote{This opens a possibility for numerically integrating
the equations by making a logarithmic substitution of the variables
after each cycle \cite{Zotov97}}. The
revolutions go on forever and the system approaches the singularity
via an infinite sequence of more and more violent oscillation cycles.
At the beginning of each cycle $N$ is very close to zero, such that
the system exhibits an infinite sequence of ``almost'' inner horizons,
but the true Cauchy horizon never appears.
Note that the oscillatory nature of the solution resembles somewhat
the well-known situation in some cosmological models \cite{Belinskii70}.
However, the solutions
are {\it not chaotic}. This follows from the existence of the effective
description in terms of the two-dimensional dynamical system,
since chaos cannot occur in two dimensions.
Inside black holes, a similar oscillatory behaviour was observed in
\cite{Page91}
for a homogeneous mass inflation model with radial null radiation.

The values of amplitudes and periods
of the oscillations can be estimated analytically \cite{Donets97a}.
The essential
quantities for this are $x_k=(r_k/R_k)^2\gg1$,
where $r_k$ and $R_k$ correspond to the beginning of the $k$-th cycle
and the $k$-th plateau, respectively. One can show that the $x_k$'s
obey the following recurrence relation
\be                                                          \label{7.5}
x_{k+1}=\frac{e^{x_k}}{x_{k}^3},
\ee
thus constituting an exponentially diverging sequence. In terms
of $x_k$ one has
\be
\frac{r_{k+1}}{r_k}=x_k\, e^{-x_k/2},
\ee
which can also be viewed as the ratio of the neighboring
oscillation periods, since $r_k\gg r_{k+1}$. The increase in $M_k$
and decrease in $\sigma_k$ can be expressed as
\be
\frac{M_k}{M_{k-1}}=\frac{e^{x_k/2}}{x_k},\ \ \ \ \ \ \ \ \ \ \ \ \
\frac{\sigma_{k+1}}{\sigma_{k}}=e^{-x_k/2},
\ee
respectively. One can see that the ratios of the oscillation amplitudes
and frequencies for the neighbouring cycles are exponentially large, where
the argument of the exponent grows exponentially for each subsequent cycle.

Note finally that the exterior structure of the
EYM solutions does not change considerably after adding some additional
matter, like a dilaton or a Higgs fields, say. However, the solutions
in the interior region
change completely
\cite{Donets97a,Breitenlohner97,Zotov97,Galtsov97,Galtsov97b,Galtsov97c,%
Sarbach97} and exhibit a regular power-law behaviour
near the singularity.
In this sense  the
oscillatory character of the interior solutions described above
distinguishes the EYM black holes among all other
known hairy black holes. Note also that, although the
EYM black hole solutions in the exterior region are unstable, their interior
oscillatory structure is stable with respect to non-linear
spherically symmetric perturbations
\cite{Donets98}.

\subsection{Non-Abelian dilaton black holes}

The EYM non-Abelian black holes can be generalized to include a dilaton
\cite{Donets93,Lavrelashvili93,Torii93}.
The field equations are then given by
(\ref{3:6:3})--(\ref{3:6:6}).
The boundary conditions at infinity are specified by (\ref{3:6:8a}),
while at the horizon one has
\bea                                             \label{bhdil}
\f=\f_h+\f'_h x+O(x^2),\
\phi=\phi_h+\phi_h'x+O(x^2),\
N=N'_h x+O(x^2)
\eea
with $x=r-r_h$, and $\f_h'$, $\phi_h'$, $N_h'$
being determined by $r_h$, $\f_h$  and $\phi_h$.
In the black hole case
the scale transformations (\ref{3:6:7}) change $r_h$. However,
since the solutions exist for any $r_h$ and $\phi_h$, one can
set $\phi_h=0$ without loss of generality.
The numerical integration gives for
any $\gamma<\infty$ a family of black hole solutions
parameterized by $(r_h,n)$ as in the $\gamma=0$ case.
The behaviour of $\f$, $m$, and $\sigma$ is
qualitatively the same as for the EYM solutions, while the
dilaton is a monotone function.

The conservation law  following from the scaling symmetry
exists in the black hole case too, but now the integration constant
in (\ref{3:6:9}) does not vanish and is given by
$C=\gamma\sigma_h N_h'r_{h}^2$. As a result,
the relation (\ref{3:6:11}) no longer holds, but instead
one finds
\be                                       \label{d3c.1}
D=\gamma \left( M-\frac{1}{4\pi}\,\kappa_{sg}{\cal A}\right),
\ee
where $\kappa_{sg}=\sigma_h N_h'/2$ is the surface gravity
and ${\cal A}=4\pi r_{h}^2$ is the area of the event horizon.
%The area is vanishes when $n\rightarrow\infty$,
%in which case the relation (\ref{3:6:11}) is reached.

The thermodynamics of dilaton black holes depends on value of the
dilaton coupling constant $\gamma$ \cite{Torii93}.
For small values of $\gamma$
there is a region in the  parameter space for which the specific
heat of the solutions is positive. For large $\gamma$
this region shrinks to zero. The stability behaviour of the solutions,
on the other hand, do not depend on $\gamma$: the number of
negative modes is the same for all values of $\gamma$.

What is entirely different from the EYM case
is the behaviour of solutions in the interior region.
First of all, for $\gamma\neq 0$
inner horizons cannot exist even after fine tuning
of the parameters \cite{Sarbach97}.
Indeed, the expression in (\ref{3:6:9})
reduces at the event horizon $r_h$ to $C=\gamma\sigma_h N_h'r_{h}^2$,
which is positive. At the inner horizon $r_{-}$
one has $C=\gamma\sigma_{-} N_{-}'r_{-}^2$, which is negative.
This contradicts the fact that $C$ is constant, and hence inner
horizon cannot exist.
The numerical analysis reveals the following characteristic picture
in the interior region, which is generic for all $\gamma\neq 0$:
the mass function $m$,
after some oscillations near the horizon, exhibits the power-law
behaviour all the way down to the singularity  \cite{Galtsov97,Sarbach97}.
The power-series solution in the vicinity of the singularity for
$\gamma\neq 0$ is generic:
\begin{equation} \label{family}
      w = w_0 + a r^{2 (1 - \lambda)}, \quad
      m = \mu r^{- \lambda^2}, \quad
      \phi   = c + \ln \left( r^{- \lambda} \right),\quad
      \sigma=\sigma_1 r^{\lambda^2},
\end{equation}
since it contains the maximal number of free parameters:
$w_0$, $a$, $c$, $\mu$, $\lambda$, and $\sigma_1$.
As $\gamma$ decreases, the solutions develop more and more
oscillation cycles in the interior region. However, as long
as $\gamma\neq 0$, the oscillations are always
replaced by the power-law behaviour for $r\to 0$.
This can be illustrated
as follows \cite{Sarbach97}. For $\gamma=0$
the interior oscillating solution can effectively be described by the
two-dimensional dynamical system (\ref{7.3}), whose trajectories
evolve in the plane $(x,y)$.
For $\gamma\neq 0$ there is an
additional degree of freedom due to the dilaton,
and the reduced system becomes
three-dimensional. As a result, after a number
of revolutions in the vicinity of the plane, which corresponds
to several oscillation cycles,  the trajectory drifts away
towards a critical point
of the equations that lies outside the plane.
This changes the character of the solution.

The EYMD black holes can be generalized in the context of the theory
with the Gauss-Bonnet term \cite{Kanti97a,Torii97}.
The string theory value of the coupling constant in (\ref{3:6:1}),
$\beta=1$, is allowed for black holes.
The solutions are parameterized by three numbers:
$(\phi_h,r_h,n)$. This is because the event horizon for $\beta\neq 0$
exists only if $r_h\geq r_0>0$,
which implies that one cannot use the scale symmetry to set $\phi_h=0$.
One has $r_0\sim\beta\exp(2\gamma\phi_h)$. For $n=0$, when
the YM field strength vanishes, the solutions describe black holes
with dilaton hair \cite{Kanti96,Kanti97,Alexeyev97}.

\subsection{Higher gauge groups and winding numbers}

The SU(2) non-Abelian black holes 
admit generalizations to higher gauge groups
\cite{Galtsov92,Donets93a,Donets93b,Kunzle94,Kleihaus95a,%
Kleihaus96,Kleihaus96a,%
Kleihaus97c,Kleihaus97e,Kleihaus98a,Sood97}. 
These exist for all values of the dilaton coupling 
constant $\gamma$.
In the maximal SU(N) case,
when all the gauge field amplitudes $\f_j$ are non-trivial,
the solutions are neutral and can be parameterized by $(r_h,n_j)$,
where $n_j$ is the node number of $\f_j$.
The regular limit is recovered for $r_h\to 0$.
% they tend to the regular SU(N) solutions.
The solution space exhibits bifurcations in this limit, since
for $r_h\to 0$ there exist
different solutions with the same node structure
which merge as $r_h$ grows. If some of the $n_j$'s
tend to infinity, the solutions approach the RN (for $\gamma=0$)
or charged dilaton (for $\gamma\neq 0$) black hole solutions.

One of the new features is that when some of the $\f_j$'s vanish identically,
the U(1) charges arise.
The non-Abelian baldness theorem hence
does not generalize to higher gauge groups.
For example, setting in the SU(3) equations (\ref{3:7:1})--(\ref{3:7:3})
$\f_1=0$, the solutions for $\f_2$ show the usual nodal structure.
However, since $\f_1$ does not obey the
boundary condition in (\ref{3:7:4a}),  there is a
magnetic charge $P=\sqrt{3}$ for all $n_2$.
This is reflecting in the behaviour of $N$
for $r\to\infty$: $N=1-2M/r+P^2/r^2+O(1/r^3)$.
For higher gauge groups one can
obtain in this way a variety of charged solutions, in particular,
solutions with an electric charge (for a more general ansatz
for the gauge field). For charged EYM black holes
the event horizon radius is bounded  from below,
$r_h\geq P$, since otherwise the singularity is naked.
The limit where $r_h\to 0$ for the charged solutions is therefore
impossible.
 Solutions with $r_h=P$ are extreme black holes with degenerate
horizon.
 For charged solutions with dilaton the $r_h\to 0$ limit is allowed,
but this leads to the extreme dilaton black holes and not to
regular particle-like solutions. For such extreme dilaton black holes
the metric-dilaton relation
$N\sigma^2= {\rm e}^{2(\phi-\phi_\infty)/\gamma}$ is restored.
At the origin these configurations approach the
extreme Abelian solutions with the same value of charge \cite{Gibbons82},
while showing the usual nodal structure for $r\geq 1$.

The SU(2) EYM(D) black holes admit also axially symmetric
generalizations \cite{Kleihaus97b,Kleihaus97d}.
By far, these give the only known explicit example of static and
non-spherically symmetric black holes with non-degenerate horizon.
The solutions
can be obtained by extending  the spherically symmetric black hole
configurations
to the higher values of the winding number $\nu$, similarly to the
procedure in the regular case.
The solutions are asymptotically flat and have a regular event horizon,
but for $\nu>1$
they are not spherically symmetric and the event horizon
is not a sphere but a prolate ellipsoid.
Specifically, expressing coordinates $\rho ,z$ used in
the axial line element ({\ref{2:axial}) in terms of
$r,\vartheta$ as $z=r\cos\vartheta$, $\rho=r\sin\vartheta$,
the event horizon is defined by the relation $r=r_h$,
where $g_{00}(r_h)=0$. The horizon hence has the $S^2$ topolgy,
but geometrically it is not a sphere, since its circumference
along the equator turns out to be different from that along a meridian.
The surfaces of constant energy density are either ellipsoidal
or torus-like. It is interesting that the energy density is not constant
at the horizon and depends on $\vartheta$.
However, the surface gravity $\kappa_{sg}$, where
\be
\kappa^{2}_{sg}=-\frac14\, g^{00}g^{ik}\,\partial_i g_{00}\,\partial_k g_{00},
\ee
is constant at the horizon.
If a dilaton is included, then the relation (\ref{d3c.1}) between
mass, dilaton charge and the event horizon parameters holds
also for $\nu>1$. The solutions are characterized
by $(r_h,\nu,n)$. Here  $\nu$ and $n$ are the winding number
and the node number of the amplitude $\f$ in (\ref{2:axial1}),
respectively.
The deviation from spherical symmetry increases with
growing $\nu$, but decreases for large $n$. As
$n\to\infty$ the solutions tend to the Abelian spherically
symmetric black hole solutions. In the limit of shrinking
horizon, $r_h\to 0$,
the solutions reduce to the regular axially symmetric EYM(D) solitons.
More details and interesting pictures can be found in \cite{Kleihaus97d}.

\section{Stability analysis of EYM solutions}
\setcounter{equation}{0}

The stability issue for the BK solitons and EYM black holes has
been extensively studied 
\cite{Straumann90,Straumann90a,Zhou91,Zhou92,Volkov95,Volkov95a,VolkovDUBNA,%
Lavrelashvili95,Brodbeck93,Brodbeck94a,Brodbeck96a},
both perturbatively and at the non-linear level.
A novel feature arising in this analysis is the fact that,
unlike the situation in the (electro)-vacuum case,
a non-trivial temporal dynamics for EYM fields exists
already in the spherically symmetric sector.
The Birkhoff theorem thus does not apply
\cite{Bartnik88,Brodbeck93}, and the physical
reason for this can be deduced
from the ``spin from isospin'' phenomenon \cite{Jackiw76a,Hasenfratz76}.
The spin and isospin of the gauge field
can combine together giving zero value for the sum,
in which case the vector YM field effectively
behaves as a scalar field. As a result, spherical waves
exist in the theory.  It turns out then that all known regular
and black hole solutions in the EYM theory with the gauge group
SU(2) are unstable with respect to small
spherically symmetric perturbations \cite{Volkov95}.
This is true also for the solutions for higher gauge groups
\cite{Brodbeck94a,Brodbeck96a}. The numerical analysis indicates
that the non-linear instability growth results in the complete dissipation
of the initial equilibrium configuration \cite{Zhou91}.
The energy is then partially
carried away to infinity by spherical waves and partially
collapses to the center.

It turns out that the $n=1$ BK solution can play the role
of the intermediate attractor in the gravitational collapse
of the YM field \cite{Choptuik96}. Near the boundary in the initial
data space between data which form black holes and data which do not
there is a region characterized by a mass gap. The minimal
black hole mass for data from this region is equal to the mass
of the $n=1$ BK solution. This is reminiscent of a first order phase
transition. There is also another region in the parameter space
characterized by the echoing phenomenon
and the scaling behavior for the black hole mass,
which can now be arbitrarily small \cite{Choptuik96}.
This is explained by the existence of another intermediate
attractor in the problem. The corresponding solution to the EYM
equations is time-dependent. It describes the situation
when there is a constant ingoing flux of the YM radiation
which comes from infinity,
completely reflects from the origin, and then goes back to infinity.
The system balances just on the verge of collapse but it does
not actually collapse due to fine tuning of the parameters.
The approximate description of this solution was found in
\cite{Gundlach97}.

In what follows we shall consider only
small, spherically symmetric perturbations for the BK solitons
and non-Abelian black holes. The complete set of perturbations
splits into two parity groups.
The even-parity sector is obtained via
perturbing the non-vanishing
gauge and metric amplitudes in Eqs.(\ref{2.40}) and
(\ref{2.41}):
\be                                          \label{8.1}
\f\rightarrow\f+\delta\f(t,r),\ \
m\rightarrow m+\delta m(t,r),\ \
\sigma\rightarrow\sigma+\delta\sigma(t,r).
\ee
The odd-parity group consists of perturbations of
those amplitudes in the Witten ansatz (\ref{2.5})
that vanish for static, purely
magnetic background solutions. These are $a_0$, $a_r$, and Im\,$\f$:
\be                                          \label{8.2}
a_0\rightarrow\delta a_0(t,r),\ \
a_r\rightarrow\delta a_r(t,r),\ \
{\rm Im}\,\f\rightarrow\delta( {\rm Im}\,\f).
\ee
The parity of the modes is determined by
Eq.(\ref{2.8}).  Since the background
solutions are parity-invariant, the two groups
of perturbations decouple at the linearized level.

\subsection{Even-parity perturbations}

Substituting (\ref{8.1}) into the EYM equations
(\ref{2:33})--(\ref{2:37}) and linearizing with respect to
$\delta\f$, $\delta m$, and $\delta\sigma$ gives the perturbation
equations. It turns out that
the gravitational amplitudes $\delta m$ and $\delta\sigma$
can be expressed in terms of $\delta\f$, which leads to
a single Schr\"odinger-like equations for $\delta\f$.
The procedure is  as follows.
Linearizing Eq.(\ref{2:36}) one
arrives at $\delta\dot{m}=2N\f'\delta\dot{\f}$ with $N=1-2m/r$,
whose solution with the appropriate boundary conditions is
\be                                                 \label{8.3}
\delta m=2N\f'\delta\f .
\ee
Similarly, linearizing Eq.(\ref{2:35}) one obtains
\be                                                \label{8.4}
\delta\left(\sigma'/\sigma\right)=4\f'\delta\f'/r.
\ee
Finally, the Yang-Mills equation (\ref{2:34})  
containt after the linearization $\delta\f$,
$\delta m$ and $\delta\left(\sigma'/\sigma\right)$, where the
last two terms can be expressed
using (\ref{8.3}) and (\ref{8.4}). As a result, one arrives at
\cite{Straumann90}
\be                                               \label{8.5}
\left\{-\frac{d^2}{d\rho^2}+\sigma^2 N\frac{3w^2-1}{r^2}+
2\left(\frac{\sigma^{\prime}_{\rho}}{\sigma}\right)^{\prime}_{\rho}
\right\}\eta=\omega^2\eta,
\ee
where $\delta\f(t,r)=\exp(i\omega t)\, \eta(\rho)$, and the      
functions $\f$, $N$, and $\sigma$ refer to the background
solution under consideration. The ``tortoise''
radial coordinate $\rho$ is defined by
\be                                              \label{8.6}
\frac{d\rho}{dr}=\frac{1}{\sigma N},
\ee
where $\rho\in [0,\infty)$ for solitons and
$\rho\in (-\infty,\infty)$ for black holes.
It is worth noting that Eq.(\ref{8.5}) can also be obtained
via computing the second variation of the mass functional
$M(1,\pi/2)$ in Eq.(\ref{5.18}) with respect to $\f$
\cite{Volkov94,Volkov94b,Brodbeck96c}.
The first variation vanishes on shell,
while the second one is
\be
\delta^2 M=\int\delta\f(\rho)\left(-\frac{d^2}{d\rho^2}+U
\right)\delta\f(\rho)\, d\rho,
\ee
where the potential $U$ coincides with the one in Eq.(\ref{8.5}).

Eq.(\ref{8.5}) determines a regular
Schr\"odinger eigenvalue problem on a line or semi-line.
The potential is bounded for
black holes, while for solitons it can be represented
as a sum of a bounded piece and
the $p$-wave centrifugal term $2/\rho^2$ (see Fig.\ref{fig8:U}).
The numerical analysis
reveals the existence of $n$ bound state solutions to Eq.(\ref{8.5})
with $\omega^2<0$ for the $n$-th BK
or EYM black hole background (with any $r_h$) \cite{Straumann90,Straumann90a}.
This implies that the solutions are unstable.
The non-linear instability growth for the
BK solitons has been analyzed numerically  \cite{Zhou91,Zhou92}.
It has been found that a part of the initial static configuration
collapses  forming a small Schwarzschild black hole,  while the
rest radiates away to infinity. It is worth noting that
for the EYM black holes with $n=1$ the negative eigenmode gives rise to
the gauge field tensor $\delta F$ that is
unbounded at the horizon \cite{Bizon91}. However, it turns
out that this mode is nevertheless physically acceptable, because
the divergence can be suppressed by making wave packets
with real frequency eigenmodes \cite{Bizon91a,Wald92}.

\subsection{Odd-parity modes}

The odd parity perturbation sector also contains negative modes,
whose existence can be established 
\cite{Galtsov92a,Boschung94,Brodbeck94a,Volkov95,Brodbeck96a}
without even resorting to numerical analysis.
The number of such modes can also be
determined analytically \cite{Volkov95a,VolkovDUBNA}.
For the BK solitons
the existence of the odd-parity negative modes directly relates to the
sphaleron interpretation of the solutions.
This is due to the fact that the Chern-Simons number of the gauge field
changes for pulsations of the type (\ref{8.2}).
The even-parity perturbations,
on the other hand, preserve the Chern-Simons number.
In order to see this it is instructive to compute the
$\ast FF$ invariant for the spherically symmetric gauge field (\ref{2.5}):
\bea                                           
\frac{\sigma r^2}{2}\, {\rm tr}\ast F_{\mu\nu}F^{\mu\nu}&=&
\left( a_1 (|\f|^2-1 )+
p_2 p_{1}^{\prime}-p_1 p_{2}^{\prime}\right)^{\displaystyle .} \nonumber \\
&-&\left(a_0 (|\f|^2-1)+
p_2 \dot{p}_{1}-p_1 \dot{p}_{2}\right)^{\prime},               \label{8.6a}
\eea
where $p_1\equiv{\rm Re}\,\f$ and $p_2\equiv{\rm Im}\,\f$.
This vanishes if $a_0=a_r={\rm Im}\,\f=0$.

The perturbation equations in the odd-parity sector
are obtained by linearizing the Yang-Mills
equations (\ref{2:33}),(\ref{2:34}) 
with respect to $\delta a_0$, $\delta a_r$ and
$\delta({\rm Im}\,\f)$,
which gives a system of three linear equations.
For time-dependent perturbations
one can impose the temporal gauge condition $\delta a_0=0$.
As a result,
the two independent perturbation equations can be represented
in the Schr\"odinger form
\be                                          \label{8.7}
\hat{H}\Psi=\omega^2\Psi.
\ee
Here, introducing $\hat{p}\equiv-id/d\rho$
and $\gamma_\ast\equiv\sqrt{2N}\sigma/r$,
the Hamiltonian $\hat{H}$ and the wavefunction $\Psi$ are given by
\be                                          \label{8.8}
\hat{H}=
\left(\begin{array}{cc}
\f^2\gamma_\ast^2 & \gamma_\ast\f^{\prime}_{\rho}
-\gamma_\ast\f\, i\hat{p} \\
\gamma_\ast\f^{\prime}_{\rho}+ i\hat{p}\, \gamma_\ast\f\ \ \  &
\hat{p}^2+\frac{1}{2}\gamma_\ast^2(\f^2-1)
\end{array}\right),\ \ \
\Psi=
\left(\begin{array}{c}
\zeta \\
\xi
\end{array}\right),
\ee
with
$\delta a_r (t,r)=\exp(i\omega t)\sqrt{2/r^2 N}\, \zeta(\rho)$ and
$\delta({\rm Im}\,\f(t,r))=\exp(i\omega t)\, \xi(\rho)$.
The radial variable $\rho$ is defined by Eq.(\ref{8.6}).
The third perturbation equation, which is
the linearized Gauss constraint (Eq.(\ref{2:33}) with $\beta=0$), is
a differential consequence of (\ref{8.7}).

%%%%%%%%%%%%%%%%%%%%%%%%%%%%%%%%%%%% 
\begin{figure} 
\hbox to\hsize{%\hss 
  \epsfig{file=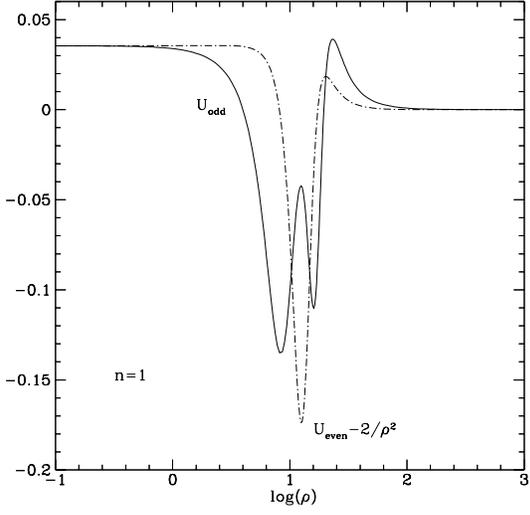,width=0.48\hsize,% 
      bbllx=1.8cm,bblly=5.5cm,bburx=20.0cm,bbury=20.0cm}\hss 
  } 
\caption{Potentials $U_{\rm even}$ and $U_{\rm odd}$ in the 
Schr\"odinger eigenvalue problems (\ref{8.5})
and (\ref{8.12}), respectively. 
} 
\label{fig8:U} 
\vspace{3 mm}
\end{figure} 
%%%%%%%%%%%%%%%%%%%%%%%%%%%%%%%%%%%%% 

The Hamiltonian in (\ref{8.8}) is formally self-adjoint
and real, which ensures the existence of self-adjoint extensions.
In order to show that the spectrum has a negative part, it
suffices to find a function $\Psi$ such that
\be                                           \label{8.9}
\omega^2(\Psi)\equiv\frac{\langle\Psi|\hat{H}|\Psi\rangle}
{\langle\Psi|\Psi\rangle}<0.
\ee
Here the scalar product
$\langle\Psi|\Phi\rangle=\int\Psi^\dagger\Phi d\rho$.
For solitons the function $\Psi$ with the required properties
is specified by \cite{Boschung94}
\be                                          \label{8.10}
\zeta=\f^{\prime}_{\rho}/\gamma_\ast,\ \ \ \ \ \ \xi=\f^2-1.
\ee
Note that this can be obtained by differentiating the
vacuum-to-vacuum interpolating family (\ref{5.6})
with respect to the
parameter $\lambda$ at the sphaleron position,
$\lambda=\pi/2$. The direct computation gives
\be                                         \label{8.11}
\langle\Psi|\hat{H}|\Psi\rangle=
-\int\left(\f^{\prime 2}_{\rho}
+\frac{\gamma_\ast^2}{2}(\f^2-1)^2\right)d\rho <0,
\ee
while the norm $\langle\Psi|\Psi\rangle$
is finite. An expression similar to (\ref{8.10})
can be found also in the black hole case \cite{Volkov95}.
As a result, all BK solitons and EYM black holes are unstable.

The above variational argument can be generalized
to arbitrary gauge groups. The conclusion is:
all static, spherically symmetric and purely magnetic regular or
black hole solutions to the EYM equations
for arbitrary gauge groups are unstable \cite{Brodbeck94a,Brodbeck96a}.
It is interesting that this conclusion can be reached without
explicit knowledge of the possible equilibrium solutions.

The actual number of the odd-parity negative modes for the $n$-th
BK or EYM black hole background is $n$.
In order to see this one makes use of the fact
\cite{Volkov95a,VolkovDUBNA} that for $\omega\neq 0$
the eigenvalue problem in (\ref{8.7}), (\ref{8.8})  can be mapped
to the equivalent {\it one-channel} Schr\"odinger problem
\be                                               \label{8.12}
\left\{-\frac{d^2}{d\rho^2}+\frac{\gamma_\ast^2}{2}\, (3w^2-1)+
2\left(\f^2 Z\right)^{\prime}_{\rho}
\right\}\psi=\omega^2\psi.
\ee
Here
\be                                               \label{8.13}
\psi=\left(\frac{d}{d\rho}+\frac{\f^{\prime}_{\rho}}{\f}
+\f^2 Z\right)\frac{\zeta}{\gamma_\ast\f},
\ee
and $Z$ is a solution of an auxiliary nonlinear differential
equation
\be                                               \label{8.14}
Z^{\prime}_{\rho}=\f^2 Z^2-\gamma_\ast^2.
\ee
The direct verification shows that for $\omega=0$ Eq.(\ref{8.12})
has a solution
\be                                               \label{8.15}
\psi_0=\f\exp\left(\int_{\rho_0}^{\rho}\f^2 Zd\rho\right).
\ee
One can show that $Z$ can be chosen in such
a way that the potential in (\ref{8.12}) is everywhere bounded,
both for solitons and black holes (see Fig.\ref{fig8:U}),
while $\psi_0$ is normalizable. Due to the factor $\f$,
the number of nodes of $\psi_0$ is $n$.
The well-known theorem of quantum mechanics guarantees then
that the eigenvalue problem (\ref{8.12}) has exactly $n$ negative
energy eigenstates%
\footnote{For multi-channel Schr\"odinger problems
bound states can be counted by using the nodal theorem
recently proven in  \cite{Amann95}.}.

\begin{table}
\caption{The eigenvalues of the negative modes for the
$n=1$ EYM solutions.}
\vglue 0.4 cm
\begin{tabular}{|c|c|c|} \hline
$r_{h}$    & $\omega^{2}$ (even-parity) & $\omega^{2}$ (odd-parity)\\ \hline

0 (soliton case) &  $-0.0524$   & $-0.0619$   \\ \hline

0.1              &  $-0.0502$   &  $-0.0618$  \\ \hline

0.5              &  $-0.0402$   &  $-0.0600$  \\ \hline

1                &  $-0.0268$   &  $-0.0492$  \\ \hline

5                &  $-0.0021$   &  $-0.0029$  \\ \hline

10               &  $-0.0005$   &  $-0.0007$  \\
\hline
\end{tabular}
\vspace{3 mm}
\end{table}

In order to actually find the bound states
one solves numerically
Eqs.(\ref{8.7}), (\ref{8.8}) or Eq.(\ref{8.12})
\cite{Lavrelashvili95,VolkovDUBNA}. For the $n=1$
EYM soliton and black holes
the bound state energies for both values of parity are presented
in Table 2.
Summarizing, for any $n$ and for all values of the 
event horizon radius $r_h$
the BK solitons and non-Abelian black holes are unstable with respect
to small spherically symmetric perturbations. The number
of instabilities is $2n$, of which $n$ belong to the odd-parity
sector and $n$ are parity-even. 

\section{Slowly rotating solutions}
\setcounter{equation}{0}

The difference between the non-Abelian EYM theory and
its Abelian counterpart is emphasized further still when one studies
stationary generalizations of the static BK solitons and EYM black holes
 \cite{Volkov97,Brodbeck97,Brodbeck97a}.
It turns out that rotating EYM black holes acquire
an electric charge, and not just a dipole correction to the background
gauge field which one would normally expect in the Abelian theory.
As a result, spinning up the neutral static solutions makes
them charged up. The family of stationary EYM black holes includes
in addition quite peculiar solutions which are non-static and
have non-vanishing angular velocity of the horizon,
but the total angular momentum measured at infinity is zero.
This shows that the electrovacuum staticity theorem
\cite{Carter87} does not generalize 
in a straightforward way to the EYM theory.
The regular BK solutions admit charged, stationary generalizations too.
This is even more unusual, since
solitons in other field-theoretical models, such as the
t'Hooft-Polyakov monopoles, say, cannot (slowly) rotate
\cite{Brodbeck97a,Heusler98}.

Studying stationary EYM fields presents certain difficulties.
Although solutions for static, axially symmetric
deformations of the BK solitons and EYM black holes have been
obtained numerically
\cite{Kleihaus97a,Kleihaus97b,Kleihaus98,Kleihaus97d},
the problem becomes much more involved
in the stationary, non-static case \cite{Galtsov98}. In particular,
one realizes then that the EYM equations do not imply the
Frobenius integrability conditions for the Killing vectors
\cite{HeuslerHPA}.
The standard Papapetrou ansatz can therefore be too narrow.
This shows that the Abelian circularity theorem also
does not generalize to EYM systems in a straightforward way.

Although solutions to the full stationary problem are still lacking,
the perturbative analysis based on the assumption of linearization
stability has been carried out \cite{Volkov97,Brodbeck97,Brodbeck97a}.
The basic idea is as follows.
Suppose that there is a one-parameter family of stationary
black hole or regular solutions of the EYM equations,
approaching  the static solutions
for angular momentum $J=0$.
The tangent to this family at $J=0$ satisfies
the linearized EYM equations. Conversely, it is reasonable to expect that
for a well-behaved solution of the linearized equations around the
static configurations there exists an exact one-parameter family
of stationary  solutions. Accordingly, the problem reduces to studying
the linear rotational excitations for the static BK solitons and
EYM black holes.

\subsection{Perturbation equations}

Consider small perturbations
around the BK solitons or EYM black holes:
\be                                                   \label{9.1}
g_{\mu\nu}\rightarrow g_{\mu\nu}+h_{\mu\nu},\ \ \ \
A_{\nu}\rightarrow A_{\nu}+\psi_{\nu},
\ee
where $(g_{\mu\nu},A_{\nu})$ refer to the background configuration,
and it is convenient to express the gauge field in the string gauge
(\ref{2.9}):
\be                                                 \label{9.1a}
 A=\f\, \, (\T_2\, d\vartheta - \T_1\sin\vartheta\, d\varphi)
+\T_3\cos\vartheta\, d\varphi.
\ee
The perturbation equations are obtained by  linearizing
the EYM field equations (\ref{1.7})-(\ref{1.9})
$$
-\nabla_{\sigma}\nabla^{\sigma}h_{\mu\nu}-
2\, R_{\mu\alpha\nu\beta}h^{\alpha\beta}+
R^{\sigma}_{\mu}h_{\sigma\nu}+
R^{\sigma}_{\nu}h_{\sigma\mu}=
4\, \delta T_{\mu\nu},
$$
\be                                                     \label{9.2}
-D_{\sigma}D^{\sigma}\psi_{\nu}+R^{\sigma}_{\nu}\psi_{\sigma}
-2[F_{\nu\sigma},\psi^{\sigma}]+
h^{\alpha\beta}D_{\alpha}F_{\beta\nu}+
F^{\alpha\beta}\nabla_{\alpha}h_{\beta\nu}=0.
\ee
Here $D_{\nu}\equiv \nabla_{\nu}+[A_{\nu},\ \cdot \ ]$,
and the gauge conditions
\be                                                      \label{9.3}
\nabla_{\sigma}h^{\sigma}_{\nu}=h^{\sigma}_{\sigma}=
D_{\sigma}\psi^{\sigma}=0
\ee
are imposed.
$\delta T_{\mu\nu}$ in Eqs.(\ref{9.2}) is obtained by varying
the energy-momentum tensor
\be                                                      \label{9.4}
T_{\mu\nu}=\frac{1}{2}\, {\rm tr}\left(
F_{\mu\alpha}F_{\nu\beta}\, g^{\alpha\beta}
-\frac{1}{4}g_{\mu\nu}
F_{\alpha\beta}F_{\rho\sigma}g^{\alpha\rho}g^{\beta\sigma}\right)
\ee
with respect to the metric and the gauge field.

\subsubsection{General mode decomposition}
In order to identify the most general rotational degrees of freedom,
one  determines those amplitudes in the partial wave decomposition
which can give a non-vanishing contribution to the ADM flux
integral for the total angular momentum
\be                                                        \label{9.5}
J^i=\frac{1}{32\pi}\oint_{S^2}\varepsilon_{ink}\,
(x^k\, \partial_j h^{0n}+\delta^n_j\, h^{0k})\, d^2 S^j.
\ee
In order to carry out the partial wave decomposition for
$(h_{\mu\nu},\psi_{\nu})$, it is convenient to introduce the
complex 1-form basis $\theta^\alpha$:
\be                                             \label{9.6}
\theta^0=\sigma\sqrt{N} dt,\ \
\theta^1=\frac{dr}{\sqrt{N}}, \ \
\theta^2=\frac{r}{\sqrt{2}}\,(d\vartheta-i\sin\theta\, d\varphi),\ \
\theta^3=(\theta^2)^\ast,
\ee
the non-vanishing components of the tetrad metric
$\eta^{\alpha\beta}\equiv(\theta^\alpha,\theta^\beta)$ being
$\eta^{00}=-\eta^{11}=-\eta^{23}=1$.
In addition, one introduces the new Lie-algebra basis:
${\bf L}_1=\T_1+i\T_2$,
${\bf L}_2=\T_1-i\T_2$,
${\bf L}_3=\T_3$.
The perturbations then expand as
\be                                        \label{9.7}
h_{\mu\nu}=\theta^{\alpha}_{\mu}\theta^{\beta}_{\nu}\,
{\bf H}_{\alpha\beta},\ \ \ \ \
\psi_{\mu}={\bf L}_a
{\Psi}_{\alpha}^{a}\theta^{\alpha}_{\mu},
\ee
and the complete separation of variables in the perturbation
equations (\ref{9.2}) is achieved by making the following ansatz:
\be                                       \label{9.8}
{\bf H}_{\alpha\beta}=\exp(i\omega t)\,
H_{\alpha\beta}(r)\, _sY_{jm}(\vartheta,\varphi),\ 
{\Psi}_{\alpha}^{a}
=\exp(i\omega t)\,
\Phi_{\alpha}^{a}(r)\, _sY_{jm}(\vartheta,\varphi),
\ee
where $_sY_{jm}(\vartheta,\varphi)$ are the spin-weighted
spherical harmonics \cite{Goldberg67}. Here the quantum numbers
$j$ and $m$ are the same for all amplitudes.
The spin weight $s$ can be different for
different harmonics and is given by $s=s_{\alpha\beta}$
for ${\bf H}_{\alpha\beta}$, where
$$
s_{00}=s_{01}=s_{11}=s_{23}=0,\
s_{02}=s_{12}=-s_{03}=-s_{13}=1,\
s_{22}=-s_{33}=2,
$$
and by $s=s_{\alpha}^{a}$
with
$$
s_{0}^{3}=s_{1}^{3}=s_{3}^{1}=0,\
s_{0}^{1}=s_{1}^{1}=s_{2}^{3}=-s_{0}^{2}=-s_{1}^{2}=-s_{3}^{3}=1,\
s_{2}^{1}=-s_{3}^{2}=2
$$
for the YM perturbation amplitudes
${\Psi}_{\alpha}^{a}$.
As a result, for given $\omega$ and $j\geq 0$ in (\ref{9.8}),
Eqs.(\ref{9.2}) reduce to a system of radial equations for
$H_{\alpha\beta}(r)$ and $\Phi_{\alpha}^{a}(r)$.
Due to the spherical symmetry of the background fields,
the quantum number $m$ does not enter the radial equations.

\subsubsection{Rotational modes}
Inserting (\ref{9.8}) into (\ref{9.5}) the integral is computed
by integrating   over a two-sphere at finite radius $r$
and then taking the limit $r\rightarrow\infty$.
The angular dependence of the integrand implies that
the integral vanishes for any $r$  unless $j=1$, and that
only the $h_{0\varphi}$ perturbation
component can give a non-vanishing contribution.
Now, the transformation behavior of the angular momentum
under space and time reflections (P,T) implies that only those
perturbation amplitudes are relevant which are even under P
and odd under T. For $j=1$ these appear only in
$h_{0\varphi}$ and in two isotopic component of $\psi_{0}$.
They decouple from the remaining modes because the
background solutions are P and T symmetric.
This leads finally
to the following most general
ansatz (up to global coordinate rotations)
for the stationary rotational modes:
\be                                                    \label{9.9}
h=2S(r)\sin^2\theta\,  dt\, d\varphi,\  \ \
\psi=\left(\T_{1}\frac{\chi(r)}{r}\sin\theta+
\T_{3}\frac{\eta(r)}{r}\cos\theta\right) dt\, .
\ee
The gauge conditions (\ref{9.3}) for this ansatz are fulfilled identically
and the perturbation equations (\ref{9.2}) reduce to the coupled
system for the radial amplitudes $S$, $\chi$, and $\eta$:
\bea
-r^2 N\sigma\left(\frac{S'}{\sigma}\right)'&+&
\left(2N+\frac{4(w^2-1)^2}{r^2}\right)S           \nonumber \\
&+&4Nr^2 w' \left(\frac{\chi}{r}\right)'+
\frac{4(w^2-1)}{r}\left(w\, \chi-\eta\right)=0,   \nonumber \\
-r^2 N\sigma\left(\frac{\chi'}{\sigma}\right)'&+&
\left(1+w^2-2w'^2 N\right)\chi                    \nonumber \\
&-& 2w\,  \eta- rN \left(w'S\right)'
+\left(2N w'^3 +\frac{w(w^2-1)}{r}\right)S=0,     \nonumber \\
-r^2 N\sigma\left(\frac{\eta'}{\sigma}\right)'&+&
2\left(1+w^2- w'^2 N\right)\eta
- 4w\, \chi+ \frac{2(1-w^2)}{r}\, S=0.            \label{9.10}
\eea
Here $w$, $m$, and $\sigma$ refer to the background solutions,
$N\equiv 1-2m/r$. Given a solution of these equations, the ADM
angular momentum is
\be                                                \label{9.11}
J^i=\delta^i_z\,
\lim_{r\rightarrow\infty} r^4\left(\frac{S}{6 r^2}\right)'.
\ee
Note that, with a suitable reparameterization of the variables $S$, $\chi$,
and $\eta$, Eqs.(\ref{9.10}) can be represented
in a Schr\"odinger form with a
manifestly self-adjoint Hamiltonian  \cite{Brodbeck97a,Brodbeck97}.

\subsection{Solutions}

For the Schwarzschild background, $N=1-2M/r$, $\sigma=w=1$,
Eqs.(\ref{9.10}) admit
the solution $\chi=\eta=0$, $S=-2JM/r$. This is
recognized as the linear rotational excitation of the
Schwarzschild metric. One can see that this mode is
bounded everywhere outside the horizon. However, since it
does not vanish at the horizon, it is not normalizable.
Accordingly, when looking for solutions to Eqs.(\ref{9.10})
for the non-Abelian backgrounds, one considers all modes that
are bounded and regular, but not necessarily normalizable.
The strategy then is to find all well-behaved local
solutions in the vicinity of the origin or horizon and at infinity,
and then match them in the intermediate region.

In the far field region
the most general solution to  (\ref{9.10})
that gives rise to regular perturbations in (\ref{9.9})
reads
\bea                                       
S&=&-\frac{2JM}{r}+\frac{aQ}{r^2}+O(r^{-3}),\ \ \nonumber \\
\chi&=&V r+Q+O(r^{-1}),\ \ \
\eta=-V r-Q+O(r^{-1}).\ \                   \label{9.12}
\eea
Here $a$, $M$ are the parameters entering the asymptotic expansion
(\ref{4.7}) for the background $\f$, $m$. This solution contains
{\it four} integration constants: $J$, $V$, $Q$, while 
the fourth one, $c_4$, is
contained in the higher order terms in (\ref{9.12}).

In the near zone, in the soliton case,
the most general local
solution to (\ref{9.10}) that is regular at the origin is
\be                                             \label{9.13}
S=c_1 r^2+O(r^4),\ \ \
\chi=c_2 r+O(r^3),\ \ \
\eta=c_2 r+O(r^3).
\ee
This contains {\it three} independent parameters: $c_1$,
$c_2$, while the third one, $c_3$, resides in the higher order
terms in (\ref{9.13}). Since the coefficients in (\ref{9.10})
are continuous and regular for $0<r<\infty$, the local solutions
(\ref{9.12}) and (\ref{9.13}) in the vicinity of $r=0$ and $r=\infty$
admit extensions to the semi-open  intervals $[0,\infty)$ and
$(0,\infty]$, respectively. The total solution space is six dimensional,
while the local solutions specify four and three dimensional
subspaces, whose intersection is (at least) one dimensional.
In other words, the matching conditions for $S$, $\chi$, and $\eta$
give {\em six} linear algebraic equations
for the {\em seven} coefficients $J$, $Q$, $V$, $c_1\ldots c_4$, such
that the matching is always possible.
For example, for the $n=1$ BK solution the numerical matching gives
\be                                             \label{9.14}
Q/J= 0.960,\ \ \
V/J= -0.594,
\ee
and similarly for $c_1,\ldots c_4$.
All BK soliton solutions therefore admit a {\em one} parameter
family of stationary excitations
parameterized by $J$.

For black holes the space of solutions that are regular in the
vicinity of the horizon is {\it four} dimensional
\bea
S&=&c_{0}r_h+c_{1}\, x+O(x^2),\ \                   \nonumber \\
\chi&=&-c_{0}\, w(r_h)+c_{2}\, x+O(x^2),\ \ \
\eta=-c_{0}+c_{3}\, x+O(x^2),                         \label{9.15}
\eea
where $x=r-r_h$ and $c_{0}$, $c_{1}$,   $c_{2}$, and $c_{3}$
are four independent integration constants.
As a result, all EYM black holes admit
a {\em two} parameter family of stationary excitation.

In order to clarify
the meaning of the parameters in the solutions,
one uses Eq.(\ref{9.11}), which shows that the
constant $J$ is the ADM angular momentum.
Next, after the gauge transformation with U$=\exp(i(\pi-\vartheta)\T_2)$
the gauge field perturbation in the asymptotic region becomes
\be
\psi=\T_3\, \left(V+\frac{Q}{r}\right)\, dt+O(r^{-2}).  \label{9.16}
\ee
This suggests that $Q$ should be identified with the electric charge.
The constant $V$ determines the asymptotic value of the temporal
component of the gauge field, $A_0(\infty)$. In the
Abelian theory it would be possible to gauge $V$ away.
In the non-Abelian theory, however, such a gauge transformation would render
the whole configuration time-dependent. The physical
significance of $V$ became clear already in the course of the
matching
procedure.  In addition, due to the coupling to the background gauge field,
it enters the asymptotic expansion of the field strength,
$\delta F$, whose non-vanishing components in the same gauge
as the one used in (\ref{9.16}) read
\bea
\delta F_{0r}&=&\frac{Q}{r^2}\,\T_3+O(r^{-3}),\ \
\delta F_{0\vartheta}= \frac{aV}{r}\, \T_1+O(r^{-2}),\nonumber \\
\delta F_{0\varphi}&=& -\frac{aV}{r}\, \T_2\,\sin\vartheta\cos\vartheta
+O(r^{-2}).                                    \label{9.17}
\eea
It is obvious  that $Q$ can be expressed in terms
of the flux of the electric field over a two-sphere at spatial
infinity. The constant $V$, on the other hand, gives no
contribution to the flux, since the corresponding piece
of electric field in (\ref{9.17}) is tangent to the sphere.

Summarizing, all BK soliton solutions admit slowly rotating
excitations with continuous angular momentum $J$ and
electric charge $Q$ proportional to $J$.
Note that solitons
in other field theoretical models, such as the t'Hooft-Polyakov
monopoles or boson stars, generically do not admit  
slow-rotating states \cite{Volkov97,Brodbeck97,Brodbeck97a,Heusler98}.
This might be due to the fact that
a soliton, being a solution to a non-linear boundary
value problem, can exist only for discrete values of the
parameters. The angular momentum should
therefore be quantized. In this sense the situation in the
EYM theory is exceptional. This is presumably 
because both gravitational and gauge fields are massless,
which manifests in the slow (polynomial) decay of the fields
at infinity.

The slow rotational excitations of the EYM black holes
can be parameterized by any two of the three parameters
$J$, $Q$, or $V$. Accordingly, there are three distinguished
branches of solutions. First, for solutions with $V=0$ the
charge is proportional to the angular momentum \cite{Volkov97}.
Secondly, there is an uncharged branch with $Q=0$.
Finally, there are solutions with $J=0$, which
are non-static, as can be seen already from  (\ref{9.12}).
In view of this the Abelian staticity theorem asserting that
stationary black holes with $J=0$ must be static does not apply.
The non-Abelian version of this
theorem \cite{Sudarsky93} states that  a stationary EYM black hole
is static and the electric field vanishes, provided that
the following condition holds
\be                                              \label{9.18}
\Omega_H J-VQ=0,
\ee
where $\Omega_H$ is the angular velocity of the horizon.
Notice that  this agrees with the results described above.
Indeed, (\ref{9.18}) implies
that stationary black holes with $J=0$ must be static if only
$VQ=0$, which is not the case.
On the other hand, for non-static solutions the condition
(\ref{9.18}) should be violated, which, in particular, shows
that rotating solitons must be charged up, since with $\Omega_H=0$
one has to have $VQ\neq0$.

\section{Self-gravitating   lumps }
\setcounter{equation}{0}

In this chapter we shall discuss gravitating solitons and black holes
in the theories admitting particle-like solutions
in the flat spacetime limit.
Historically, the first investigation of such systems was
motivated by a wish to understand the structure
of very heavy magnetic monopoles \cite{Nieuwenhuizen76}.
It was found that the gravitating generalizations for
flat space monopoles exist at least  for small
values of Newton's constant. More systematic investigations
of the problem were undertaken after the discovery of the BK
solutions. These have revealed a number of the typical  features.
First, it has been found that, for a large class of non-linear matter
models, the flat space solitons can be generalized to curved spacetime,
provided that the dimensionless gravitational coupling
constant $\kappa$ is small. This is not very surprising, since it is
intuitively clear that for small values of $\kappa$  perturbation theory
applies.  However, and this is far less obvious,
it turns out that under certain
conditions gravity can be treated perturbatively even for black holes,
provided that the event horizon radius $r_h$ is small.
As a result, the gravitating lumps
can be further  generalized by replacing the regular center by
a black hole with a small radius $r_h$.
Below we shall review the corresponding argument based on the implicit
function theorem \cite{Kastor92}. This is so simple that the existence
of hairy black holes might actually have been foreseen many years ago.

Secondly, in the non-linear regime, when gravity is not weak,
the fundamental lumps and black holes
admit a discrete spectrum of gravitational excitations.
The excited solutions become infinitely heavy
for $\kappa\to 0$ and, remarkably, reduce then to the
rescaled BK solutions. The excitations thus can be thought of as
gravitating solitons with small BK particles or EYM black holes
in the center.
Finally, the fundamental and excited lumps and black holes cannot
exist beyond certain maximal values of $\kappa$ and $r_h$.
The existence of a bound for $\kappa$ is easy to understand. As $\kappa$
is proportional to the ratio of the gravitational radius
of the object to its
typical size, it cannot be too large, since otherwise the
system becomes unstable with respect to the gravitational collapse.
The solutions, however, do not collapse as $\kappa$ approaches the
critical value. Instead they either become gravitationally closed
or coalesce with the excited solutions. The existence of the bound
for $r_h$ is quite interesting. It seems as if a small black hole could not
swallow up a soliton which is larger than the black hole itself.
As a result, a hairy black hole appears. However, a big black hole
can  completely absorb all non-trivial hair.

\subsection{Event horizons inside classical lumps}

One can use the implicit function theorem to argue that,
within a large class of non-linear models, and for
small values of the gravitational coupling constant
$\kappa$ and the event horizon radius $r_h$,
the flat space solitons
admit both regular and black hole gravitating generalizations.
This comes about as follows \cite{Nieuwenhuizen76,Kastor92}.
Consider a static,
spherically symmetric gravitating system with the metric
(\ref{2.41}) and the stress tensor
$T^{\mu}_{\nu}={\rm diag}\,(\rho,-p_r,-p_\theta,-p_\theta)$.
With the notation of Sec.$2$, the non-trivial Einstein equations
$G_{\mu\nu}=2\kappa T_{\mu\nu}$ are
\be                                            \label{4a.1}
m'=r^2\rho,\ \ \ \ \sigma'=\kappa\, \frac{r}{N}\,(\rho+p_r)\,\sigma,
\ee
with $N=1-2\kappa  m/r$. This has to be supplemented with the
condition $\nabla_{\mu}T^{\mu}_{\nu}=0$, which can be rewritten
with the use of (\ref{4a.1}) as the Oppenheimer-Volkoff equation
\be                                            \label{4a.2}
p_r'=-\kappa \,\frac{m+r^3p_r}{r^2N}\,(\rho+p_r)
+\frac{2}{r}\,(p_\theta-p_r).
\ee
Suppose that there is an equilibrium flat space configuration,
such that equations (\ref{4a.1}) and (\ref{4a.2}) are                       
fulfilled for $\kappa=0$. In most cases one can expect
that this configuration will also survive if
$\kappa$ is non-zero and small. Indeed, for $\kappa\ll 1$ the additional
terms in the equations are small and gravity can be treated
perturbatively. One will have then $N\approx\sigma\approx 1$.
At the same time,
it is less clear whether the corresponding black hole
generalization will also exist. For black holes $N$ vanishes at $r_h$
implying that the deviation from the flat space value $N=1$ is non-small.
This also implies that
one cannot use perturbation theory, since the coefficient
$1/N$ in (\ref{4a.1}), (\ref{4a.2}) diverges at $r=r_h$. However,
if the matter model is
such that $(\rho+p_r)\sim N$ at the horizon then the blowing up
of $1/N$ will be canceled, in which case
gravity can be treated
perturbatively even for black holes \cite{Kastor92}.

It has already been mentioned
that for all matter models considered in Sec.$2.3$
the reduced 2D matter Lagrangian in the static, purely
magnetic case can be represented in the form
\be                                               \label{4a.3}
L_m=-(NK+U),
\ee
where $K$ and $U$ depend on $r$ and on the matter field variables,
but not on the gravitational variables $N$ and $\sigma$.
For this Lagrangian one has
\be                                                \label{4a.4}
\rho=(NK+U)/r^2,\ \ \ \ \ p_r=(NK-U)/r^2,
\ee
such that $(\rho+p_r)=2NK$ vanishes at the horizon. As a result,
given a flat space soliton solution, one can expect that it
will have gravitating generalizations, both in the regular and black hole
cases. The argument is as follows.
Inserting (\ref{4a.4}) into (\ref{4a.1}) the Einstein equations
for $m$ and $\sigma$ become {\em linear}, and hence can be
integrated in quadratures for given $K$ and $U$.
For example,  one will have
\be                                              \label{4a.5}
\sigma=\exp\left(-2\kappa \int_{r}^{\infty} K\,\frac{dr}{r}\right)
\ee
and similarly for $m$. Therefore, in order to solve
the gravitating problem, it remains to determine the matter variables
entering $K$ and $U$. Let us denote these variables collectively by $\phi$.
The equations of motion for $\phi$ can be
obtained by varying the ADM mass functional \cite{Nieuwenhuizen76}.
The ADM mass is $M=m(\infty)=m(\infty)\sigma(\infty)$.
Using $(m\sigma)'=(K+U)\sigma$ one obtains
\be                                             \label{4a.6}
M[\phi,\kappa,r_h]=\frac{r_h}{2}\, \sigma(r_h)+\int_{r_h}^{\infty}dr(K+U)
\exp\left(-2\kappa \int_{r}^{\infty}K\,\frac{dr}{r}\right).
\ee
This expression takes gravity into account but depends only
on the matter variables;
notice that for $r_h=0$ it was used in Eq.(\ref{5.18}).
Varying $M$ with respect to $\phi$ gives the matter equation
of motion
with all gravitational degrees of freedom expressed using
formulas like (\ref{4a.5}). The problem
therefore reduces to studying solutions of
\be                                                 \label{4a.7}
F[\phi,\kappa,r_h]\equiv \frac{\delta M}{\delta\phi}=0.
\ee
By assumption, there exists a flat space solution $\phi_0$,
such that $F[\phi_0,0,0]=0$. Suppose that the operator
$\delta F/\delta\phi$ is invertible for this solution. This will
be the case, for example, if the solution is a local minimum
of energy, $\delta F[\phi_0,0,0]>0$.  Then the
implicit function theorem ensures that for $\kappa$ and $r_h$
sufficiently close to zero, there exists $\phi(\kappa,r_h)$
satisfying (\ref{4a.7}) such that $\phi(0,0)=\phi_0$ \cite{Kastor92}.

Summarizing, if a flat space soliton solution
fulfills the conditions specified above
then it will survive also in the  weak gravity case,
and in addition will be able to contain a small black hole inside.
This happens, for example, for 
monopoles, sphalerons, and Skyrmions considered below.             
However, not all matter models are of the type (\ref{4a.3}).
For example, systems with electric fields are not of this type
and hence are not covered by the above arguments.
Another example is provided by boson stars.
For the BK particles there are no flat space
counterparts, and the above arguments do not directly apply.
However, it follows that for $\kappa\neq 0$
the existence of regular solutions to $F[\phi,\kappa,0]=0$ implies
the existence of corresponding black hole solutions to
$F[\phi,\kappa,r_h]=0$. As a result, the existence of the BK solitons
implies the existence of the EYM black holes, at least for small $r_h$.

\subsection{Gravitating monopoles}

The regular monopole solutions were discovered by
t'Hooft and Polyakov in the YMH model (\ref{2:25})
with the triplet Higgs field \cite{t'Hooft74,Polyakov74}.
For monopoles in flat space there is the absolute
lower bound for energy
determined by the topology of the Higgs field  \cite{Bogomolny76}:
\be                                                \label{4b.1}
E\geq \frac{4\pi v}{{\sl g}}|\nu|.
\ee
Here $\nu$ is the winding number of the Higgs field, which
coincides with the magnetic charge.
Solutions saturating this bound are topologically stable.
For the hedgehog ansatz in (\ref{2:25a}) one has $\nu=1$,
and the corresponding EYMH equations following from (\ref{2:26})
in the static, purely magnetic case read
\bea
m'&=&N\f^{\prime 2}+\frac{(\f^2-1)^2}{2r^2}
+\frac{r^2}{2}N\phi^{\prime 2}
+\f^2\phi^2+r^2 V(\phi),                           \label{4b.2}    \\
\sigma'&=&\kappa\sigma\left(\frac{2 \f^{\prime 2}}{r}
+r\phi^{\prime 2}\right),                            \label{4b.3}     \\
(N\sigma\f')'&=&\sigma\left(\frac{\f(\f^2-1)}{r^2}
+\phi^2\f\right),                                   \label{4b.4}     \\
(r^2N\sigma\phi')'&=&\sigma\left(2\f^2\phi
+r^2 V'(\phi)\right)                                 \label{4b.5}
\eea
with
\be                                                  \label{4b.5b}
N=1-\frac{2\kappa m}{r},\quad V(\phi)=\frac{\epsilon}{4}\,(\phi^2-1)^2.
\ee
Here $\kappa=4\pi Gv^2$ and the length scale is
${\rm L}=1/{\sl g}v\equiv 1/M_{\rm W}$.
Note that $\sigma$ can be eliminated from the
equations in the usual way. For a given solution to these equations the
total energy is the ADM mass $M=m(\infty)$. The dimensionful energy
is $(4\pi v/g)\, m(\infty)$.
The boundary conditions for
finite energy solutions are
\be                                                   \label{4b.6}
\f=1-b\,r^2+O(r^4),\quad \phi=c\,r+O(r^3),\quad m=O(r^3),
\ee
\be                                                  \label{4b.7}
\f=B\, f_1\exp(-r),\
\phi=1+C\, f_2\exp(-\epsilon r),\  m=M+O(r^{-1})
\ee
at the origin and at infinity, respectively. Here $b$, $c$, $B$, $C$,
and $M$ are integration constants, and the leading behaviour
of the functions $f_1(r)$ and $f_2(r)$ for $r\to\infty$ is power law
\cite{Chernavskii78,Kerner80}.

\subsubsection{Flat space solutions}
The flat space limit corresponds to $\kappa\to 0$, in which case
the YMH equations (\ref{4b.4}) and (\ref{4b.5}) decouple,
while the Einstein equations (\ref{4b.2}) and (\ref{4b.3})
give $N=\sigma=1$.
The mass
of the solutions, $m(\infty)$,
 is obtained from (\ref{4b.2}), which gives after
simple rearrangements
\bea
\left. M=\int_{0}^{\infty}
\right\{\left(\f'+\f\phi\right)^2&+&
\frac{r^2}{2}\left(\phi'+\frac{\f^2-1}{r^2}\right)^2   \nonumber \\
&+&\left. \frac{\epsilon^2 r^2}{4}\,(\phi^2-1)^2
+\left(\phi(1-\f^2)\right)^\prime
\right\}dr.                                         \label{4b.8}
\eea
Using (\ref{4b.6}) and (\ref{4b.7}) the integral of the last term
in the integrand is equal to one,
while the first three terms are positive
definite. This shows that
\be                                                  \label{4b.8a}
M\geq 1,
\ee
in agreement with
the bound in (\ref{4b.1})
for $\nu=1$. In fact, (\ref{4b.1}) is obtained in a similar way
by picking a total derivative in the 3D mass functional.
Now, if $\epsilon=0$ and
\be                                                   \label{4b.9}
\f'+\f\phi=0,\quad\quad r^2\phi'+\f^2-1=0
\ee
then the first three terms in the integrand in (\ref{4b.8}) vanish
and the bound $M=1$ is saturated. The solution of the
Bogomol'nyi equations (\ref{4b.9}) describes the BPS monopole
\cite{Prasad75}:
\be                                                   \label{4b.10}
\f=\frac{r}{\sinh(r)},\quad \phi=\coth(r)-\frac1r.
\ee
By construction, this solution is stable. It has
only one non-vanishing component of the stress tensor,
$T^{0}_{0}$, and hence the condition in (\ref{3.2}) is fulfilled.

For $\epsilon>0$ the bound in (\ref{4b.8a}) is not attained and the
first order equations cannot be used.
One has to solve then the second order YMH equations
(\ref{4b.4}) and (\ref{4b.5}). Solutions
with the boundary conditions (\ref{4b.6}), (\ref{4b.7})
can be obtained numerically
\cite{Bogomolny76a,Kirkman81}. They  describe the
t'Hooft-Polyakov monopoles.
For any given value of $\epsilon$  there is one solution,
the
existence of which was shown in \cite{Tyupkin76}.
The uniqueness of the $\epsilon=0$ solution
was established in \cite{Maison81}.
The mass $M$ increases with $\epsilon$ such that
$1\leq M(\epsilon)<M(\infty)=1.96$. In the limit $\epsilon\to\infty$
the Higgs field becomes ``frozen'' and there remains only the
YM equation (\ref{4b.4}), where one sets $\phi=1$ \cite{Kirkman81}.
The t'Hooft-Polyakov monopoles are linearly stable \cite{Baake92}.
For any $\epsilon$ the monopole solutions can be generalized to include
an electric charge \cite{Julia75}.

\subsubsection{Regular gravitating monopoles}
For $\kappa\neq 0$
the energy condition (\ref{4b.8a}) no longer holds due to the
gravitational binding. In fact, for a given $\kappa$,
taking the boundary conditions into account,
the mass functional $M[\phi,\kappa,0]$ in (\ref{4a.6}) with
\be                                              \label{4b.11}
K=\f^{\prime 2}+\frac{r^2}{2}\,\phi^{\prime 2},\quad
U=\frac{(\f^2-1)^2}{2r^2}
+\f^2\phi^2+\frac{\epsilon^2}{4}\,r^2(\phi^2-1)^2,
\ee
is still bounded from below by a non-zero value \cite{Nieuwenhuizen76}.
However, it is unclear whether the
$M_{\rm min}(\kappa)\equiv\inf M[\phi,\kappa,0]$
is reached in the set of solutions of the EYMH equations.
For example, it turns out that if $\kappa$ is large, the fields
that minimize $M[\phi,\kappa,0]$ do not even belong to the space of
differentiable functions \cite{Lee92}. At the same time,
the argument of the preceding section suggests that the
gravitating monopoles exist at least for small values of
$\kappa$.  This shows that one can study these solutions
numerically, however, the issue of their stability remains open.

The numerical integration of Eqs.(\ref{4b.2})--(\ref{4b.5})
with the boundary conditions (\ref{4b.6})--(\ref{4b.7})
reveals the following picture
\cite{Ortiz92,Lee92,Breitenlohner92,Aichelburg93,Breitenlohner95a,%
Maison96}.
First, for $\kappa$ being small enough, there are
the self-gravitating monopole solutions.
These are qualitatively similar to the
flat space t'Hooft-Polyakov monopoles
and reduce to them for $\kappa\to 0$.
The amplitudes  $\f$, $\phi$, $m$, and $\sigma$  are
monotone functions, while $N$ develops
a minimum at some $r_m\sim 1$, $N(r_m)>0$.
One can call $r_m$ the monopole radius.
The solutions are characterized
by $\kappa$ and $\epsilon$.
The mass, $M(\kappa)$, decreases with growing $\kappa$
due to the gravitational binding (see Tab.$3$ and Fig.\ref{fig:MONOPOLE}).
At the same time, the monopole
radius does not change considerably. 
As a result, the ratio of the
gravitational radius of the monopole to its radius $r_m$
is proportional to $\kappa$.
For regular solutions $\kappa$ must not therefore
exceed some critical value, otherwise the system becomes unstable
with respect to the gravitational collapse.

%%%%%%%%%%%%%%%%%%%%%%%%%%%%%%%%%%%%%
\begin{figure}
\hbox to\hsize{%\hss
  \epsfig{file=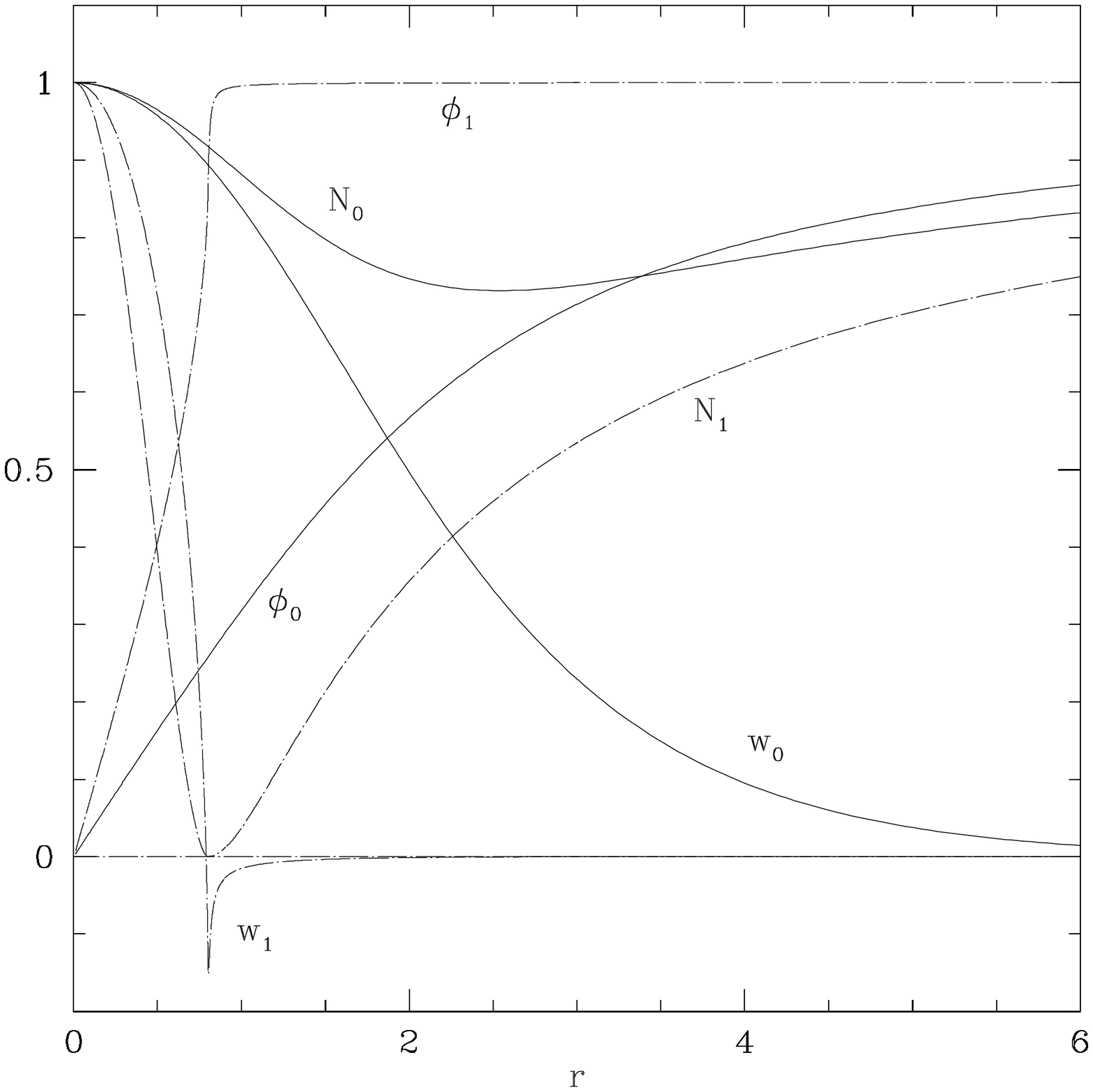,width=0.48\hsize,%
      bbllx=1.8cm,bblly=5.5cm,bburx=20.0cm,bbury=20.0cm}\hss
  \epsfig{file=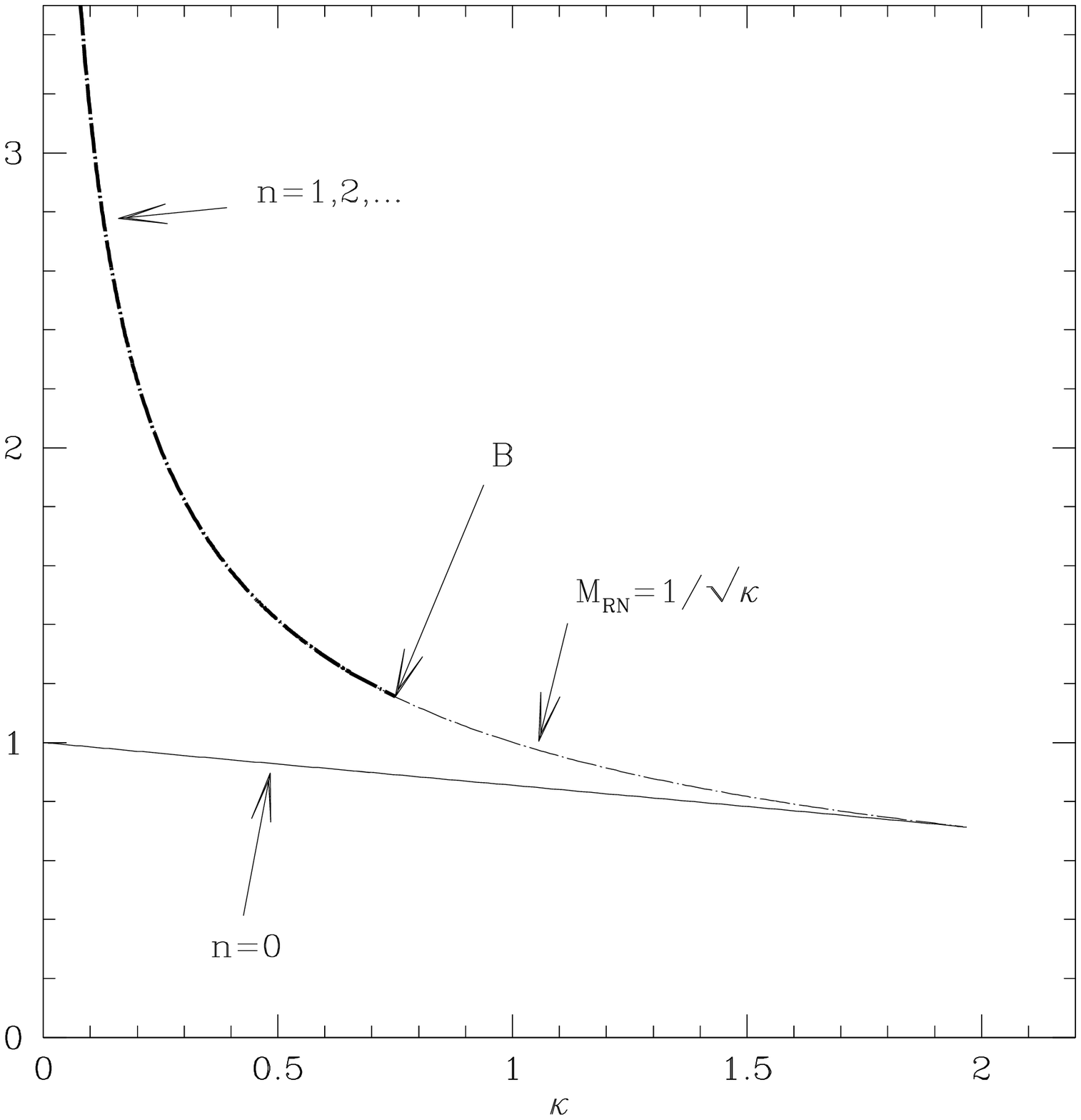,width=0.48\hsize,%
     bbllx=1.8cm,bblly=5.5cm,bburx=20.0cm,bbury=20.0cm}%\hss
  }
\caption{On the left:
Amplitudes $w_n$, $\phi_n$, and $N_n$ for the fundamental,
$n=0$,
monopole solution (solid lines)
and for its first, $n=1$, excitation (dashed lines) for
$\kappa=0.6$ and $\epsilon=0$. On the right: the ADM mass versus $\kappa$
for the $n=0$ and $n\geq 1$ regular monopole solutions with $\epsilon=0$;
and the
mass $M_{\rm RN}=1/\sqrt{\kappa}$ of the extreme RN solution. All curves for
$n\geq 1$ terminate at the point B with $\kappa=3/4$,
while staying very close
to each other and to the RN curve for $\kappa<3/4$.
 For small $\kappa$ the mass of the $n$-th solution
is $M_n/\sqrt{k}$, where $M_n$ is the mass of the $n$-th BK soliton.
The $n=0$ curve terminates at $\kappa=1.97$.
At the point B the extreme RN solution bifurcates with infinitely many
non-Abelian EYMH black hole solutions.
}
\label{fig:MONOPOLE}
\vspace{5 mm}
\end{figure}
%%%%%%%%%%%%%%%%%%%%%%%%%%%%%%%%%%%%%

In agreement with the last remark,
the self-gravitating monopoles exist only for
a finite range
$0<\kappa\leq\kappa_{\rm max}(\epsilon)$. One has
$\kappa_{\rm max}(\infty)=1/2<\kappa_{\rm max}(\epsilon)
\leq \kappa_{\rm max}(0)=1.97$ \cite{Breitenlohner95a}.
As $\kappa$ approaches the critical value,
$N(r_m)$ tends to zero (see Fig.\ref{fig:MONOPOLE}),
the functions $\f$ and $\phi$
reach their asymptotic values already at $r=r_m$, and
the proper distance
\be                                             \label{4b.12}
l=\int_{0}^{r}\frac{dr}{\sqrt{N}}
\ee
diverges for $r\to r_m$%
\footnote{For small $\epsilon$ the picture
is slightly more complicated; details
can be found in \cite{Breitenlohner92}.}.
As a result, the spatial geometry on the
hypersurface $t=$const. develops an infinite throat separating
the interior region with a smooth origin and non-trivial YMH field
from the exterior region where $\f=0$, $\phi=1$, and
the metric is extreme RN. The throat is characterized by a constant
radius, $r_o=\sqrt{\kappa_{\rm max}}$.
The metric function $\sigma(r)$, normalized by
$\sigma(0)=1$, diverges for $r\to r_m$. The limiting solution thus
splits into two independent solutions, which
is similar to the  behaviour of the BK solutions for $n\to\infty$.
The interior solution is geodesically complete.

\subsubsection{Gravitational excitations}
For the gravitating monopoles there
is an infinite sequence of radial excitations.
These exist for $\kappa$ being small enough,
but do not admit the flat space
limit, since their masses diverge  for $\kappa\to 0$.
The boundary conditions
(\ref{4b.6})--(\ref{4b.7}) are still fulfilled,
but the amplitude $\f$ now
oscillates (see Fig.\ref{fig:MONOPOLE})
and its node number ($n=1,2\ldots\, $) characterizes the
solutions. 
This similarity with the BK case is not accidental.
In fact, for $\kappa\to 0$, the excitations correspond to the
rescaled BK solutions. This comes about as follows.
When $\kappa=4\pi G v^2$ tends to zero, this can be
understood either as the weak gravity limit, $G\to 0$ with fixed $v$,
or as the limit $v\to 0$ with fixed $G$.
Notice that in the latter case the length scale
${\rm L}=1/{\sl g}v$ diverges.
Consider, however, the rescaling
\be                                               \label{4b.13}
r\to\sqrt{\kappa}\,r,\quad m\to m /\sqrt{\kappa},\quad
\phi\to \phi/\sqrt{\kappa}.
\ee
The whole effect of this on the EYMH equations is that $N$ and
$V(\phi)$ in (\ref{4b.5b}) are replaced by
\be                                                  \label{4b.14}
N=1-\frac{2 m}{r},\quad V(\phi)=\frac{\epsilon}{4}\,(\phi^2-\kappa)^2.
\ee
As a result, the limit $\kappa\to 0$ corresponds now to 
$V(\phi)$ vanishing (since $\phi^2\leq\kappa$), and the EYMH system
(\ref{4b.2})--(\ref{4b.5}) reduces to the EYM equations, which
leads to the appearance of the BK solutions. Restoring
the original length scale, the excited solutions for small $\kappa$
consist of the very small (Planck size) BK particles sitting inside
the large ($1/M_{\rm W}$) size monopole. The amplitude
$\f$ oscillates $n$ times in the short interval near the origin
and then tends to zero for $r\to\infty$. For $\kappa\to 0$ the mass
tends to $M_{n}/\sqrt{\kappa}$, where $M_n$ is the mass of the
$n$-th BK solution.  All excitations have the same value of
$\kappa_{\rm max}=\kappa_{\rm max}(\epsilon)$. One has
$\kappa_{\rm max}(0)=3/4\geq
\kappa_{\rm max}(\epsilon)>
\kappa_{\rm max}(\infty)=1/2$ \cite{Breitenlohner95a}.
For $\kappa\to\kappa_{\rm max}(\epsilon)$
the excited solutions show the same
limiting behaviour with a throat
as the fundamental solution.

\begin{table}
%\begin{center}
%\label{Tab:MONOPOLEMASS}
\caption{Masses of the fundamental monopole
and its first excitation ($\epsilon=0$).}
\vglue 0.4cm
 \begin{tabular}{|c|c|c|c|c|c|c|c|c|}
\hline
$n\setminus\kappa$
& $0$
& $0.02$
& $0.1$
&$0.5$
& $0.75$
& $1$
& $1.5$
& $1.96$ \\     \hline
$0$
&  $1$
&  $0.996$
& $0.985$
& $0.926$
& $0.887$
& $0.854$
& $0.783$
& $0.714$\\
$1$
&   $\infty$
&   $6.629$
&   $3.108$
&   $1.413$
&   $1.154$
&   $-$
&   $-$
&   $-$    \\
\hline
\end{tabular}
\vspace{3 mm}
\end{table}

The picture described above remains qualitatively the same
for any $0\leq\epsilon<\infty$. A novel feature arises for
$\epsilon\to\infty$ \cite{Breitenlohner92,Aichelburg93,Breitenlohner95a}.
The Higgs field is frozen in this case,
and the EYMH field equations reduce to those of the
gravitating gauged non-Abelian O(3) sigma-model.
Eq.(\ref{4b.5}) then should be dropped, and there remain
three equations (\ref{4b.2})--(\ref{4b.4}), where one sets $\phi=1$.
Note that these admit the first integral (\ref{4b.16}).
The boundary conditions at the origin for $\f$ and $N$
are given by
\be
\f=1+O(r^2),\ \ \ N=1-2\kappa^2+O(r^2),
\ee
where $\kappa\in[0,1/2]$. Since $N(0)\neq 1$,
there is the solid angle deficit and a conical
singularity at the origin. This is due to the fact that
the energy density $T^{0}_{0}$ contains the term $\f^2\phi^2/r^2$,
which blows up, because $\phi$ does not vanish at the origin.
For $r>0$ the solutions for $\f$ and $N$ are qualitatively
similar to those for $\epsilon<\infty$. In particular, they
are asymptotically flat $(N(\infty)=1)$
and exist for arbitrary node number $n$.
The existence of these solutions
was rigorously established in \cite{Breitenlohner95a}.

In this connection it is instructive to mention also
the gravitating global monopoles \cite{Barriola89}.
These arise in the coupled Einstein-Higgs model,
which corresponds to the EYMH theory with
dynamics of the YM field being suppressed.
The field equations in the spherically symmetric case can be obtained
from (\ref{4b.2})--(\ref{4b.5}) by ignoring the YM equation (\ref{4b.4})
and setting $\f=1$ in the remaining equations.
The boundary conditions for $\phi$ and $m$ at $r=0$ are still
given by (\ref{4b.6}) (if only $\epsilon<\infty$),
and the origin is regular: $N(0)=1$.
However, since $\phi\to 1$ for $r\to\infty$ while $\f=1$,
the term $\f^2\phi^2/r^2$ in $T^{0}_{0}$ leads to the
divergence of the energy
in such a way that $N(\infty)$=const.$<1$, which
corresponds to the solid angle deficit at infinity.
Note that the absence of finite-energy solutions in the model
agrees with the general arguments in Sec.3.

For $\epsilon\to\infty$ the Einstein-Higgs theory reduces to the
global gravitating O(3) sigma-model. In this case there remain only
two non-trivial field equations, (\ref{4b.2})--(\ref{4b.3}),
where one sets $\f=\phi=1$. The solution is $\sigma=1$,
$N=1-2\kappa$, such that there is a constant
solid angle deficit. It is worth noting that this model
admits the Bogomol'nyi bound and
non-trivial exact solutions in the case
of cylindrical symmetry \cite{Comtet88}.

\subsubsection{EYMH Black holes}
For black holes, instead of (\ref{4b.6}), one has
the boundary conditions at the regular horizon,
which are given by the expression similar to that in (\ref{bhdil})
with $\f_h'$, $\phi_h'$, and $N_h'$ being functions
of $\f_h$ and $\phi_h$.
The possibility of a degenerate horizon was discussed in
\cite{Hajicek83,Hajicek83a,Degen87,Bicak95,Breitenlohner95a}.
According to the argument of the preceding section,
one can expect the black hole generalizations of the regular
monopoles to exist at least for small values of $r_h$.
This is confirmed by the numerical analysis for the
fundamental ($n=0$) monopole and for its excitations $(n\geq 1)$.
For $r_h\ll 1$ the solutions resemble small black holes
sitting in the center of the regular lumps,
the latter being almost unaffected for $r\gg r_h$ by the
presence of the black hole. The solutions exist only
for a limited region of the $(\kappa,r_h)$ space.
The domain of $\kappa$  can be
subdivided into two parts: $0<\kappa<3/4$\,%
\footnote{The numbers are given for the case where $\epsilon=0$.},
in which
case solutions for any $n$ are possible, and
$3/4<\kappa<\kappa_{\rm max}=1.96$,
which leads only to the fundamental solutions.
 In both cases the solutions exist if only
$r_h$ is less than a maximal value,
$\tilde{r}_h(\kappa)$, where
$\tilde{r}_h(\kappa)\leq \tilde{r}_h(3/4)=\sqrt{3}/2$.
For $r_h\to 0$ the black hole solutions tend to the corresponding
regular ones. In the opposite limit, $r_h\to \tilde{r}_h(\kappa)$,
black holes with $\kappa>3/4$ develop the infinite throat
and become gravitationally closed.
The metric amplitude $N$ in this case, apart from a simple zero at $r=r_h$,
develops also a double zero at $r=\sqrt{\kappa}>r_h$.
For  $\kappa<3/4$
the picture is more complicated. In this case
for some values of $\kappa$ and $r_h$
one finds more than one non-Abelian black hole solution.
In addition, for
$r_h\geq\sqrt{\kappa}$ and $M\geq 1/\sqrt{\kappa}$
there exist Abelian RN solutions:
\be                                                     \label{4b.15}
\f=0,\quad \phi=1,\quad \sigma=1,\quad
N=1-\frac{2\kappa M}{r}+\frac{\kappa}{r^2}.
\ee
As a result, the no-hair conjecture is violated.
When $r_h$ changes, different non-Abelian solutions
bifurcate with each other or with the RN solution
\cite{Breitenlohner92}. This
leads to the existence of the upper bound for $r_h$.
The value  $\tilde{r}_h(3/4)=\sqrt{3}/2$ is the least upper
bound for $r_h$ for all solutions (for any $n$ and $\epsilon$)
with $\kappa=3/4$.
All of them tend for $r_h\to\sqrt{3}/2$ to the same
limiting configuration, which is the extreme RN solution.
A more detailed discussion can be found in
\cite{Breitenlohner92,Breitenlohner95a}.

Note that the existence of an upper bound for the black hole
radius is quite typical for gravitating lumps \cite{Nunez96}.
One can think that
a small black hole is unable to swallow up a monopole whose radius
is much larger than $r_h$, however a bigger black hole can do this,
such that all non-Abelian structures disappear inside the horizon.
This phenomenon does not exist, however, in the pure EYM theory,
where there is no energy scale other than Planck's mass.
The non-Abelian EYM black holes can be arbitrarily large.
It is worth noting also that the fundamental and excited
gravitating monopoles and black holes can be generalized to include
an electric charge \cite{Brihaye98}.

\subsubsection{Deformed EYMH black holes}
The fundamental gravitating monopoles and black holes
are stable, while the excited solutions are unstable with respect
to small spherically symmetric perturbations.
This was shown in \cite{Hollmann94} with the use of the Jackoby
criterion. The RN solution (\ref{4b.15}), which is stable
in the Abelian theory, becomes unstable in the non-Abelian case
for $\kappa<3/4$
\cite{Lohiya82,Bizon91a,Lee92a,Breitenlohner92,Aichelburg93,Breitenlohner95a}.
This comes about as follows. Perturbing the 
amplitudes $\f$, $m$, and $\sigma$ of the RN solution (\ref{4b.15})
as in Eq.(\ref{8.1}), while $\phi\to\phi+\delta\phi(t,r)$,
and linearizing the full system of EYMH equations,
the linearized YM equation decouples from the rest of the system.
It is convenient to rescale $r$ and $M$ in (\ref{4b.15})
according to (\ref{4b.13}), which leads to the following equation
for $\delta\f(t,r)=\exp(i\omega t)\, \eta(\rho)$:
\be                                               \label{RNstabil}
\left\{-\frac{d^2}{d\rho^2}+N
\left(\kappa -\frac{1}{r^2}\right)\right\}\eta=\omega^2\eta,
\ee
where $N=1-2M/r+1/r^2$ and $d\rho=dr/N$.
Notice that for $\kappa\to 0$, which now corresponds to the limit
where the Higgs field decouples, this equation reduces to
that in (\ref{8.5}). As was observed in \cite{Breitenlohner95a},
in the extreme RN case, where $M=1$, this equation admits infinitely
many bound states for $\kappa<3/4$, because the
corresponding potential for $\rho\to-\infty$ behaves like
$(\kappa-1)/\rho^2$. For $\kappa>3/4$ there are
no bound states \cite{Breitenlohner95a}.
This corresponds to the fact that
for  $\kappa=3/4$ the extreme RN solution
bifurcates with infinitely many non-Abelian black hole solutions
\cite{Breitenlohner95a}%
\footnote{The exterior region of the extreme RN solution can also coalesce
with the corresponding part of the limiting solutions
with the throat. This happens for $\kappa>3/4$ (see Fig.\ref{fig:MONOPOLE}).
However, since the interior parts are different, this does not
lead to the change of stability, and the RN solution for
$\kappa>3/4$ is stable.}. For non-extreme RN solutions the
situation is more complicated, but the result is the same:
the instability arises only for small enough values of $M$ and $\kappa$.

The instability of the RN solution within the context
of the non-Abelian EYMH theory has important implications,
since it can be viewed as indication of the existence of new
non-Abelian solutions
\cite{Lee94,Ridgway95b,Ridgway95,Ridgway95a,Weinberg95}.
This basic idea is as follows.
First of all, notice that the RN solution in (\ref{4b.15})
can be generalized to arbitrary integer values of the magnetic charge.
For this one uses the ansatz (\ref{3:8:1a}) for higher
winding numbers $\nu$, while the Higgs field is $\Phi=\phi\,\T_3$.
The solution is given by
\be                                                     \label{4b.15a}
a=\f=0,\quad \phi=1,\quad \sigma=1,\quad
N=1-\frac{2\kappa M}{r}+\frac{\kappa\nu^2}{r^2},
\ee
and the gauge field strength is
$F=\T_3\,\nu\sin\vartheta\, d\vartheta\wedge d\varphi$.
For any given $\nu$ this RN solution can be either stable or unstable --
depending on values of $\kappa$ and $M$ \cite{Lee94,Ridgway95a}.
Suppose that $\kappa$ and $M$ are chosen such that the
solution
is close to the border of instability and is only ``barely'' unstable.
One can expect then that there is a nearby stable
non-Abelian solution with lower mass that
bifurcates with the RN solution when $\kappa$ and $M$ approach
the border. It is plausible  that
 this new solution
differs only slightly from the RN one, and can be approximated
with the use of linear  perturbation theory
\cite{Ridgway95b,Ridgway95,Weinberg95}%
\footnote{The corresponding perturbation equations are not quite the same as
in the standard perturbation analysis \cite{Ridgway95b}.}.
Now, for $\nu=1$ the static perturbations around the RN solution are
spherically symmetric and approximate the
non-Abelain EYMH black holes described above.
However, for $\nu>1$ the solutions of the perturbation
equations  are no longer
spherically symmetric%
\footnote{It has already been mentioned that all
spherically symmetric YM fields with
$\nu>1$ are necessarily embedded Abelian. It follows then that
non-Abelian fields for $\nu>1$ cannot be spherically symmetric.}
and exhibit a complicated dependence
on spherical angles.  As a result, assuming that the
linear modes indeed approximate some new solutions, these
new black holes
turn out to be not even axially symmetric.
Their shape can be quite complicated,
exhibiting discrete, crystal-type symmetries
(see \cite{Ridgway95b} for interesting pictures).

It is worth noting that the deformed black holes have actually
been described within a more general theory, which
includes the EYMH model as a special case
\cite{Lee94,Ridgway95b,Ridgway95,Ridgway95a,Weinberg95}.
In this theory the third isotopic components of the gauge field
is regarded as an
Abelian vector fields, ${\cal A}_\mu$, while the first two component
are viewed as a complex vector field, $W_\mu$.
The gravitational part of the action is standard,
while the matter Lagrangian is
\bea
{\cal L}=&-&\frac14\,
{\cal F}_{\mu\nu}{\cal F}^{\mu\nu}
-\frac12\, H_{\mu\nu}^\ast H^{\mu\nu} +a\,
d_{\mu\nu}{\cal F}^{\mu\nu}                    \nonumber \\
&-&b\, d_{\mu\nu}d^{\mu\nu}
+m^2(\phi)\,W^{\ast}_{\mu}W^\mu
+\frac12\,\partial_\mu\phi\, \partial^\mu\phi-V(\phi). \label{weinberg}
\eea
Here ${\cal F}_{\mu\nu}=\partial_\mu {\cal A}_\nu-\partial_\nu{\cal A}_\mu$
and $H_{\mu\nu}=D_\mu W_\nu-D_\nu W_\mu$, while
$d_{\mu\nu}=i(W^{\ast}_{\mu}W_{\nu}-W^{\ast}_{\nu}W_{\mu})$.
The covariant derivative is $D_\mu W_\nu\equiv
(\partial_\mu-i{\cal A}_\mu)W_\nu$,  and $a$, $b$ are parameters.
For $a=1/2$, $b=4$,
and $m(\phi)=\phi$ the theory reduces to the triplet EYMH model
in the string gauge (\ref{2.9})
with $\Phi=\phi\,\T_3$.  For a range of the parameters
the model admits soliton and black hole solutions, and,
in particular, the deformed black holes described above.
Together with the results obtained in the pure EYM theory
\cite{Kleihaus97b,Kleihaus97d}, this suggests that the
existence of deformed static black holes is typical for the
gravitating non-Abelian models.

\subsubsection{The Bogomol'nyi bound}
Unfortunately, the flat space Bogomol'nyi equations (\ref{4b.9})
do not generalize to curved spacetime. The existence of the
Bogomol'nyi bound in flat space
has a very deep physical reason: the YMH theory for $\epsilon=0$
can be obtained via truncation of a
supersymmetric model. On the other hand, the EYMH theory is not a
truncation of a supergravity model, and no
Bogomol'nyi bound exists in this case.
As a result, a search for Bogomol'nyi equations for $\kappa\neq 0$
leads to an almost trivial result: one can find first order equations,
but they admit only  the extreme
RN solution \cite{Comtet84,Balakrishna92}.
For $\epsilon\to \infty$, when the Higgs field is frozen,
the Bogomol'nyi equations are less trivial \cite{Forgacs94,Braden98}%
\footnote{The existence of these equations has no explanation at present.
Note that in flat space there are no Bogomol'nyi equations for
$\epsilon\to \infty$.}:
\be                                                   \label{4b.16}
\kappa=1,\quad
N\f'=\f,\quad r^2(\sigma N)'=\sigma(\f^2-1),\quad
rN+\f^2-1=\pm r.
\ee
However, all their solutions, apart from the extreme RN one, have naked
singularities. In order to obtain a physically meaningful
Bogomol'nyi bound in curved space \cite{GibbonsHull},
one should extend the EYMH theory
to make it a part of a supergravity model. Such an extension was
considered in \cite{Gibbons94a,Gibbons95a}. In addition to the EYMH
fields it contains a dilaton and an abelian vector field,
which is typical for toroidally
compactified string theory models. The BPS solutions in this case
can be obtained analytically.

Sometimes gravitating monopoles are associated with
inflation \cite{Linde94,Vilenkin94}. This is due to the fact that, since
the Higgs field vanishes for $r\to 0$, there is a false vacuum
in the monopole core. The results of the numerical analysis of the
time-dependent problem suggest that monopoles indeed inflate
if $\kappa$ exceeds the critical value, $\kappa>\kappa_{\rm max}$
\cite{Sakai96a,Sakai96}.

Other related topics we would like to mention include solitons for
the EYMH system with the ghost Higgs field \cite{Clement81},
monopole solutions on the cosmological background \cite{Melnikov84},
and the Rubakov-Callan effect for the colored black holes \cite{Galtsov88}.

\subsection{Gravitating YMH sphalerons}

Consider the EYMH model (\ref{2:27}) with the doublet Higgs field.
The spherically symmetric ansatz (\ref{2:27a}) contains two
independent amplitudes, $\phi$ and $\xi$. One can impose
the on-shell condition
\be                                                   \label{4c.1}
\xi=\pi,
\ee
which implies that the configuration is parity-even.
As a result, the field equations following from the reduced
Lagrangian (\ref{2:28}) in the static, purely magnetic case
are given by
\bea
m'&=&N\f^{\prime 2}+\frac{(\f^2-1)^2}{2r^2}
+r^2N\phi^{\prime 2}
+\frac12\phi^2(\f+1)^2+r^2 V(\phi),               \label{4c.2}    \\
\sigma'&=&2\kappa\sigma\left(\frac{\f^{\prime 2}}{r}
+r\phi^{\prime 2}\right),                            \label{4c.3}     \\
(N\sigma\f')'&=&\sigma\left(\frac{\f(\f^2-1)}{r^2}
+\frac12\,\phi^2(\f+1)\right),                         \label{4c.4}     \\
(r^2N\sigma\phi')'&=&\sigma\left(\frac12(\f+1)^2\phi
+r^2 V'(\phi)\right),                                 \label{4c.5}
\eea
where $N$ and $V(\phi)$ are the same as in (\ref{4b.5b}).
The boundary conditions are
\be                                                   \label{4c.6}
\f=\pm( 1-b\,r^2)+O(r^4),\quad \phi=c\,r+O(r^3),\quad m=O(r^3),
\ee
\be                                                  \label{4c.7}
\f=-1+B f_1\exp(-r),\
\phi=1+C f_2\exp(-\epsilon r),\  m=M+O(r^{-1})
\ee
at the origin and at infinity, respectively,
where the notation is the same as in (\ref{4b.7}).

Let us mention the following important differences with the
triplet EYMH case. First, (\ref{4c.7}) implies that
the solutions are neutral. Secondly, the vacuum manifold for
the Higgs field is now three-sphere, and
since $\pi_2(S^3)$ is trivial, no topologically stable solutions exist.
However, as was argued in \cite{Manton83,Klinkhamer84},
the non-triviality of $\pi_3(S^3)$
can lead to the existence of  saddle point solutions
in the flat space YMH theory, which relate
to the top of the potential barrier between the topological vacua.
The corresponding argument is essentially the same as the one
presented in Sec.3.2 for the BK sphalerons. First, one finds
sequences of static fields $\{A[\lambda],\Phi[\lambda]\}$ interpolating
between the distinct vacua.
Identifying the end points, they become
non-contractible loops. Second, for each loop one computes the maximal
value of energy, $E_{\rm max}$, which is then minimized over all
loops to obtain $E_s\equiv\inf\{E_{\rm max}\}$. Finally, it is
assumed that {\bf a)} $E_s$ exists and positive; {\bf b)}
there is a loop whose
$E_{\rm max}$ is equal to $E_s$. Even though these assumptions
have not been rigorously proven, they seem to be plausible in the
case under consideration. This implies the existence of a saddle
point solution called sphaleron.

The sphaleron solution to (\ref{4c.2})--(\ref{4c.5}) in the
flat space limit was found numerically in \cite{Dashen74,Boguta83}.
This corresponds to the choice of the plus sign in (\ref{4c.6}) and
the gauge amplitude $\f$ interpolating between
the values one and minus one --
very much similar to  the $n=1$ BK solution.
The Higgs field amplitude is a monotone function.
The existence of this solution was established in \cite{Burzlaff84}.
For small $\epsilon$
there is only one solution for a given value of $\epsilon$.
For large $\epsilon$ there are additional solutions for which
the amplitude $\xi$ is non-trivial,
these are called deformed sphalerons  \cite{Kunz89,Yaffe90}.
Sphalerons are unstable.
The instability is associated with the
dynamics of the field $\xi$.

%%%%%%%%%%%%%%%%%%%%%%%%%%%%%%%%%%%%%
\begin{figure}
\hbox to\hsize{%\hss
  \epsfig{file=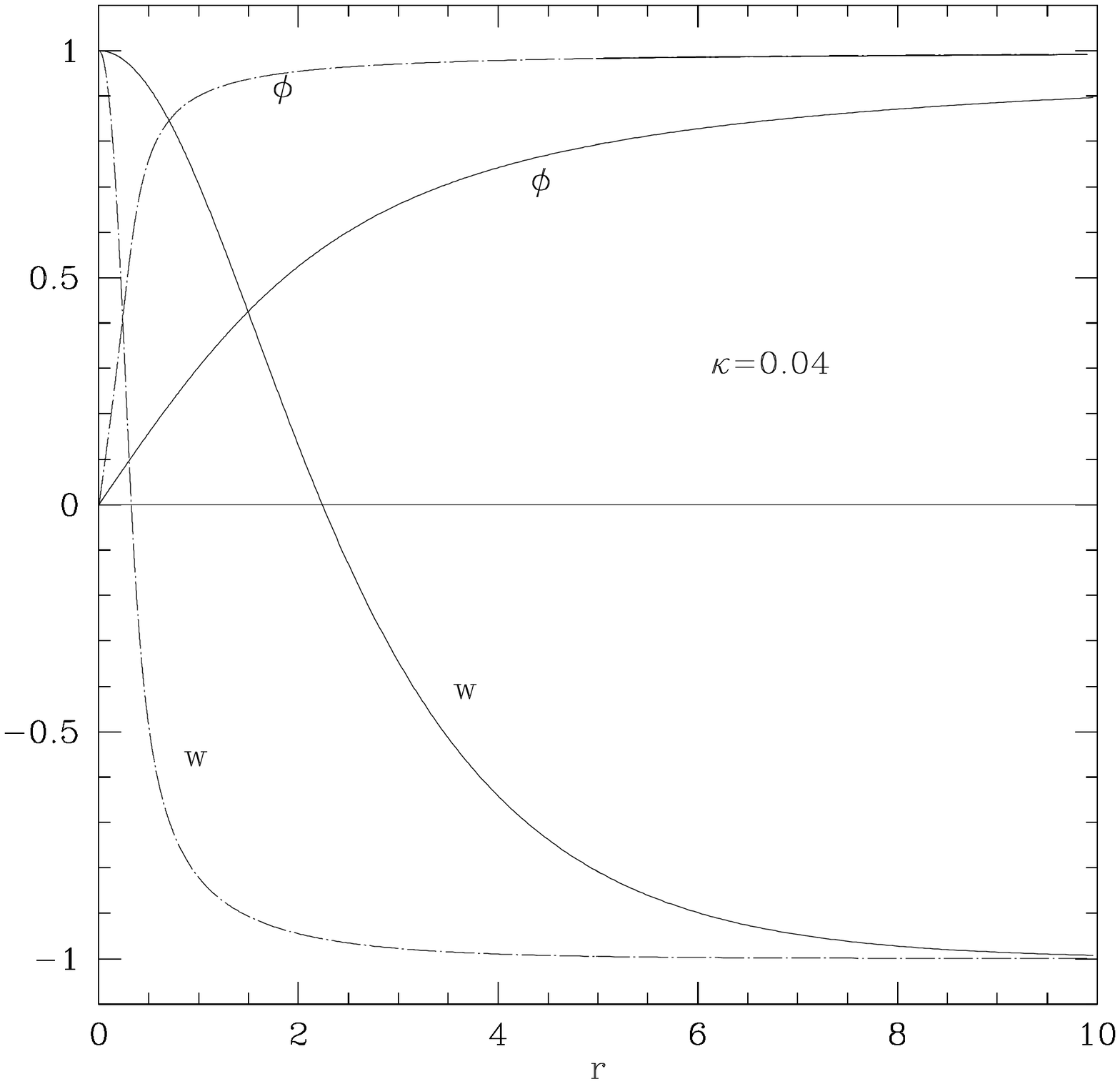,width=0.48\hsize,%
      bbllx=1.8cm,bblly=5.5cm,bburx=20.0cm,bbury=20.0cm}\hss
  \epsfig{file=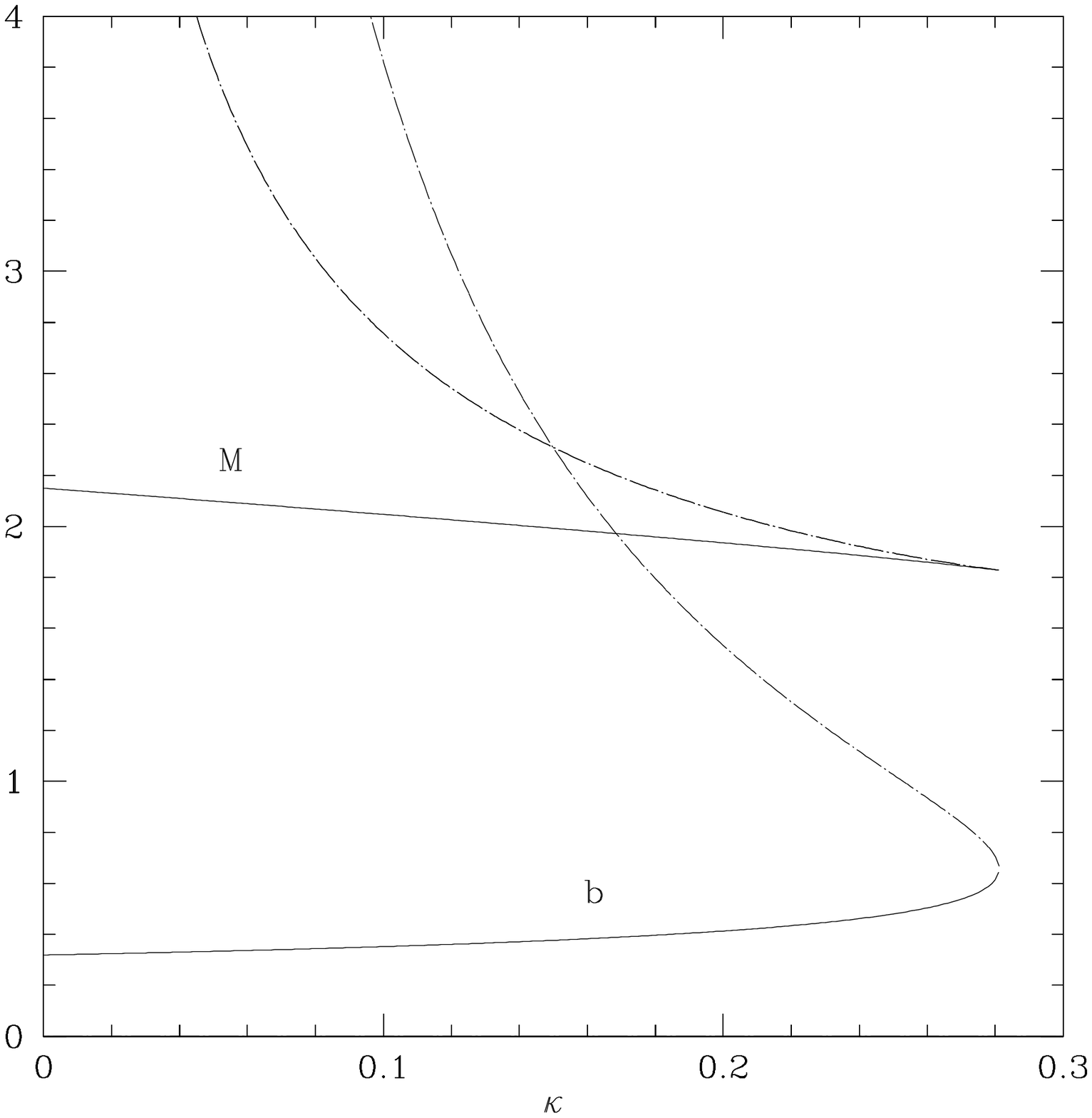,width=0.48\hsize,%
     bbllx=1.8cm,bblly=5.5cm,bburx=20.0cm,bbury=20.0cm}%\hss
  }
\caption{On the left:
Amplitudes $w$ and $\phi$ for the fundamental
EYMH sphaleron solution (solid lines)
and for its first excitation (dashed lines) for
$\kappa=0.04$ and $\epsilon=0$. On the right: the ADM mass $M$ and the
parameter $b$ in (\ref{4c.6}) versus $\kappa$
for the fundamental and the first excitation branches.
The solid and dashed curves merge at the critical value
$\kappa_{\rm max}=0.281$. The curves for $c(\kappa)$, where $c$ is
the third parameter of the solutions in (\ref{4c.6}),
merge in a similar way.
}
\label{fig:sphal}
\end{figure}
%%%%%%%%%%%%%%%%%%%%%%%%%%%%%%%%%%%%%

Although unstable, flat space YMH sphalerons admit
gravitating generalizations
\cite{Greene93}. First,
for small values of $\kappa$ there is a fundamental branch of regular
solutions that admit the limit
$\kappa\to 0$. These are qualitatively similar to the flat spacetime
sphaleron.  Next, there is
a branch of excited solutions with large mass, which also look similar.
In particular,
the amplitude $\f$ still has one node (see Fig.\ref{fig:sphal}).
However, the excitations do not admit the flat space limit and for
$\kappa\to 0$ reduce to the
rescaled $n=1$ BK solution -- exactly in the same manner as for
monopoles. The excited solutions can be thought of as the BK
sphalerons sitting inside the YMH sphalerons.
Finally, there are solutions with higher
node numbers. For even values of $n$ one chooses the minus sign in
(\ref{4c.6}). Similar to the $n=1$ solutions, those for $n>1$ also
exist in pairs, each pair containing two solutions with the same
nodal structure but with different masses.
For $\kappa\to 0$ the $n$-th solution with lower mass reduces
to the $n$-th rescaled BK solution, while the one with larger mass --
to the $n+1$-th rescaled BK solution. The change of the node
number then arises as follows: for $\kappa\to 0$ the field
configurations shrink
and the amplitudes $\f$ with different node numbers become
practically identical. For example, the amplitude $\f$
for the $n=1$ excited solution shown in  Fig.\ref{fig:sphal}
coincides for $\kappa\to 0$ with the one for the $n=2$ solution,
for which one has $\f(0)=-1$.

For all $n$ the solutions
exist only for a finite range of the parameter $0\leq\kappa\leq
\kappa_{\rm max}(n,\epsilon)$. The limiting behaviour
for $\kappa\to\kappa_{\rm max}$ is different from that for
monopoles. Notice that the RN solution (\ref{4b.15}) does not
fulfill the equations in the doublet case. As a result, the
solutions cannot have a throat connecting an interior region and
the exterior RN part. What happens instead is that the two
solutions with the same $n$ coalesce for $\kappa=\kappa_{\rm max}$,
at which point all their parameters coincide
(see Fig.\ref{fig:sphal}).
For example, the two solutions shown in Fig.\ref{fig:sphal}
get closer and closer as $\kappa$ increases and finally
merge for $\kappa_{\rm max}(1,0)=0.281$ becoming one solution.
The critical values $\kappa_{\rm max}$ thus correspond to
bifurcation points.
It is worth noting that, unlike the situation in the monopole case,
the limiting solutions for all $n$ are
perfectly smooth and regular. In particular,
the minimal value of $N$ is quite large.
No regular solutions exist for $\kappa>\kappa_{\rm max}$.

For any $n$ the regular EYMH sphalerons can be generalized
to include a small black hole inside. This is possible
for $0\leq\kappa<\kappa_{\rm max}(n,\epsilon)$ and for
the event horizon radius $r_h$ being bounded from above by
some value $\tilde{r}_h(\kappa)$, which tends to zero for
$\kappa\to\kappa_{\rm max}$. For a given $n$ and $r_h< \tilde{r}_h(\kappa)$
there are again two different solutions, which coalesce as
$r_h\to \tilde{r}_h(\kappa)$. For  $r_h> \tilde{r}_h(\kappa)$
only the Schwarzschild solution is possible.
All known regular and black hole solutions of the doublet
EYMH theory are unstable \cite{Boschung94,Winstanely95,Mavromatos96}.

The picture described above remains qualitatively the same for any
$\epsilon<\infty$. For $\epsilon\to\infty$ the Higgs field
becomes frozen, $\phi=1$, and, if the phase of the Higgs field
is still fixed according to (\ref{4c.1}), one obtains a simpler theory of
gravitating massive non-Abelian gauge field.
This is sometimes called the Einstein-non-Abelian-Proca model.
\cite{Greene93,Maeda94,Torii95,Tachizawa95}.
The conical singularity at the origin for regular
solutions can be avoided in this case by choosing the minus sign
in (\ref{4c.6}). As a result, solutions for even values of the
node number $n$ are globally regular. For odd values of $n$ solutions
contain the conical singularity, as in the monopole case.
All regular solutions for $r>0$, as well as their black hole counterparts,
look qualitatively similar to those for $\epsilon<\infty$.
In particular, they exhibit the same bifurcation picture.
This model has also been studied in the context
of the Brans-Dicke theory \cite{Tamaki97}.

If the constraint (\ref{4c.1}) is abandoned then
the Higgs field is allowed to rotate in the internal space
spanning a unit sphere $S^3$. The EYMH theory reduces then
to the gravitating gauged O(4) sigma-model, whose solutions
are called local textures.
The equations of motion in the static, spherically symmetric and
purely magnetic case can be obtained by varying the reduced
Lagrangian (\ref{2:28}) with $\phi=1$.
Note that for $\xi'\neq 0$ one cannot set to zero
the radial component of the gauge field $a_r$.
We are unaware of any curved space solutions in this case, while
those without gravity were studied in \cite{Kunz89,Yaffe90}.
For {\em global} gravitating textures static solutions with finite
energy cannot exist, however, there are non-trivial time-dependent
solutions \cite{Durer91}, which can be important in the theory of
fluctuations of cosmic microwave background \cite{Turok90}.

\subsection{Gravitating Skyrmions}

The gravitating Skyrme model provided one of the first indications
of the existence of hairy black holes.
In \cite{Luckock86} it was found that the equation for the chiral field
on the Schwarzschild background admitted a regular solution. This
can be regarded as approximately describing a black
hole with Skyrme hair.
In \cite{Luckock87} the self-consistent gravitating problem
was treated numerically, which work for some reasons remained
almost unknown. In \cite{Glendenning88}
self-gravitating Skyrmions with higher winding numbers were
studied as candidates for soliton stars, but no stable solutions
were found. The first systematic investigation of the problem
was undertaken in \cite{Droz91,Heusler91,Heusler92,HeuslerHPA93}.
This revealed the existence of {\em stable} regular
gravitating Skyrmions and Skyrme black holes. The solution
space of the models was studied also in \cite{BizonSkyrm}.
The thermodynamics of Skyrme black holes was considered in \cite{Torii93}.

The equations of motion of the Einstein-Skyrme model following from the
2D Lagrangian (\ref{2:31}) read
\bea
m'=N\left(\frac{r^2}{2}+\sin^2\chi\right)\chi^{\prime 2}
&+&\left(r^2+\frac12\sin^2\chi\right)\frac{\sin^2\chi}{r^2},   \label{4d.1}\\
\sigma'&=&\kappa\sigma\left(r+\frac2r\,
\sin^2\chi\right)\chi^{\prime 2},                            \label{4d.2}\\
\left(\sigma N\left(r^2+2\sin^2\chi\right)\chi'\right)'
&=&\sigma\left(1+N\chi'^2
+\frac{\sin^2\chi}{r^2}\right)\sin 2\chi,                    \label{4d.3}
\eea
where $N=1-2\kappa m/r$ and $\kappa=4\pi Gf^2$. The length scale is given by
${\rm L}=1/ef$,
and the dimensionful energy is $(4\pi f/e)M$ with $M=m(\infty)$.
The function $\sigma$ can be eliminated from the equations.
The boundary conditions for the regular solutions read
\be                                                         \label{4d.4}
\chi=\pi\nu-b\,r+O(r^3),\quad 2m=b^2(1+b^2)\,r^3+O(r^5),
\ee
\be                                                        \label{4d.5}
\chi=a\, r^{-2}+O(r^{-3}),\quad m=M+O(r^{-3}),
\ee
at the origin and at infinity, respectively.
Here $b$, $a$, and $M$ are free parameters and $\nu$ is integer.

The flat space Skyrme model can be regarded
as an effective theory in the low energy
limit of QCD \cite{Skyrme61,Skyrme62,Witten83}.
The particle-like solutions are interpreted as baryons,
and the integer parameter $\nu$ in (\ref{4d.4})
then plays the role of the baryon number.
Note that $\nu$ has the meaning of the topological winding
number. This is because the description of the chiral field in terms
of the SU(2) valued function U leads to the mapping
$S^3\to SU(2)$, provided that the asymptotic condition (\ref{5.2})
is imposed. The winding number is then given by Eq.(\ref{5.3}),
$\nu={\bf k}[{\rm U}]$,
which formula reduces in  the spherically symmetric case to
$\nu=\{\chi(0)-\chi(\infty)\}/\pi$.
For $\nu=1$ the flat space solution
was obtained numerically in \cite{Jackson83}. This solution is stable.
Solutions for higher winding numbers are also known, but they turn out
to be unstable. There are also stable solutions for $\nu>1$, but these
are not spherically symmetric.
It is worth mentioning the role  of the higher order term $F^2$ in
the Skyrme Lagrangin in (\ref{2:29}). Due to this term, the theory contains
both attractive and repulsive interactions, which leads to the
existence of non-trivial solutions.

%%%%%%%%%%%%%%%%%%%%%%%%%%%%%%%%%%%%%
\begin{figure}
\hbox to\hsize{%\hss
  \epsfig{file=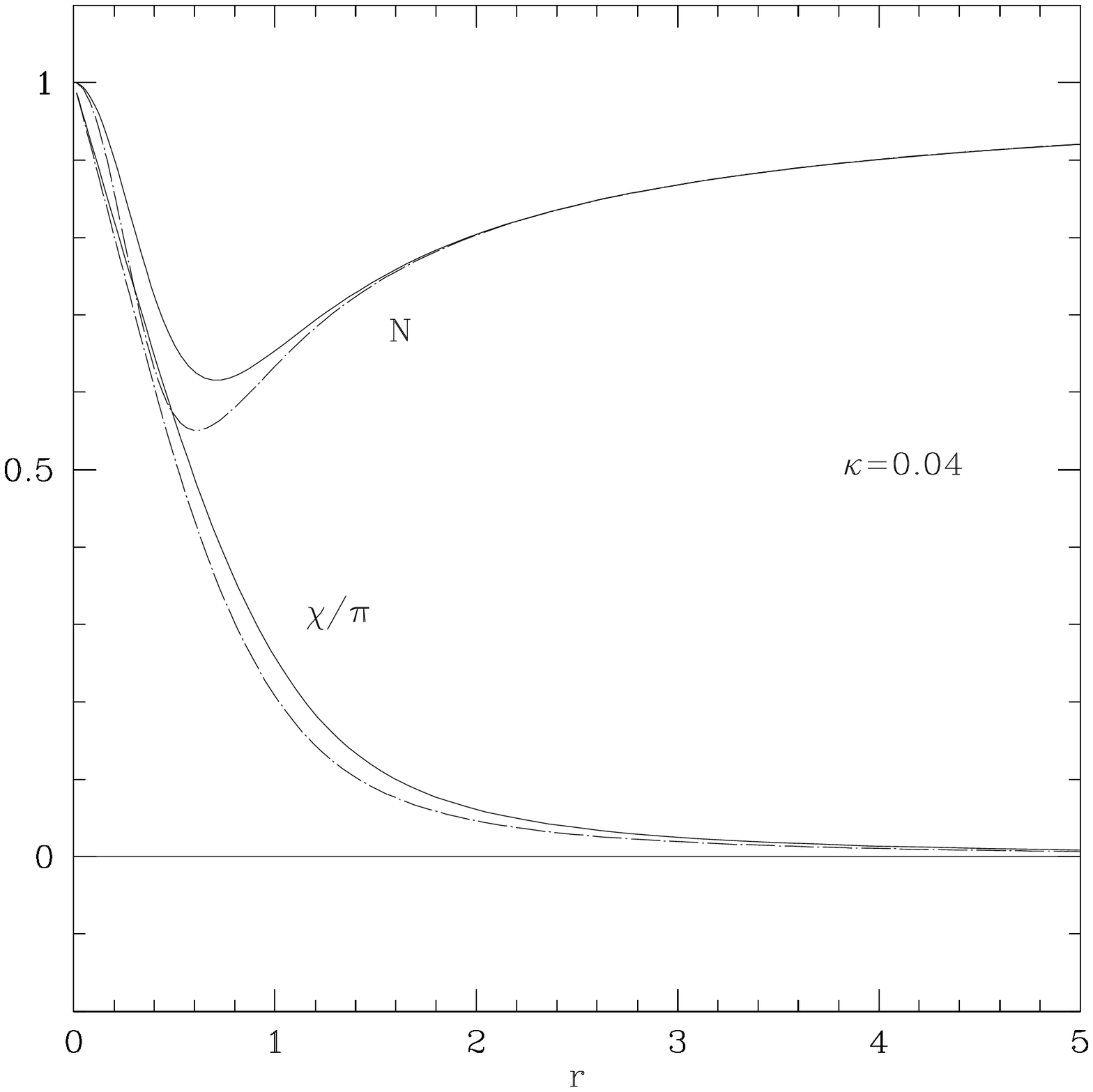,width=0.48\hsize,%
      bbllx=1.8cm,bblly=5.5cm,bburx=20.0cm,bbury=20.0cm}\hss
  \epsfig{file=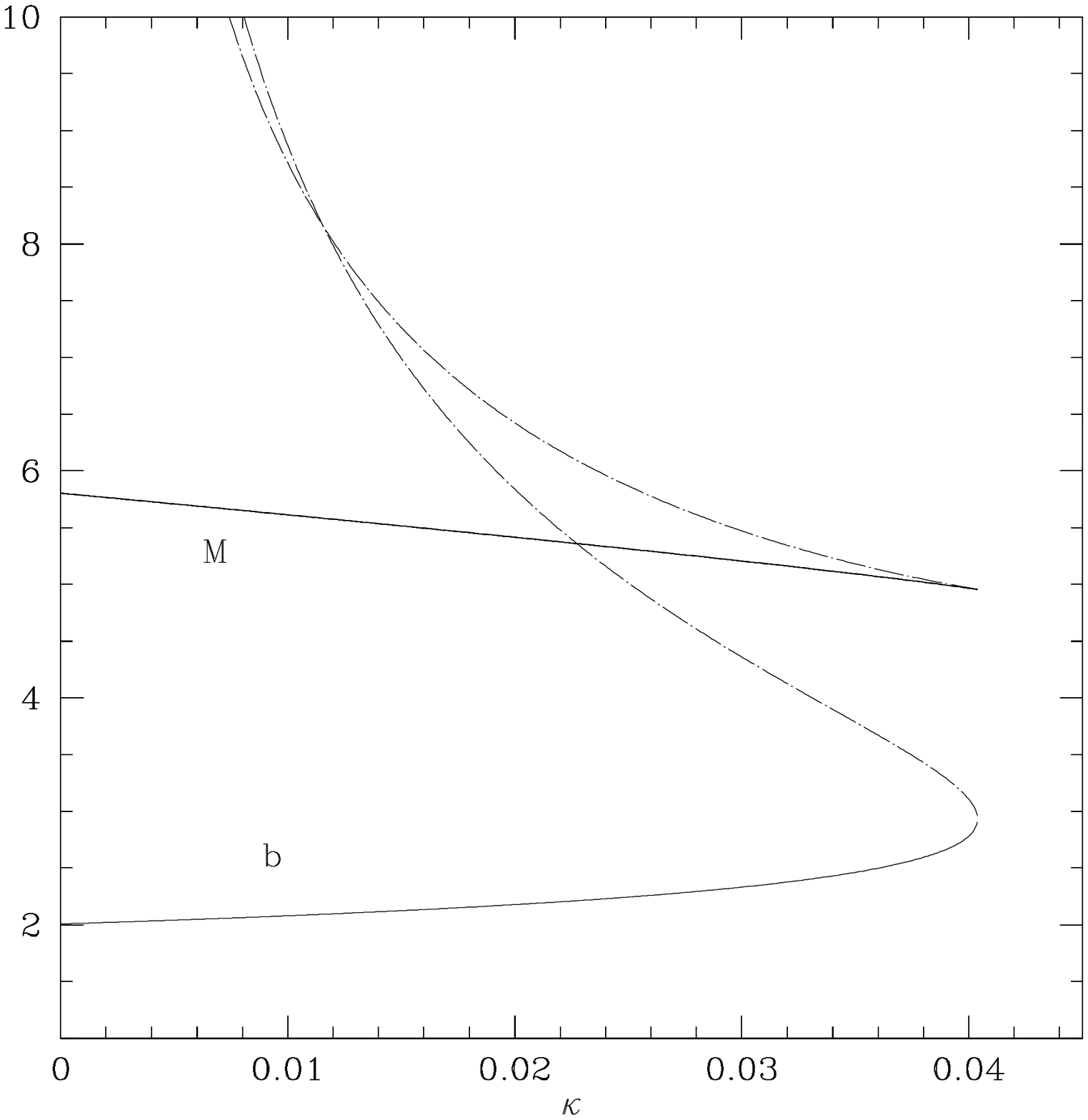,width=0.48\hsize,%
     bbllx=1.8cm,bblly=5.5cm,bburx=20.0cm,bbury=20.0cm}%\hss
  }
\caption{The $\nu=1$ fundamental Skyrmion (solid lines)
and its excitation (dashed lines).
On the left: the chiral field $\chi(r)/\pi$ and the metric function
$N(r)$ for the two solutions with $\kappa=0.04$. For
$\kappa\to \kappa_{\rm max}=0.0437$ the solid and dashed curves
merge. On the right: the mass $M$ and the parameter $b$
in Eq.(\ref{4d.4})  versus $\kappa$
for the lower and upper branches of solutions.
}
\label{fig:SKYRMEON}
\end{figure}
%%%%%%%%%%%%%%%%%%%%%%%%%%%%%%%%%%%%%

Consider solutions to (\ref{4d.1})--(\ref{4d.3}) for
$\kappa\neq 0$. The numerical analysis reveals that the
gravitating Skyrmions exhibit the same typical features
as the solutions of the EYMH models
\cite{Droz91,HeuslerHPA93,BizonSkyrm,Torii93}.
In particular, since
Skyrmions have no charge, their behaviour is very similar to that
for the EYMH sphalerons. For small values of $\kappa$
there is the fundamental branch of solutions with $\nu=1$
that reduce to the flat space Skyrmion as $\kappa\to 0$ \cite{Droz91}.
For these solutions $\chi$, $m$, and $\sigma$ are monotone
functions,  while $N$ develops one minimum at some $r_m$.
These solutions are stable \cite{Heusler91}.
Let us call this branch of solutions lower branch.
Next, for $\nu=1$ there are also excited solutions with larger mass,
which look qualitatively similar
(see Fig.\ref{fig:SKYRMEON}) \cite{BizonSkyrm,HeuslerHPA93}.
In particular, $\chi$ is still monotone.
For $\kappa\to 0$ the excitations become infinitely heavy and,
remarkably, similar to the situation in the EYMH case,
reduce to the rescaled BK solution with $n=1$.
Specifically, under
\be                                               \label{4d.6}
r\to\sqrt{\kappa}\,r,\quad m\to m /\sqrt{\kappa},\quad
\cos\chi\to\f
\ee
the system (\ref{4d.1})--(\ref{4d.3}) reduces in the limit
$\kappa\to 0$ to the EYM field equations, whose solutions are
the BK sphalerons. Let us call the branch of excited solutions
upper branch.

The lower and upper branches for $\nu=1$ exist only for
$\kappa\leq\kappa_{\rm max}$. In the limit
$\kappa\to\kappa_{\rm max}$ the two branches
coalesce, exactly in the same way as this happens for the EYMH sphalerons.
No regular solutions exist for $\kappa>\kappa_{\rm max}$.
At the bifurcation point the stability behaviour of the solutions
changes: all solutions of the upper branch are unstable.
This change of stability agrees with the general considerations
based on the catastrophe theory
\cite{Torii93,Maeda94,Torii95,Tachizawa95}.
For higher winding numbers, $\nu>1$,
the picture is essentially the same: there are two
branches of solutions merging at some $\kappa_{\rm max}(\nu)$.
The value
$\kappa_{\rm max}(\nu)$ decreases rapidly with growing $\nu$.
For $\kappa\to 0$ the lower branch reduces to the flat space
Skyrmion with $\nu>1$, while the upper branch again tends to the rescaled
BK solution; more details can be found in \cite{BizonSkyrm,HeuslerHPA93}.
All solutions with $\nu>1$ are unstable. It is unknown whether
the lower branch solutions with $\nu>1$ admit
stable and non-spherically symmetric generalizations.

All solutions described above have black hole analogues.
These exist for $\kappa<\kappa_{\rm max}(\nu)$
and if the event horizon
radius $r_h$ does not exceed a maximal value $\tilde{r}_h(\kappa)$.
The lower and upper branches of
solutions exist in the black hole case too, and
reduce to those of the regular case for $r_h\to 0$.
In the opposite limit, $r_h\to \tilde{r}_h(\kappa)$,
the two branches coalesce. No black holes
apart from the Schwarzschild solution exist  for
$r_h>\tilde{r}_{h}(\kappa)$.
The maximal value $\tilde{r}_h(\kappa)$ reduces to zero as $\kappa$
tends to $\kappa_{\rm max}(\nu)$.

To recapitulate, the critical values $\kappa_{\rm max}(\nu)$
for the regular solutions and $\tilde{r}_h(\kappa)$ with
$\kappa<\kappa_{\rm max}(\nu)$ for black holes
correspond to bifurcation
points where the lower and upper branches merge.
These values determine the region of the parameter space
$(\kappa,r_h)$
for which non-trivial solutions exist. A similar bifurcation
picture arises in the doublet EYMH theory. It is worth noting
that the solutions specified by the critical values
are perfectly regular. In particular, the minimal
value of $N$ is non-zero. In the monopole case, on the other hand,
the limiting solutions exhibit
special geometrical structures%
\footnote{The charged EYMH black holes with $\kappa<3/4$
show the usual bifurcation picture.}.

Note finally that, apart from the solutions described above,
for which $\chi$ is monotone,
there are also those for which $\chi$
oscillates around the value $\pi/2$. These exist for very
small values of $\kappa$ \cite{BizonSkyrm,HeuslerHPA93}.
The solutions of the gravitating SU(2)
Skyrme model have been generalized to the gauge group SU(3),
in which case their essential features remain the same \cite{Kleihaus95}.
The critical collapse of Skyrmions was considered in \cite{Bizon98}.

\section{Concluding remarks}
\setcounter{subsection}{1}
\setcounter{equation}{0}

We have given a fairly complete account of known up-to-date
solitons and black holes in the basic gravity-coupled models with
gauge and/or scalar fields which respect the non-Abelian symmetries.
These solutions show that some of the fundamental concepts
of the Abelian Einstein-Maxwell theory do not extend to the
non-Abelian case. Instead,  a number of new typical features arise.
One of such features
is the appearance of the characteristic discrete structures
due to the interaction of the YM field with gravity. These
manifest in the BK solutions and their various generalizations.
For those solitons which exist in the flat spacetime limit,
one observes the spectrum of gravitational BK-type excitations.
Another typical phenomenon, which arises
in theories containing a length scale other than the Planck scale,
is the existence of the upper bound
for the gravitational coupling constant $\kappa$
and for the event horizon radius $r_h$.
Non-trivial solutions can exist only for small values of
$\kappa$ and $r_h$. For those approaching the bound,
the solutions either become gravitationally closed
or develop bifurcations, where different
branches of solutions merge. Such a merging is accompanied
by the change of the stability properties, which can be interpreted
in terms of catastrophe theory.

The existence of static and non-spherically
symmetric solutions, so far demonstrated at the non-linear level only  in the
EYM theory, is probably also typical for gravitating non-Abelian
models. This is supported by the perturbative analysis
for the quite general non-linear model (\ref{weinberg})
carried out in \cite{Ridgway95b}.
Such deformed solutions can be regarded as the limit of
multi-soliton/black hole configurations when
the separation tends to zero.

A generic feature associated with the  the interior structure of non-Abelian
black holes is the absence of Cauchy horizons. Let us remind in this
connection the basic idea behind the mass-inflation scenario
\cite{Poisson90}. The Cauchy horizon
inside a Kerr-Newman black hole, unlike the event horizon, exists
solely due to the high symmetry
of the solution. In the generic situation, when deviations from
the strict axial symmetry are allowed, inner horizons do not exist.
Following this idea, one can argue in the same spirit that
an equally efficient way to get rid of
Cauchy horizons is to add into the system more general matter.
This is confirmed by the examples of the non-Abelian black holes.

For the sake on completeness, let us mention briefly also some
other important solutions for gravitating Yang-Mills fields
which remained outside the main text, because they are
not of soliton or black hole type.
Most of these are obtained in the pure EYM theory.

\subsubsection{EYM cosmologies}
The basic idea is as follows. For the
SO(4)-invariant spacetime metric
\be                                                  \label{10.1}
ds^2=a^2(t)\{dt^2-dr^2
-\sin^2 r\, (d\vartheta^2+\sin^2\vartheta\, d\varphi^2)\},
\ee
with $r\in[0,\pi]$, one can find the SO(4)-invariant YM field.
This can be obtained either as a solution of the symmetry
conditions (\ref{2.4}) \cite{Henneaux82} or by requiring that
the stress tensor for the spherically symmetric gauge field (\ref{2.5})
respect the SO(4) symmetry \cite{Galtsov91}. The result is
\be                                                       \label{10.2}
A=
i\, \frac{1-\f(t)}{2}\, {\rm  {U}}\, d\, {\rm  {U}}^{-1},
\ \ \ {\rm where}\ \ \
{\rm  {U}}=\exp(ir\, \T_r).
\ee
The YM equations reduce to
\be                                                    \label{10.3}
\dot{\f}^2+(\f^2-1)^2={\cal E},
\ee
where ${\cal E}$ is an integration constant.
In view of the conformal invariance, the conformal
factor of the metric, $a(t)$, does not enter the YM equations.
For any solution to (\ref{10.3}) the stress tensor has the
perfect fluid structure with $\rho=3p=3{\cal E}/2a^4$,
which allows one 
\cite{Cervero78,Henneaux82,Hosotani84,Sinzinkayo85,Sinzinkayo86,Galtsov91}
to solve the Einstein equations.

The above solutions can be generalized to spatially open and spatially flat
cosmological models \cite{Henneaux82,Galtsov91}, as well as to
higher gauge groups \cite{Bertolami91a,Moniz91,Moniz93,Rudolph97}.
The corresponding superspace quantization was considered,
for example, in \cite{Bertolami91,Cavaglia94}.
Since the metric in (\ref{10.1}) is conformally flat and the YM
equations are conformally invariant, Eq.(\ref{10.3}) gives rise also
to the flat space solutions for the YM field describing
collapsing shells of the non-linear YM radiation
\cite{Luscher77,Farhi93,Gibbons95}.
The homogeneous and anisotropic EYM cosmological models were considered in
\cite{Darian96,Darian97,Barrow98},
in which case the solutions turn out to be chaotic.
The inhomogeneous and spherically symmetric case
was considered in \cite{Shchigolev97}.

\subsubsection{Cosmological sphaleron}
Eq.(\ref{10.3}) describes a particle in the double-well
potential $V(\f)=(\f^2-1)^2$.
For ${\cal E}=1$ there is a non-trivial static solution
for the particle sitting on the top of the barrier separating the two
wells, $\f=0$ \cite{Hosotani84}.
This solution was considered in \cite{Gibbons94,Ding96} as a model
of collapsing BK solitons. Since the bottoms of the well,
$\f=\pm 1$, correspond to pure gauge fields with different
winding numbers, which is obvious from (\ref{10.2}), the
static configuration with  $\f=0$ can be naturally interpreted as
sphaleron \cite{Gibbons94,Ding94}. This sphaleron solution
is distinguished  by the fact
that it consists of the pure gauge field alone, which is possible
due to the interaction with the background gravitational field.
Another characteristic feature of the solution is its high
(SO(4)) symmetry. This implies that the configuration is not localized
and fills the whole universe.

Such a sphaleron solution can be used
to evaluate the fermion production rate in a closed universe.
This problem was considered in \cite{Volkov96,VolkovHPA} in the
approximation when the universe described by the metric
(\ref{10.1}) is static.
It was assumed that the universe is filled with (quazi)-thermal YM quanta
corresponding to excitations over the trivial YM vacuum with $\f=1$,
while the sector with unit winding number is empty.
As a result, there is a net diffusion of the field modes between
the two sectors, which is accompanied by the change in the fermion
number due to the axial anomaly. The corresponding diffusion rate
was computed in \cite{Volkov96,VolkovHPA} at the one-loop level.

\subsubsection{EYM instantons}
When continued to the imaginary time, the
 cosmological solutions described above fulfill the Euclidean EYM
equations
\cite{Verbin89,Hosoya89,Verbin90,Bertolami91a,%
Yoshida90,Rey90,Donets92}.
The metric (\ref{10.1}) then becomes conformal to the standard
metric on $S^4$, while the potential $V(\f)$
in the YM equation (\ref{10.3})
changes sign. Solutions with finite Euclidean action are
interpreted as tunneling geometries leading to the creation of
baby universes. Note that, unless the constant  ${\cal E}$
in Eq.(\ref{10.3}) vanishes,
the YM field is non-self dual.
In the self-dual case
one has $F_{\mu\nu}=\ast\! F_{\mu\nu}$ implying that the stress
tensor is zero. As a result, in order to obtain EYM instantons
with  self-dual gauge fields one can start from a
vacuum gravitational instanton and use it as a fixed
background on which the YM self-duality equations are solved.
Since topology of gravitational instantons can be quite arbitrary,
solutions for self-dual YM fields can differ substantially from those
in flat space
\cite{Charap77,Charap77a,Pope78,Duff78,Boutaleb79,Boutaleb79a,%
Boutaleb80,Boutaleb80a,Boutaleb80b,Chakrabarti87}.
It was shown in \cite{Witten79} that the  self-duality
equations for the YM field in axially symmetric case
in flat spacetime are equivalent to the Ernst equations in GR.

The results mentioned above, together with those discussed
in the main text, almost exhaust the
list of 4D solutions for gravitating YM fields.
Relatively
little is known about solutions with symmetries other than spherical.
At the same time, we do not discuss here solutions for $D\neq 4$,
since a separate review would be
necessary for this
(see \cite{Deser84,Kunzle93,Brindejonc98} for some results in $D=3$).
A large number of such solutions can be obtained
from the ten-dimensional heterotic five-brane via various
compactifications \cite{Duff95}.

\vspace{2 mm}

\begin{ack}
M.S.V. would like to thank Othmar Brodbeck,
Marcus Heusler,
and Norbert Straumann for
discussions and the reading of the manuscript.
His work was supported by the
Swiss National Science Foundation and by the Tomalla Foundation.
D.V.G. thanks the Yukawa Institute for Theoretical Physics,
where a part of the work was done, for hospitality
and acknowledges the support of the COE and
the RFBR Grant 96--02--18899.
\end{ack}

%\newpage
%\bibliography{j}
\end{document}